\documentclass[prd,nofootinbib,onecolumn,preprint,11pt]{revtex4}

\usepackage{graphicx}
\usepackage{bm}
\usepackage{float}
\usepackage{subfigure}
\usepackage{amsmath}
\usepackage{amsfonts}
\usepackage{color}
\usepackage{slashed}
\usepackage{fancyhdr}

\newcommand{\GN}{G_{\rm N}}
\newcommand{\gb}{\bar{g}}

\newcommand{\Tr}[1]{\text{Tr}\left[ #1 \right]}

\newcommand{\sbar}{\bar{\sigma}}

\newcommand{\dd}{\text{d}}

\newcommand{\Laplacian}{\vec{\nabla}^2}

\begin{document}

\title{More On Cosmological Gravitational Waves And Their Memories}

\author{Yi-Zen Chu}
\affiliation{
Department of Physics, University of Minnesota, \\
1023 University Dr., Duluth, MN 55812, USA
}

\begin{abstract}
\noindent We extend recent theoretical results on the propagation of linear gravitational waves (GWs), including their associated memories, in spatially flat Friedmann--Lema\^{i}tre--Robertson--Walker (FLRW) universes, for all spacetime dimensions higher than 3. By specializing to a cosmology driven by a perfect fluid with a constant equation-of-state $w$ -- conformal re-scaling, dimension-reduction and Nariai's ansatz may then be exploited to obtain analytic expressions for the graviton and photon Green's functions, allowing their causal structure to be elucidated. When $0 < w \leq 1$, the gauge-invariant scalar mode admits wave solutions, and like its tensor counterpart, likely contributes to the tidal squeezing and stretching of the space around a GW detector. In addition, scalar GWs in 4D radiation dominated universes -- like tensor GWs in 4D matter dominated ones -- appear to yield a tail signal that does not decay with increasing spatial distance from the source. We then solve electromagnetism in the same cosmologies, and point out a tail-induced electric memory effect. Finally, in even dimensional Minkowski backgrounds higher than 2, we make a brief but explicit comparison between the linear GW memory generated by point masses scattering off each other on unbound trajectories and the linear Yang-Mills memory generated by color point charges doing the same -- and point out how there is a ``double copy" relation between the two.
\end{abstract}

\maketitle

\tableofcontents

\section{Introduction and Motivations}
\label{Section_Introduction}
Humanity has, after several decades of effort, finally achieved the direct detection of gravitational waves \cite{Abbott:2016blz, Abbott:2016nmj}. Given that we expect to detect gravitational wave (GW) events from astrophysical systems located at cosmological distances -- even the first two events, GW150914 and GW151226, were produced at non-trivial redshifts $0.05 \lesssim z \lesssim 0.13$ -- it is desirable to examine from first principles how GWs propagate over cosmological spacetimes, by solving Einstein's equations perturbatively about the background spatially flat Friedmann--Lema\^{i}tre--Robertson--Walker (FLRW) universe we reside in. Such a strategy is to be contrasted against the asymptotically flat spacetime calculations from which GW observables are usually extracted. To be very scrupulous when predicting GW events at cosmological distances, as already pointed out in \cite{Chu:2015yua}, one has to match the `far zone' predictions of these asymptotically flat calculations onto a near/intermediate zone cosmological one, before evolving the wave solutions out to infinity. Furthermore, one may also wonder if these GWs journeying over a size-able fraction of our observable universe could experience the expansion of space itself, as well as interact with the intervening inhomogeneities -- thereby picking up additional features that could in turn provide us with novel probes of dark energy and dark matter.

In this paper, we will continue our effort initiated in \cite{Chu:2013xca} to understand the causal structure of gravitational (and electromagnetic) signals in cosmological geometries of all spacetime dimensions greater or equal to four. We will do so by reducing the problem to $1+1$ and $2+1$ dimensions, so that the portion of waves that travel strictly on the null-cone can be cleanly separated from the part that does so less than unit speed (also known as the ``tail"). Specifically, we will work in a $(d \geq 4)$--dimensional background cosmology sourced by a perfect fluid with a constant equation-of-state (EoS) $w$, so that the resulting geometry is
\begin{align}
\label{SpatiallFlatFLRW_Metric}
\gb_{\mu\nu} = a[\eta]^2 \eta_{\mu\nu} , \qquad\qquad
\eta_{\mu\nu} = \text{diag}[1,-1,\dots,-1] ,
\end{align}
with Cartesian-like coordinates $x^\mu \equiv (\eta,x^i) \equiv (\eta,\vec{x})$. The scale factor describing the size of the universe is, in turn,
\begin{align}
\label{SpatiallFlatFLRW_Metric_w}
a[\eta] \equiv \left( \frac{\eta}{\eta_0} \right)^{ \frac{2}{q_w} }, \qquad\qquad q_w \equiv (d-3) + (d-1)w .
\end{align}
Because we wish to consider expanding universes, whenever $w > -(d-3)/(d-1)$ (i.e., $2/q_w > 0$),
\begin{align}
\eta \in (\eta_\circ,\eta_\text{future}) \equiv (0,\infty) ;
\end{align}
and whenever $w < -(d-3)/(d-1)$ (i.e., $2/q_w < 0$),
\begin{align}
\eta \in (\eta_\circ,\eta_\text{future}) \equiv (-\infty,0) . 
\end{align}
The $w=-(d-3)/(d-1)$ case will not be considered.

We will then place an isolated astrophysical system in such a geometry, and seek the metric perturbations it engenders, by solving the resulting linearized Einstein's equations in terms of appropriate Green's functions convolved against components of its energy-momentum-shear-stress. In other words, we shall solve for the $\chi_{\mu\nu}$ in the perturbed metric
\begin{align}
\label{SpatiallFlatFLRW_PerturbedMetric}
g_{\mu\nu} = a[\eta]^2 \left( \eta_{\mu\nu} + \chi_{\mu\nu} \right) 
\end{align}
due to the presence of this astrophysical system. Since we are working only to first order in metric perturbations, it is physically fruitful to employ the well-developed gauge-invariant cosmological linear perturbation theory\footnote{Cosmological perturbation theory was first tackled by Lifshitz \cite{Lifshitz1946}. See Mukhanov's \cite{Mukhanov:2005sc} and Weinberg's \cite{Weinberg:2008zzc} textbooks for pedagogical treatments.} because once the solutions are non-trivial, they cannot be rendered trivial simply by transforming to a ``nearby" coordinate system -- and therefore ought to play key roles in characterizing physical observables such as the strain measured by a GW detector. In \cite{Chu:2015yua}, where we stumbled upon a tail-induced GW memory effect (to be explained shortly) in de Sitter and matter dominated universes, only the 4-dimensional (4D) gauge-invariant tensor mode was analyzed in detail. In the follow-up work of \cite{Chu:2016qxp}, we solved the linear metric perturbations in all de Sitter backgrounds higher than $3$ dimensions, but did not employ gauge-invariant variables. More recently, Tolish and Wald \cite{Tolish:2016ggo} have extended the 4D GW memory analysis to the full set of gauge-invariant scalar, vector and tensor metric variables. In this paper, while we specialize to a constant$-w$ cosmology from the outset, we shall tackle not only the gauge-invariant scalar-vector-tensor perturbations but also provide analytic solutions for the relevant Green's functions in all dimensions higher than three.

One key GW observable is the fractional displacement between a pair of test masses that comprises part of a hypothetical experiment. It is convenient to assume that these test masses are freely falling and co-moving in some given spacetime, so their spatial coordinates $\vec{Y}_0$ and $\vec{Z}_0$ are time independent. As we demonstrate in appendix \eqref{Section_GeodesicDistance},\footnote{The derivation in \cite{Chu:2016qxp} suffers from deficiencies, which we fix in appendix \eqref{Section_GeodesicDistance}.} if we denote the proper spatial distance between the masses at a fixed time as $L[\eta]$, and if we began our GW experiment at time $\eta'$, then the fractional displacement $(\delta L/L_0)[\eta] \equiv (L[\eta]-L[\eta'])/L[\eta']$ is given by the following integral of the total change in the spatial metric expressed in synchronous gauge:\footnote{By synchronous gauge, we mean the coordinate system such that the perturbations are purely spatial, namely $\chi_{\mu\nu} \dd x^\mu \dd x^\nu \to \chi_{ij} \dd x^i \dd x^j$.}
\begin{align}
\label{FractionalDistortion_AtConstantCosmicTime_FirstOrderHuman}
\left(\frac{\delta L}{L_0}\right)[\eta > \eta'] 
= - \frac{\widehat{n}^i \widehat{n}^j}{2} \int_{0}^{1} \Delta\chi_{ij} \dd \lambda 
+ \mathcal{O}\left[ (\chi_{mn})^2,(\eta-\eta') \dot{a}[\eta']/a[\eta'] \right],
\end{align}
where here and throughout the rest of the paper an overdot always denotes a derivative with respect to the time coordinate being used; $\widehat{n} \equiv (\vec{Z}_0 - \vec{Y}_0)/|\vec{Z}_0 - \vec{Y}_0|$ is the unit radial vector pointing from one mass to the other; and
\begin{align}
\label{FractionalDistortion_AtConstantCosmicTime_Deltachiij}
\Delta \chi_{ij}
\equiv 
\chi_{ij}\left[ \eta, \vec{Y}_0 + \lambda \left(\vec{Z}_0-\vec{Y}_0\right) \right] 
- \chi_{ij}\left[ \eta', \vec{Y}_0 + \lambda \left(\vec{Z}_0-\vec{Y}_0\right) \right] .
\end{align}
(The argument $\vec{Y}_0 + \lambda (\vec{Z}_0-\vec{Y}_0)$ describes the Euclidean straight line joining $\vec{Y}_0 \leftrightarrow \vec{Z}_0$.)

Suppose the primary wave train of some GW event has just passed our detector. GW memory is the phenomenon where $\chi_{ij}$ does not return to its original value at $\chi_{ij}[\eta',\vec{x}]$, but instead to a `DC shift' such that $\Delta \chi_{ij} \to C_{ij}$, a constant matrix, in eq. \eqref{FractionalDistortion_AtConstantCosmicTime_Deltachiij}. In such a scenario, eq. \eqref{FractionalDistortion_AtConstantCosmicTime_FirstOrderHuman} then informs us that the fractional distortion becomes permanent:
\begin{align}
\label{GWMemory_Def}
\left(\frac{\delta L}{L_0}\right)_{\text{memory}} \approx - \frac{\widehat{n}^i \widehat{n}^j}{2} C_{ij} .
\end{align}
By placing pairs of test masses with different orientations $\{\widehat{n}\}$, and by comparing the pair-separations before and after the passage of the GW signal, we may thus detect whether such a permanent fractional distortion of space itself has occurred.\footnote{Recently, a new gravitational ``spin memory" effect has been uncovered by Pasterski, Strominger and Zhiboedov in \cite{Pasterski:2015tva}, where the passage of a GW induces a difference in travel time between a light beam circulating on some closed path vs. a counter-orbiting one on the same path.}

The GW memory effect is not the only reason motivating us to solve the perturbed cosmological system at hand. Since \cite{Chu:2013xca} we have been motivated by the possibility that future-generation GW detectors such as LISA may hear GWs emanating from the strong field regime of super-massive black holes residing within the central core of galaxies. While the unperturbed Kerr black hole (BH) does not radiate, compact objects that orbit and plunge into its horizon do stir up vibrations of spacetime that carry energy-momentum to infinity. There is some hope that, when the compact object is much less massive than the BH itself, this ``Extreme-Mass-Ratio-Inspiral" (EMRI) system can be tackled analytically by expanding in powers of $m/M_{\text{BH}}$, where $m$ and $M_\text{BH}$ are respectively the compact body and BH masses. In the strong gravity region near the black hole horizon, and at least within the Lorenz gauge condition, compact bodies exert a tail-induced self-force that needs to be modeled properly for GW detectors to successfully track EMRI systems over $\mathcal{O}[M_\text{BH}/m]$ orbital cycles \cite{Poisson:2011nh}. Additionally, GWs from compact binary systems, modeled by the well-established weak field post-Newtonian/Minkowskian (PN/PM) framework, are also known to scatter off their own spacetime curvature and develop tails. Therefore, in this new era of GW astronomy, the tail phenomenon in 4D curved spacetimes is no longer a mere intellectual curiosity; but understanding it properly could lead to observable consequences. To this end, we hope it will prove instructive to first understand thoroughly the gravitational and electromagnetic Green's functions in the simpler context of cosmology.

It is worth pointing out that such an indirect detection of the tail effect -- through its impact on the compact bodies' dynamics, which in turn would alter the resulting EMRI/PN/PM GW-forms -- is a test of our understanding of classical field theory in curved spacetime. For, since the pioneering work of Hadamard \cite{Hadamard}, it is known that Green's functions in 4D curved spacetimes, at least whenever the observer at $x$ and the spacetime-point-source at $x'$ can be joined by a unique geodesic, take the generic form
\begin{align}
\label{Hadamard}
G[x,x'] = U[x,x'] \delta_+[\sigma_{x,x'}] + V[x,x'] \Theta_+[\sigma_{x,x'}] ,
\end{align}
where the Dirac $\delta$-function term describes the strictly-null signals and the (step-) $\Theta$-function term the tail signals.\footnote{The assumption that observer and source can always be linked via a unique geodesic in fact breaks down in black hole spacetimes. Null geodesics zipping about near the black hole horizon admit caustics and do not have unique solutions. As such, the causal structure of BH Green's functions will be richer than that in eq. \eqref{Hadamard} -- see, for instance, Casals and Nolan \cite{Casals:2016qyj} for recent work.} (Here, $\sigma_{x,x'}$ is proportional to the square of the geodesic distance between $x$ and $x'$.) The key insight is that: while understanding the tail function $V$ requires the full wave equation to be solved, the null-cone parts $U$ and $\sigma$ only require the geodesics of the background spacetime to be known. This lends itself to the physical interpretation that classical wave tails arise from the long wavelength modes, because it is these low frequency waves that are sensitive to the background spacetime curvature.

We begin, in \S\eqref{Section_LiteratureReview}, with a review of and commentary on the extant literature on the GW memory effect in cosmology. Following that, in \S\eqref{Section_Setup}, we describe in detail the cosmological setup we are considering in this paper, including the background metric and fluid solutions upon which we are going to carry out perturbation theory. The technical core of the paper lies in \S\eqref{Section_PT}, where we will carry out first order gauge-invariant perturbation theory by first computing the equations-of-motion (EoM) of the metric variables before presenting their analytic solutions in terms of convolution of appropriate Green's functions against the astrophysical source. Potential GW memory effects, in both the scalar and tensor sectors, will be pointed out. In \S \eqref{Section_GaugeInvariantElectromagnetism}, we carry out an analogous gauge-invariant analysis for electromagnetic fields in cosmological spacetimes, and work out potential electric memory effects. We summarize our findings and discuss future directions in \S \eqref{Section_Summary}. In appendix \S \eqref{Section_LinearizedEinstein}, Einstein's equations are linearized about flat $(d \geq 4)$-dimensional spacetime and solved in both the gauge-invariant and de Donder gauge formalism. The latter leads us to the Li\'{e}nard-Wiechert metric perturbation due to a point mass in motion, and its pure tail nature in odd dimensions is highlighted. In \S \eqref{Section_EMandLinearizedYM}, a similar line of pursuit is carried out for electromagnetism and linearized Yang-Mills theory -- culminating in the comparison between the linear gravitational memory due to point masses scattering off each other on unbound trajectories and the linear Yang-Mills memory due to point color charges doing the same. For completeness we also solve Maxwell's equations in a generalized Lorenz gauge in constant$-w$ spatially flat cosmologies of all dimensions $d \geq 4$. In \S \eqref{Section_GeodesicDistance}, we re-derive the fractional displacement formula in eq. \eqref{FractionalDistortion_AtConstantCosmicTime_FirstOrderHuman}, and attempt to carefully lay out the assumptions that lead up to it. In \S \eqref{Section_PDESolution}, the solution to the primary partial differential equation encountered in this paper is reviewed, and a caution is issued regarding its global incompleteness whenever $q_w > 0$ (cf. eq. \eqref{SpatiallFlatFLRW_Metric_w}). Finally, in \S \eqref{Section_DimRaisingOperator}, we explain from an algebraic perspective why $-(2\pi r)^{-1}\partial_r$ can be viewed as a dimension-raising operator in flat spacetime.

\section{Brief Literature Review on Linear GW Memory in Cosmology}
\label{Section_LiteratureReview}
In this section we shall attempt to provide, in chronological order, a brief review of and commentary on the extant literature on linear GW memory in a spatially flat FLRW cosmology.

{\bf Chu (2015)} \qquad While GW memory around a flat geometric background has been studied since the works of Zel'dovich and Polnarev \cite{Zel’Dovich1974}, Braginsky and Grishchuk \cite{Braginsky1985}, Braginsky and Thorne \cite{BraginskyThorne}, Christodoulou \cite{Christodoulou:1991cr}, Thorne \cite{Thorne1992}, etc. -- as far as we are aware -- it was our recent work in \cite{Chu:2015yua} that first examined it in the cosmological context. In particular, we uncovered a novel contribution to linear memory arising from GWs propagating inside the null cone. This latter ``tail" phenomenon finds no counterpart in linear massless fields propagating in asymptotically flat 4D spacetimes. Specifically, in 4D de Sitter spacetime, because the Green's function tail of the spin$-2$ graviton (i.e., the transverse-traceless portion of the spatial metric) is a spacetime constant, we pointed out that this lead to a tail-induced linear memory that does not decay with increasing distance nor increasing time from the source(s)' world-tube in spacetime. Moreover, our work in \cite{Chu:2015yua} showed that the light cone portion of the spin$-2$ GWs takes on the following universal form regardless of the details of cosmic evolution (cf. eq. (28) of \cite{Chu:2015yua}):
\begin{align}
\label{GW_NullCone}
D^{(\gamma)}_{ij}[\eta,\vec{x}] 
= 4 \GN \int_{\mathbb{R}^3} \dd^3 \vec{x}' a[\eta_r]^3 \frac{\Pi_{ij}^{\text{(T)}}[\eta_r,\vec{x}']}{a[\eta] |\vec{x}-\vec{x}'|} .
\end{align}
Here, we have employed Weinberg's notation \cite{Weinberg:2008zzc}, where $D_{ij}$ denotes the transverse-traceless (TT) part of $\chi_{ij}$; $\Pi_{ij}^{\text{(T)}}$ is the (negative) TT part of the GW source's shear-stress tensor written in a co-moving orthonormal frame; and $\eta_r \equiv \eta - |\vec{x}-\vec{x}'|$ is the retarded time. We went on to argue that the $1/a$ redshift in eq. \eqref{GW_NullCone} is the dominant effect cosmic expansion has on gravitons propagating in a spatially flat FLRW universe; because the GW tail amplitude is suppressed relative to its null-cone counterpart of eq. \eqref{GW_NullCone} by the duration of the GW source to the age of the universe, as well as the observer-source spatial distance to the size of the observable universe.

{\bf Bieri-Garfinkle-Yau (2015)} \qquad A few months after \cite{Chu:2015yua} was posted on the arXiv, Bieri, Garfinkle and Yau \cite{Bieri:2015jwa} tried to employ the techniques Bieri and Garfinkle developed in \cite{Bieri:2013ada} to study the GW memory effect in de Sitter spacetime. The bulk of \cite{Bieri:2015jwa} consisted of solving the linearized Weyl tensor sourced by a null matter stress-energy tensor. Following that, they integrated a certain leading piece of the electric portion of the Weyl tensor to obtain the memory effect. Because the Weyl tensor vanishes in any conformally flat spacetime such as de Sitter, this means the Weyl tensor is gauge-invariant at first order in perturbations. Bieri, Garfinkle and Yau \cite{Bieri:2015jwa} had probably hoped that this would thus provide a gauge-invariant formalism to investigate linear GW memory in cosmology. However, they did not justify why the Minkowski background and zero cosmological constant formalism of \cite{Bieri:2013ada} could be carried over to cosmological backgrounds. In particular, why the Weyl tensor alone is sufficient to determine GW memory was not explained.

The standard treatment of GW memory appeals to the geodesic deviation equation. Along a timelike geodesic whose tangent vector we shall denote as $U^\mu$, the displacement vector $\epsilon^\mu$ joining it to an infinitesimally close-by geodesic obeys
\begin{align}
\label{GeodesicDeviation_Riemann}
U^\alpha U^\beta \nabla_\alpha \nabla_\beta \epsilon^\mu = - R^\mu_{\phantom{\mu}\nu \alpha\beta} U^\nu \epsilon^\alpha U^\beta ,
\end{align}
where $R^\mu_{\phantom{\mu}\nu \alpha\beta}$ is the Riemann tensor.\footnote{Our Riemann tensor is $R^\alpha_{\phantom{\alpha}\beta \mu\nu} = \partial_{[\mu} \Gamma^\alpha_{\phantom{\alpha}\nu]\beta} + \Gamma^\alpha_{\phantom{\alpha}\sigma[\mu} \Gamma^\sigma_{\phantom{\sigma}\nu]\beta}$, Ricci tensor is $R_{\beta\nu} = R^\sigma_{\phantom{\alpha}\beta \sigma\nu}$, Ricci scalar is $\mathcal{R} = g^{\sigma\rho} R_{\sigma\rho}$, and the Christoffel symbols are $\Gamma^\mu_{\phantom{\mu}\alpha\beta} = (1/2)g^{\mu\sigma} (\partial_{\{\alpha} g_{\beta\}\sigma} - \partial_\sigma g_{\alpha\beta})$. Einstein summation convention is in force; Latin/English alphabets run over the spatial indices $1$ through $d-1$ while Greek ones run from $0$ (time) through $d-1$. Symmetrization of indices are denoted as, for e.g., $T_{\{\alpha\beta\}} \equiv T_{\alpha\beta}+T_{\beta\alpha}$; and anti-symmetrization as $T_{[\alpha\beta]} \equiv T_{\alpha\beta}-T_{\beta\alpha}$.} To arrive at eq. \eqref{GeodesicDeviation_Riemann} one makes use of
\begin{align}
[\nabla_\mu,\nabla_\nu] \epsilon^\alpha = R^\alpha_{\phantom{\alpha}\beta \mu\nu} \epsilon^\beta, \qquad\qquad
U^\sigma \nabla_\sigma U^\mu = 0 \qquad\qquad \text{ and } \qquad\qquad [U,\xi] = 0 .
\end{align}
Integrating eq. \eqref{GeodesicDeviation_Riemann} allows one to solve for the fractional change in the spatial geodesic distance between a given pair of timelike geodesics. GW memory occurs when this fractional change becomes permanent after the passage of the primary GW train. Now, the Riemann tensor $R_{\alpha \beta \mu \nu}$ itself can be decomposed into its trace-free (i.e., the Weyl tensor $C_{\alpha \beta \mu \nu}$) plus trace-parts, namely
\begin{align}
\label{Riemann_To_Weyl}
R_{\alpha \beta \mu \nu} &= C_{\alpha \beta \mu \nu} +
\frac{1}{d-2} \left( R_{\alpha [\mu} g_{ \nu] \beta } - R_{\beta [\mu} g_{ \nu]
	\alpha } \right) - \frac{1}{(d-1)(d-2)} g_{\alpha [\mu} g_{\nu] \beta}
\mathcal{R} .
\end{align}
In eq. \eqref{Riemann_To_Weyl}, if we employ Einstein's equation with a non-zero cosmological constant $\Lambda$ -- namely, $R_{\mu\nu} - (g_{\mu\nu}/2) \mathcal{R} = \Lambda g_{\mu\nu} + 8\pi\GN T_{\mu\nu}$ -- to re-express the Ricci-tensors $R_{\alpha\beta}$ and Ricci-scalar $\mathcal{R}$ into terms involving only the cosmological constant $\Lambda$ and the energy-momentum-shear-stress tensor of matter $T_{\alpha\beta}$, the Riemann tensor itself will be translated into
\begin{align}
\label{Riemann_WeylCCMatter}
R^\alpha_{\phantom{\alpha}\beta \mu\nu}
&= C^\alpha_{\phantom{\alpha}\beta \mu \nu} 
- \frac{2 \Lambda}{(d-2)(d-1)} \delta^{\alpha}_{\phantom{\alpha}[\mu} g_{\nu]\beta} \nonumber \\
&\qquad\qquad
+ \frac{8 \pi \GN}{d-2} \left( T^{\alpha}_{\phantom{\alpha}[\mu} g_{\nu]\beta} - T_{\beta[\mu} \delta_{\nu]}^{\alpha} - \delta^\alpha_{[\mu} g_{\nu]\beta} \frac{2 T}{d-1} \right) ,
\qquad\qquad T \equiv g^{\rho\sigma} T_{\rho\sigma} .
\end{align}
By inserting the form of the Riemann tensor in eq. \eqref{Riemann_WeylCCMatter} into the geodesic deviation eq. \eqref{GeodesicDeviation_Riemann}, we see that -- even after taking into account the dynamics of General Relativity -- tidal forces are not encoded solely within the Weyl tensor in generic geometries. In the 4D de Sitter spacetime Bieri-Garfinkle-Yau \cite{Bieri:2015jwa} was considering, while $T_{\alpha\beta}=0$ as long as the GW detectors are assumed to be far away from the GW source(s), the cosmological constant $\Lambda$ is crucial for its very existence and therefore cannot be zero. In generic spatially flat FLRW cosmologies, say a matter or radiation dominated universe, the evolution of the scale factor is driven by a non-trivial background perfect fluid, so that even if one neglects $\Lambda$, the $T_{\alpha\beta}$ terms in eq. \eqref{Riemann_WeylCCMatter} cannot be discarded.

It is possible that the cosmological constant $\Lambda$ and matter $T_{\alpha\beta}$ terms in eq. \eqref{Riemann_WeylCCMatter} are sub-dominant relative to the Weyl tensor term in the cosmological scenarios of interest -- but this we advocate, ought to be justified more carefully before the results in \cite{Bieri:2015jwa},\cite{Bieri:2013ada} can be meaningfully applied/extended to cosmological GW memory.

{\bf Kehagias-Riotto (2016)} \qquad Earlier this year, Kehagias and Riotto \cite{Kehagias:2016zry} extended the Strominger and Zhiboedov \cite{Strominger:2014pwa} work in 4D asymptotically flat spacetimes, to draw a connection between the Bondi-van der Burg-Metzner-Sachs (BMS) symmetries at null infinity in decelerating but expanding spatially flat FLRW universes to GW memories in the same cosmologies. They used the retarded Green's functions derived by Iliopoulos et al. \cite{Iliopoulos:1998wq} to first compute, in a matter dominated universe, the TT tensor mode produced by $n$ point masses that collide at a spacetime point before moving apart on separate trajectories; finding (like we did in \cite{Chu:2015yua}) a light cone part that, apart from the $1/a$ redshift in eq. \eqref{GW_NullCone}, is very similar to its Minkowski counterpart -- as well as a space-independent tail signal. Then, they went on to integrate a variant of the geodesic deviation equation to find the fractional distortion between two test masses after the passage of a non-trivial GW ``Bondi news function". The latter could, in turn, be gotten via a BMS transformation of the perturbed spatially flat FLRW metric near null infinity.

{\bf Chu (2016)} \qquad In \cite{Chu:2016qxp} the theoretical observation in \cite{Chu:2015yua}, that the gauge-invariant tensor GW-tail of 4D de Sitter (dS) is a spacetime constant, was extended to all higher even dimensional dS$_{4+2n}$. Building on the work by Ashtekar, Bonga, and Kesavan \cite{Ashtekar:2015lxa} as well as Date and Hoque \cite{Date:2015kma}, we expressed the tensor waveform in terms of the mass and pressure quadrupole moments of its isolated astrophysical source, allowing us to measure the magnitude of the tail-induced memory effect directly in terms of the latter's net change over the course of the GW generation process.\footnote{Ashtekar, Bonga, and Kesavan \cite{Ashtekar:2015lxa} did state that ``{\it \dots the presence of the tail term opens a door to a new contribution to the `memory effects' associated with gravitational waves}," but it is somewhat puzzling to us why they did not go on to quantify this in the rest of the paper, since they were dealing with essentially the same set of equations as ours.} On a technical level, we also confirmed that conformal re-scaling, dimension reduction, and Nariai's ansatz allow the $(d \geq 4)$-dimensional Green's functions of the full set of linear de Sitter graviton fields, i.e., not just the $D_{ij}$, to be derived explicitly.

The result in eq. \eqref{GW_NullCone} was reiterated in \cite{Chu:2016qxp} to argue that all known asymptotically 4D Minkowskian GW memory effects should carry over to the asymptotically spatially flat FLRW case, with the additional feature that these on-the-null-cone spin$-2$ GWs will be further diluted as the universe expands, due to the scale factor in the $1/(a[\eta]|\vec{x}-\vec{x}'|)$ fall-off. Furthermore, in a matter dominated spatially flat FLRW spacetime, the spin$-2$ graviton tail is constant in space but decays in time over cosmological timescales; this leads to a corresponding linear GW memory that is approximately spacetime constant over timescales of human experiments.

{\it Remark} \qquad This is an appropriate place to highlight that the portion of the GW memory effect that propagates on the null-cone, really arises from a non-zero difference between the asymptotic matter configuration after and before the time period of active GW production. For concreteness, let $\eta_\star$ denote the cosmic time when GW production peaked. Referring to eq. \eqref{GW_NullCone}, and assuming the GW detector is located far enough from the system such that $|\vec{x}-\vec{x}'| \approx |\vec{x}|$ -- with the spatial coordinate axes centered within the astrophysical system at $\eta_\star$ -- we then see that its TT contribution to the GW memory formulas of equations \eqref{FractionalDistortion_AtConstantCosmicTime_FirstOrderHuman} and \eqref{FractionalDistortion_AtConstantCosmicTime_Deltachiij} is
\begin{align}
\label{GWMemory_NullCone}
\Delta D^{(\gamma)}_{ij}[\eta,\vec{x}] 
\approx \frac{4 \GN}{a[\eta] |\vec{x}|} \Delta \int_{\mathbb{R}^3} \dd^3 \vec{x}' a^3 \Pi_{ij}^{\text{(T)}} ,
\end{align}
where the $\Delta (\dots)$ terms refer to the net change in the spatial-total of (the TT part of) the shear-stress of the GW source:
\begin{align}
\Delta \int_{\mathbb{R}^3} \dd^3 \vec{x}' a^3 \Pi_{ij}^{\text{(T)}}
&\equiv 
\lim_{\eta_r \gg \eta_\star} \int_{\mathbb{R}^3} \dd^3 \vec{x}' a[\eta_\star]^3 \Pi_{ij}^{\text{(T)}}[\eta_r,\vec{x}'] 
- \lim_{\eta_r \ll \eta_\star} \int_{\mathbb{R}^3} \dd^3 \vec{x}' a[\eta_\star]^3 \Pi_{ij}^{\text{(T)}}[\eta_r,\vec{x}'] .
\end{align}
\footnote{These formulas assume the matter source $\Pi_{ij}^{\text{(T)}}$ is sufficiently localized that the far zone limit may be taken; this is could be an issue because $\Pi_{ij}^{\text{(T)}}$, being the TT portion of the (local) stress-energy tensor of the same matter source, is a non-local function of the latter.}Here and throughout the rest of the paper -- by $\eta_r \approx \eta-|\vec{x}| \gg \eta_\star$ we really mean some retarded time well after peak GW production, but not long enough for the scale factor to have changed appreciably, consistent with the timescales of human experiments; likewise, $\eta_r \approx \eta-|\vec{x}| \ll \eta_\star$ means some retarded time well before the GW event itself, but not long enough that the scale factor itself can still be approximated as $a[\eta_\star]$.

{\bf Tolish-Wald (2016)} \qquad Most recently, Tolish and Wald \cite{Tolish:2016ggo} examined potential GW memory effects in a generic spatially flat FLRW spacetime due to a setup similar to the one in Kehagias and Riotto \cite{Kehagias:2016zry}. Specifically, they considered $N$ incoming geodesic point particles that meet at a single spacetime event, followed by $N'$ outgoing geodesic point particles emerging from that same event. (Their $N$ is not necessarily equal $N'$.) After examining the partial differential equations for the first order gauge-invariant scalar, vector and tensor metric variables, they proceeded to argue that only the (TT) tensor metric perturbation would give rise to a derivative of a Dirac $\delta$-function in the Riemann curvature tensor linearized about eq. \eqref{SpatiallFlatFLRW_Metric}. This, they asserted, shows that only the tensor mode exhibits GW memory in their setup. They also argued that GW tails do not yield memory effects in their setup. While this $\delta'$-criterion is understandably mathematically convenient, we feel it obscures the physics behind GW memory that propagates on the null/acoustic-cone: as emphasized above, it is due to the difference between the GW source configuration in the asymptotic future versus its configuration in the asymptotic past; whereas, the detailed process of active GW production is relevant only through its impact on the asymptotic future trajectories of the astrophysical system.

\section{Setup: Background Spatially Flat Cosmology with Constant EoS $w$}
\label{Section_Setup}
Throughout this paper, we will consider an isolated astrophysical system moving about in a background spatially flat Friedmann-Lema\^{i}tre-Robertson-Walker (FLRW) universe driven by a perfect fluid of constant equation-of-state $w$. We wish to examine the GW signatures produced by this astrophysical system as detected by a hypothetical experiment located at distances large compared to the system's characteristic size. The total action of such a system is
\begin{align}
\label{Setup_TotalAction}
S_{\text{Einstein-Hilbert}} + S_{\text{Fluid}} + S_{\text{Astro}} .
\end{align}
The gravitational dynamics is governed by the Einstein-Hilbert action
	\begin{align}
	\label{Setup_EinsteinHilbertAction}
	S_{\text{Einstein-Hilbert}} \equiv -\frac{1}{16 \pi \GN} \int \dd^d x \sqrt{|g|} \mathcal{R} ,
	\end{align}
	with $\GN$ being Newton's constant in $d$ spacetime dimensions, $\sqrt{|g|}$ is the square root of the absolute value of the determinant of the spacetime metric $g_{\mu\nu}$, and $\mathcal{R}$ is the Ricci scalar of the same geometry.\footnote{Since every term in the Ricci scalar contains exactly $2$ derivatives and the action is dimensionless, $\GN$ itself must be of dimensions [Length]$^{d-2}$ or equivalently [1/Mass]$^{d-2}$. We have, evidently, also set $\hbar = c = 1$. The Levi-Civita symbol $\epsilon_{\sigma_1 \dots \sigma_d}$ is zero whenever there are repeated indices, but otherwise returns the sign of the permutation bringing $\{0,1,\dots,d-1\}$ to $\{ \sigma_1 , \dots, \sigma_d \}$; in particular, $\epsilon_{(0)(1)\dots(d-1)}=1$. The covariant Levi-Civita tensor is defined as $\widetilde{\epsilon}_{\sigma_1 \dots \sigma_d} \equiv \sqrt{|g|} \epsilon_{\sigma_1 \dots \sigma_d}$. In what follows, the ``square" of spatial vectors always means its Euclidean dot product with itself; for e.g., $\vec{k}^2 \equiv \delta^{ij} k_i k_j$, where $\delta^{ij}$ is the Kronecker delta. \\ Even though it is increasingly irresponsible to do so -- due to the mounting astrophysical/cosmological observations that point to its existence -- we have set the cosmological constant to zero in this paper. For, having too many scales in a given problem makes analytic solutions difficult to come by.}

The perfect fluid dynamics is described by a Lagrangian density that coincides with (negative of) its energy density $\rho_\text{f}$,
	\begin{align}
	\label{Setup_FluidAction}
	S_{\text{Fluid}} \equiv - \int \dd^d x \sqrt{|g|} \rho_{\text{f}}\left[n^2/2\right] , \qquad\qquad n^2 \equiv n^\mu n_\mu .
	\end{align}
	\footnote{In this paper we have adopted the effective field theory (Lagrangian-space) description of fluids in Dubovsky et al. \cite{Dubovsky:2011sj}, which was then applied to the cosmological context by Ballesteros and Bellazzini \cite{Ballesteros:2012kv}. Alternatively, our approach to perfect fluid dynamics can be found pedagogically explained in Andersson and Comer \cite{Andersson:2006nr}. In any case, the Lagrangian density in eq. \eqref{Setup_FluidAction} is to be viewed as the first term in an infinite-series gradient-expansion.}The prime denotes a derivative with respect to the argument; while the entropy current $n^\mu$ is built out of $(d-1)$--scalar fields $\{ \varphi^\text{I} \vert \text{I} = 1,2,3,\dots,d-1 \}$ that labels the elements of the fluid itself in its co-moving frame. If $\widetilde{\epsilon}^{\mu_1 \dots \mu_d}$ is the covariant Levi-Civita tensor,
	\begin{align}
	\label{Setup_EntropyCurrent}
	n^\mu \equiv \widetilde{\epsilon}^{\mu \sigma_1 \dots \sigma_{d-1}} \partial_{\sigma_1} \varphi^1 \dots \partial_{\sigma_{d-1}} \varphi^{d-1} .
	\end{align}
	By construction, $\nabla_\mu n^\mu = 0 = \nabla_\mu \varphi^{\text{I}} n^\mu$ are identities. 
	
	That the Lagrangian density in eq. \eqref{Setup_FluidAction} governs the dynamics of a perfect fluid can be seen by obtaining the expected form of its stress energy tensor, which we shall denote as $\,^{\text{(f)}}T^{\alpha\beta}$. By varying $S_{\text{Fluid}}$ with respect to the metric, i.e., replace $g_{\mu\nu} \to g_{\mu\nu} + \delta g_{\mu\nu}$, $\,^{\text{(f)}}T^{\alpha\beta}$ is then the coefficient of $-(1/2) \sqrt{|g|} \delta g_{\alpha\beta}$:
	\begin{align}
	\label{PerfectFluid_StressTensor_Raw}
	\,^{\text{(f)}}T^{\alpha\beta}
	= g^{\alpha\beta} \left( \rho_{\text{f}}\left[n^2/2\right] - \rho_{\text{f}}'\left[  n^2/2\right] n^2 \right) 
	+ \rho_{\text{f}}'\left[n^2/2\right] n^\alpha n^\beta ,
	\end{align}
	with the prime on $\rho_\text{f}$ denoting a derivative with respect to its argument. In the co-moving frame, and provided $n^\mu$ is timelike, $n^{\widehat{\mu}}/\sqrt{n^2} = \delta^\mu_0$ and
	\begin{align}
	\label{PerfectFluid_CoMovingStressTensor}
	\,^{\text{(f)}}T^{\widehat{0}\widehat{0}} = \rho_{\text{f}}\left[n^2/2\right], \qquad\qquad
	\,^{\text{(f)}}T^{\widehat{0}\widehat{i}} = 0 , \qquad\qquad
	\,^{\text{(f)}}T^{\widehat{i}\widehat{j}} = \delta^{ij} \left( \rho_{\text{f}}'\left[  n^2/2\right] n^2 - \rho_{\text{f}}\left[n^2/2\right] \right) ;
	\end{align}
	allowing us to identify the fluid pressure density $p_\text{f}$ as
	\begin{align}
	\label{PerfectFluid_Pressure}
	p_\text{f} = \rho_{\text{f}}'\left[  n^2/2\right] n^2 - \rho_{\text{f}}\left[n^2/2\right] .
	\end{align}
	Here, because pressure depends only on the energy density and none other thermodynamic variables, we are neglecting other possible conserved quantities that might exist in a more general fluid.

	If the perfect fluid has a constant pressure to energy density ratio $w$, we will shortly explain why the energy/mass density then takes the form
	\begin{align}
	\label{PerfectFluid_PowerLawLagrangian}
	\rho_\text{f}[n^2/2] = \frac{(d-2)(d-1)}{4\pi\GN (q_w \eta_0)^2} \left(n^2\right)^{\frac{w+1}{2}} .
	\end{align}
	The stress tensor of the fluid in eq. \eqref{PerfectFluid_StressTensor_Raw}, in turn, simplifies to
	\begin{align}
	\label{PerfectFluid_StressTensor}
	\,^{\text{(f)}}T_{\alpha\beta} 
	= \left( - g_{\alpha\beta} \cdot w + (w+1) \frac{n_\alpha n_\beta}{n^2} \right) \rho_\text{f}[n^2/2] .
	\end{align}
	When $w=-1$, notice from equations \eqref{PerfectFluid_PowerLawLagrangian} and \eqref{PerfectFluid_StressTensor} that the perfect fluid stress-tensor takes the form of the cosmological constant term in Einstein's equations. Specifically, $8\pi\GN \,^{\text{(f)}}T_{\alpha\beta} = \Lambda g_{\alpha\beta}$, with $\Lambda = (d-2)(d-1)/(2 \eta_0^2) > 0$. This means our linearized solutions for $w=-1$ below should really be viewed as the first-order gauge-invariant form of the linearized de Sitter solutions we obtained in \cite{Chu:2016qxp} by fixing a gauge.
	
Finally, we will be largely agnostic about the details of the astrophysical system, as encoded within $S_{\text{Astro}}$.

{\bf Equations-of-Motion (EoM)} \qquad By varying the total action in eq. \eqref{Setup_TotalAction} with respect to the metric, one obtains Einstein's equations for the cosmological system:
\begin{align}
\label{Setup_EinsteinEquations_Full}
G_{\mu\nu}[g] - 8\pi\GN \,^{(\text{f})}T_{\mu\nu} = 8\pi\GN \,^{(\text{a})}T_{\mu\nu} ,
\end{align}
where $G_{\mu\nu} = R_{\mu\nu} - (g_{\mu\nu}/2) \mathcal{R}$ is Einstein's tensor, $\,^{(\text{f})}T_{\mu\nu}$ is the energy-momentum-shear-stress tensor of the fluid in eq. \eqref{PerfectFluid_StressTensor_Raw} and $\,^{(\text{a})}T_{\mu\nu}$ is that of the astrophysical system described by $S_{\text{Astro}}$. 

The fluid dynamics is gotten by varying its action in eq. \eqref{Setup_FluidAction} with respect to each and every scalar $\{ \varphi^\text{I} \}$. This yields the $(d-1)$ equations, with $\text{I} \in \{ 1,2,3,\dots,d-1 \}$:
\begin{align}
\label{Setup_PerfectFluidEOM}
\partial_\sigma \left( \rho_\text{f}'[n^2/2] n_\mu \epsilon^{\mu \alpha_1 \alpha_2 \dots \alpha_{\text{I}-1} \sigma \alpha_{\text{I}+1} \dots \alpha_{d-1}} 
\partial_{\alpha_1} \varphi^1 \dots \partial_{\alpha_{\text{I}-1}} \varphi^{\text{I}-1} \partial_{\alpha_{\text{I}+1}} \varphi^{\text{I}+1} \dots \partial_{\alpha_{d-1}} \varphi^{d-1} \right) = 0,
\end{align}
where the $\epsilon^{\mu \alpha_1 \alpha_2 \dots \alpha_{\text{I}-1} \sigma \alpha_{\text{I}+1} \dots \alpha_{d-1}}$ is the Levi-Civita symbol with no dependence on the metric and $\epsilon^{(0)(1)(2)\dots(d-1)} \equiv (-)^{d-1}$.

In principle, the astrophysical system $S_{\text{Astro}}$ obeys its own equations-of-motion as well.

{\bf Background Solutions: Constant EoS $w$} \qquad Since we are primarily interested in how GWs generated by isolated astrophysical systems propagate over cosmological distances, we shall begin by ignoring the astrophysical contribution in eq. \eqref{Setup_EinsteinEquations_Full} and solving the ``background" Einstein-perfect fluid equations
\begin{align}
\label{Setup_EinsteinEquations_Background}
G_{\mu\nu}[\gb] - 8\pi\GN \,^{(\text{f})}T_{\mu\nu}[\gb] = 0 .
\end{align}
Because we are seeking analytic solutions to GW systems, in this paper, we shall simplify the cosmological perturbation theory below by studying a perfect fluid with a constant equation-of-state (EoS) $w$. Eq. \eqref{PerfectFluid_CoMovingStressTensor} tells us
\begin{align}
\label{PerfectFluid_EoS}
\frac{\,^{\text{(f)}}T^{\widehat{1}\widehat{1}}}{\,^{\text{(f)}}T^{\widehat{0}\widehat{0}}} 		= \dots = \frac{\,^{\text{(f)}}T^{\widehat{(d-1)}\widehat{(d-1)}}}{\,^{\text{(f)}}T^{\widehat{0}\widehat{0}}}
= - 1 + 2 \frac{\rho_{\text{f}}'\left[  n^2/2\right]}{\rho_{\text{f}}\left[n^2/2\right]} \frac{n^2}{2}
= \frac{p_\text{f}}{\rho_\text{f}} \equiv w \qquad\qquad \text{(constant)} .
\end{align}
The general solution of this first order ordinary differential equation for $\rho_\text{f}$ in terms of $n^2/2$ is
\begin{align}
\label{PerfectFluid_PowerLawLagrangian_draft}
\rho_\text{f}[n^2/2] = \rho_0 \left(\frac{n^2}{2}\right)^{ \frac{w+1}{2} }, \qquad\qquad (\rho_0 \text{ constant}) .
\end{align}
On the other hand, eq. \eqref{Setup_EinsteinEquations_Background} tells us the ratio of the diagonal space-space component of Einstein's tensor to its $00$ component must also equal $w$. If we insert the spatially flat, homogeneous and isotropic geometry of eq. \eqref{SpatiallFlatFLRW_Metric} (but for a generic scale factor $a[\eta]$) into the background Einstein-fluid eq. \eqref{Setup_EinsteinEquations_Background},
\begin{align}
\frac{G_{11}}{G_{00}} = \dots = \frac{G_{(d-1)(d-1)}}{G_{00}} 
= - \frac{(d-5) \dot{a}^2 + 2 a \ddot{a}}{(d-1)\dot{a}^2}
= \frac{p_\text{f}}{\rho_\text{f}} = w .
\end{align}
The general solution is $a[\eta] = ((\eta-\eta_\odot)/\eta_0)^{2/q_w}$, where $\eta_\odot$ corresponds to the cosmic time where the universe has zero size if $q_w > 0$ or infinite size if $q_w < 0$. For technical convenience, we will set it to zero, $\eta_\odot = 0$. This recovers the cosmic evolution in eq. \eqref{SpatiallFlatFLRW_Metric_w}.

Through a direct calculation, one may verify the background solution $\overline{\varphi}^\text{I}$ of the fluid eq. \eqref{PerfectFluid_EoS} to be
\begin{align}
\label{PerfectFluid_BackgroundSoln}
\overline{\varphi}^\text{I} = x^\text{I} .
\end{align}
For, when eq. \eqref{PerfectFluid_BackgroundSoln} holds, the entropy current eq. \eqref{Setup_EntropyCurrent} evaluated on equations \eqref{SpatiallFlatFLRW_Metric} is
\begin{align}
\label{PerfectFluid_Backgroundn}
\bar{n}^\mu = (-)^{d-1} a^{-d} \delta^\mu_0 ,
\end{align}
and $n^2 = n_\mu n^\mu$ evaluated on the same gives
\begin{align}
\label{PerfectFluid_Backgroundn2}
n^2 \to \bar{n}^2 \equiv \gb_{\mu\nu} \bar{n}^\mu \bar{n}^\nu = a^{2(1-d)} .
\end{align}
With equations \eqref{PerfectFluid_BackgroundSoln}, \eqref{PerfectFluid_Backgroundn} and \eqref{PerfectFluid_Backgroundn2}, the quantity inside the derivative $\partial_\sigma(\dots)$ in eq. \eqref{Setup_PerfectFluidEOM} then becomes independent of spatial coordinates and proportional to $\delta^\sigma_\text{I}$ because of the fully anti-symmetric nature of the Levi-Civita symbol. Eq. \eqref{Setup_PerfectFluidEOM} is thus satisfied. Plugging eq. \eqref{PerfectFluid_Backgroundn2} into eq. \eqref{PerfectFluid_PowerLawLagrangian_draft} then allows us to fix the $\rho_0$ there through the background Einstein's equation \eqref{Setup_EinsteinEquations_Background}; namely, setting the Einstein tensor built from equations \eqref{SpatiallFlatFLRW_Metric} and \eqref{SpatiallFlatFLRW_Metric_w} to be equal to $8\pi\GN$ times the stress tensor of eq. \eqref{PerfectFluid_PowerLawLagrangian_draft} in eq. \eqref{PerfectFluid_StressTensor} then hands us eq. \eqref{PerfectFluid_PowerLawLagrangian}.

\section{First Order Gauge-Invariant Cosmological Perturbation Theory In Real Space}
\label{Section_PT}
With the background metric in eq. \eqref{SpatiallFlatFLRW_Metric} and fluid solution in eq. \eqref{PerfectFluid_BackgroundSoln} understood, we will now introduce the astrophysical system $\,^{(\text{a})}T_{\mu\nu}$ in eq. \eqref{Setup_EinsteinEquations_Full} as a first order perturbation. Expand the spacetime metric about its background one in eq. \eqref{SpatiallFlatFLRW_Metric}, as in eq. \eqref{SpatiallFlatFLRW_PerturbedMetric}; and expand the fluid scalar fields away from their equilibrium values in eq. \eqref{PerfectFluid_BackgroundSoln},
\begin{align}
\label{PerfectFluid_ExpansionFromEq}
\varphi^\text{I} = x^\text{I} + \Pi^{\text{I}} .
\end{align} 
In this section, we will proceed to tackle the linearized gravitational system
\begin{align}
\label{Setup_EinsteinEquations_FirstOrder}
\Delta_{\mu\nu} \equiv \left(G \vert 1\right)_{\mu\nu} - 8\pi\GN \,^{(\text{f})}\left(T \vert 1\right)_{\mu\nu} = 8\pi\GN \,^{(\text{a})}T_{\mu\nu} ,
\end{align}
with $\left(G \vert 1\right)_{\mu\nu}$ and $\,^{(\text{f})}\left(T \vert 1\right)_{\mu\nu}$ representing the parts of Einstein's tensor and stress energy of the perfect fluid that are linear in the metric $\{ \chi_{\mu\nu} \}$ and fluid perturbation $\{ \Pi^\text{I} \}$ variables. The fluid eq. \eqref{Setup_PerfectFluidEOM} will be similarly linearized about its background one evaluated on eq. \eqref{PerfectFluid_BackgroundSoln}.

As already advertised in the introduction, we shall employ the first order gauge-invariant formalism commonly used in the physically important case of 4D cosmology. For our purposes, gauge-invariance means the quantities we are computing do not alter their form under an infinitesimal change in the coordinate system. This ensures that, whatever physical effect we associate with some metric/fluid variable -- once it is found to be non-zero in one coordinate system, it cannot be made trivial merely by switching to a ``nearby" one. 

On a more technical level, let us make the key observations that, since the left hand side of Einstein's equations in eq. \eqref{Setup_EinsteinEquations_Full} is zero at the background level -- because we are treating the astrophysical system as a first order perturbation -- that means its first order counterpart $\Delta_{\mu\nu}$ in eq. \eqref{Setup_EinsteinEquations_FirstOrder} must be gauge-invariant. (This will be elaborated upon below.) Likewise, the left hand side of the fluid EoM in eq. \eqref{Setup_PerfectFluidEOM} is, too, zero at the background level $\overline{\varphi}^\text{I} = x^\text{I}$ and must also be gauge-invariant at linear order in $\{\chi_{\mu\nu}\}$ and $\{\Pi^\text{I}\}$. By identifying appropriate metric $\{ D_{ij}, V_i, \Phi, \Psi \}$ and fluid $\{ \Xi^{\text{I}}, \Xi \}$ perturbation variables that are gauge-invariant, we are guaranteed that the linearized Einstein and fluid equations will be expressed solely in terms of them -- thus providing us with not only physical equations to solve from the outset, but also a check of the intermediate calculations.

Hence, before we embark on perturbation theory proper, let us take a detour to lay out the scalar-vector(-tensor) decomposition that underlies the gauge-invariant formalism; followed by delineating the gauge-invariant metric and fluid perturbation variables that we will be solving for. We will spend \S \eqref{Section_SVT} and \S \eqref{Section_Bardeen} doing so, before tackling the first order EoM and their GW solutions in sections \S \eqref{Section_PT_EoMandSoln} and \S \eqref{Section_PT_GWsAndMemory}.

\subsection{Scalar-Vector(-Tensor) Decomposition in $(d-1)$-space}
\label{Section_SVT}
In order to exploit the background SO$_{d-1}$ symmetry of the cosmological Einstein-fluid system, the first step to obtaining a gauge-invariant formulation of the equations governing the metric perturbations $\{\chi_{\mu\nu}\}$ and that of the fluid $\{\Pi^\text{I}\}$ is to perform an irreducible scalar-vector-tensor decomposition of the former and a similar scalar-vector one of the latter.

{\bf Scalar-Vector Decomposition} \qquad We will first tackle the scalar-vector decomposition. Any $(d-1)$-tuple of spacetime functions $\{ \mathcal{F}_\text{I}[\eta,\vec{x}] \vert \text{I} = 1,2,3,\dots,d-1 \}$ can be decomposed as
\begin{align}
\label{ScalarVector_RealSpace}
\mathcal{F}_\text{I}[\eta,\vec{x}] = \partial_{\text{I}} \mathcal{E} + \mathcal{E}_{\text{I}} ,
\end{align}
where $\mathcal{E}_\text{I}$ is divergence-less $\partial_{\text{I}} \mathcal{E}_\text{I} = 0$.\footnote{We are using capital Latin/English alphabets to de-emphasize any spatial coordinate transformation properties these scalar-vector(-tensor) decompositions may exhibit.} This scalar-vector decomposition of eq. \eqref{ScalarVector_RealSpace} holds whenever the spatial Fourier transform of $\mathcal{F}_\text{I}$, namely
\begin{align}
\mathcal{F}_{\text{I}}[\eta,\vec{x}] 
\equiv \int_{\mathbb{R}^{d-1}} \frac{\dd^{d-1}\vec{k}}{(2\pi)^{d-1}} \widetilde{\mathcal{F}}_\text{I}[\eta,\vec{k}] e^{ik_{\text{I}} x^\text{I}} ,
\end{align}
exists\footnote{Here and below, we will denote the Fourier transform of a function $f$ as $\widetilde{f}$.} and is non-singular in the long-distance limit $\vec{k} \to \vec{0}$. For, in Fourier space, we may write
\begin{align}
\label{ScalarVector_FourierSpace_IofIII}
\widetilde{\mathcal{F}}_\text{I}[\eta,\vec{k}] 
= i k_\text{I} \frac{k_\text{J}}{i \vec{k}^2} \widetilde{\mathcal{F}}_\text{J}[\eta,\vec{k}] 
+ \left( \delta_{\text{I}\text{J}} - \frac{k_\text{I} k_\text{J}}{\vec{k}^2} \right) \widetilde{\mathcal{F}}_\text{J}[\eta,\vec{k}] ,
\qquad\qquad
\vec{k}^2 \equiv \delta^{\text{I}\text{J}} k_\text{I} k_\text{J} ,
\qquad\qquad
\vec{k} \neq \vec{0} .
\end{align}
Notice that partial spatial derivatives are replaced with momentum vectors in Fourier space, i.e., $\partial_{\text{I}} \to i k_{\text{I}}$, so that the spatial divergence turns into a ``dot product," namely $\partial_\text{I} \mathcal{F}_{\text{I}\text{J}} \to i k_\text{I} \widetilde{\mathcal{F}}_{\text{I}\text{J}}$. These allow us to read off the ``gradient piece" of eq. \eqref{ScalarVector_RealSpace} in Fourier space as the first term (from the left) of eq. \eqref{ScalarVector_FourierSpace_IofIII}:
\begin{align}
\label{ScalarVector_FourierSpace_IIofIII}
\widetilde{\mathcal{E}}[\eta,\vec{k}] = \frac{k_\text{J}}{i \vec{k}^2} \widetilde{\mathcal{F}}_\text{J}[\eta,\vec{k}] ;
\end{align}
whereas the ``divergence-less piece" of eq. \eqref{ScalarVector_RealSpace} in Fourier space is the second term (from the left) of eq. \eqref{ScalarVector_FourierSpace_IofIII} 
\begin{align}
\label{ScalarVector_FourierSpace_IIIofIII}
\widetilde{\mathcal{E}}_{\text{I}}[\eta,\vec{k}] 	&= P_{\text{I}\text{J}} \widetilde{\mathcal{F}}_\text{J}[\eta,\vec{k}] , \\
\label{Transverse_Projector}
P_{\text{I}\text{J}} 								&\equiv \delta_{\text{I}\text{J}} - \frac{k_\text{I} k_\text{J}}{\vec{k}^2} ,
\end{align}
because the symmetric projector, which obeys $P_{\text{I}\text{J}} = P_{\text{J}\text{I}}$ and $P_{\text{I}\text{K}} P_{\text{K}\text{J}} = P_{\text{I}\text{J}}$, is transverse to the momentum vector $\vec{k}$: $k_\text{I} P_{\text{I}\text{J}} = 0$.

In what follows, it is also very useful to remember that, any relation of the form
\begin{align}
\label{ScalarVector_FourierSpace_v2_IofII}
\partial_\text{I} \mathcal{U}[\eta,\vec{x}] + \mathcal{U}_\text{I}[\eta,\vec{x}] = 0 ,
\end{align}
with $\partial_\text{I} \mathcal{U}_\text{I} = 0$, implies that 
\begin{align}
\label{ScalarVector_FourierSpace_v2_IIofII}
\mathcal{U}[\eta,\vec{x}] = \mathcal{U}_\text{I}[\eta,\vec{x}] = 0 ,
\end{align}
as long as $\mathcal{U}[\eta,\vec{x}]$ and $\mathcal{U}_\text{I}[\eta,\vec{x}]$ admit spatial Fourier transforms that are well-behaved in the ``IR" limit, $\vec{k} \to \vec{0}$. This follows from first taking the divergence of eq. \eqref{ScalarVector_FourierSpace_v2_IofII} in Fourier space:
\begin{align}
-\vec{k}^2 \widetilde{\mathcal{U}}[\eta,\vec{k}] = 0 ,
\end{align}
which yields, for $\vec{k}^2 \neq 0$, $\widetilde{\mathcal{U}}[\eta,\vec{k}]=0$. This in turn means eq. \eqref{ScalarVector_FourierSpace_v2_IofII} in Fourier space now reads $\widetilde{\mathcal{U}}_\text{I}[\eta,\vec{k}] = 0$.

{\bf Scalar-Vector-Tensor Decomposition} \qquad The scalar-vector-tensor decomposition of any $(d-1) \times (d-1)$ symmetric matrix of spacetime functions $\{ \mathcal{F}_{\text{I}\text{J}} = \mathcal{F}_{\text{J}\text{I}} \vert \text{I},\text{J} = 1,2,3,\dots,d-1 \}$ proceeds in a similar spirit as the scalar-vector one above. It is given by
\begin{align}
\label{ScalarVectorTensor_RealSpace}
\mathcal{F}_{\text{I}\text{J}}[\eta,\vec{x}] 
= \mathcal{E}_{\text{I}\text{J}} + \partial_{\{\text{I}} \mathcal{E}_{\text{J}\}} + \frac{\mathcal{E}}{d-1} \delta_{\text{I}\text{J}}
+ \left( \partial_{\text{I}} \partial_{\text{J}} - \frac{\delta_{\text{I}\text{J}}}{d-1} \vec{\nabla}^2 \right) \mathcal{W} ,
\end{align}
where $\vec{\nabla}^2 \equiv \delta^{\text{I}\text{J}} \partial_\text{I} \partial_\text{J}$ and the following constraints are in force,
\begin{align}
\mathcal{E}_{\text{I}\text{J}} = \mathcal{E}_{\text{J}\text{I}}, \qquad\qquad
\partial_{\text{I}} \mathcal{E}_{\text{I}\text{J}} = \delta^{\text{I}\text{J}} \mathcal{E}_{\text{I}\text{J}} = 0, \qquad\qquad 
&\text{(symmetric, transverse, traceless)} ; \\
\partial_{\text{I}} \mathcal{E}_{\text{I}} = 0, \qquad\qquad &\text{(transverse)} .
\end{align}
The scalar-vector-tensor decomposition of eq. \eqref{ScalarVectorTensor_RealSpace} holds whenever the spatial Fourier transform of $\mathcal{F}_{\text{I}\text{J}}$ exists and is non-singular in the long-distance limit $\vec{k} \to \vec{0}$. For, in Fourier space, we may write
\begin{align}
\label{ScalarVectorTensor_FourierSpace_IofII}
\widetilde{\mathcal{F}}_{\text{I}\text{J}}[\eta,\vec{k}] 
&= P_{\text{I}\text{J} \text{A}\text{B}} \widetilde{\mathcal{F}}_{\text{A}\text{B}}[\eta,\vec{k}] 
+ i k_{\{\text{I}} \left( P_{\text{J}\}\text{K}} \frac{\widetilde{\mathcal{F}}_{\text{K}\text{L}} k_{\text{L}}}{i \vec{k}^2} \right)
+ \frac{\delta^{\text{K}\text{L}} \widetilde{\mathcal{F}}_{\text{K}\text{L}}}{d-1} \delta_{\text{I}\text{J}} \nonumber\\
&\qquad\qquad
- \left( k_{\text{I}} k_{\text{J}} - \frac{\delta_{\text{I}\text{J}}}{d-1} \vec{k}^2 \right) 
	\left( -\frac{d-1}{d-2} \frac{1}{\vec{k}^4} \left\{ \widetilde{\mathcal{F}}_{\text{K}\text{L}} \cdot k_{\text{K}} k_{\text{L}} - \frac{\vec{k}^2}{d-1} \left( \widetilde{\mathcal{F}}_{\text{K}\text{L}} \delta^{\text{K}\text{L}} \right) \right\} \right) ,
\end{align}
whenever $\vec{k} \neq \vec{0}$. The divergence-less projector $P_{\text{J}\text{K}}$ is the one in eq. \eqref{Transverse_Projector}, while its higher rank counterpart reads
\begin{align}
P_{\text{I}\text{J} \text{M}\text{N}}[\vec{k}] &\equiv
\frac{\delta_{\text{I}\{ \text{M}} \delta_{\text{N}\} \text{J}}}{2}
- \frac{\delta_{\text{I}\text{J}} \delta_{\text{M}\text{N}}}{d-2}
+ \frac{\delta_{\text{I}\text{J}} k_\text{M} k_\text{N} + \delta_{\text{M}\text{N}} k_\text{I} k_\text{J}}{(d-2) \vec{k}^2} \nonumber\\
&\qquad\qquad
- \frac{1}{2 \vec{k}^2} \left( \delta_{\text{I} \{\text{M}} k_{\text{N}\}} k_\text{J} + \delta_{\text{J} \{\text{M}} k_{\text{N}\}} k_\text{I} \right)
+ \frac{d-3}{d-2} \frac{k_\text{I} k_\text{J} k_\text{M} k_\text{N}}{\vec{k}^4} .
\end{align}
It enjoys the following symmetric, transverse, and trace-free properties,
\begin{align}
P_{\text{I}\text{J} \text{M}\text{N}} = P_{\text{M}\text{N} \text{I}\text{J}}, \qquad\qquad
P_{\text{I}\text{J} \text{M}\text{N}} = P_{\text{J}\text{I} \text{M}\text{N}}, \qquad\qquad
k_\text{I} P_{\text{I}\text{J} \text{M}\text{N}} = 0, \qquad\qquad
P_{\text{I}\text{J} \text{M}\text{N}} \delta^{\text{I}\text{J}} = 0 ;
\end{align}
and the projector property
\begin{align}
P_{\text{I}\text{J} \text{M}\text{N}} P_{\text{M}\text{N} \text{K}\text{L}} = P_{\text{I}\text{J} \text{K}\text{L}} .
\end{align}
That eq. \eqref{ScalarVectorTensor_FourierSpace_IofII} is an identity can be verified through a direct calculation. Taking into account the properties of the projectors $P_{\text{I}\text{J} \text{M}\text{N}}$ and $P_{\text{I}\text{J}}$, comparison of the Fourier-space scalar-vector-tensor decomposition in eq. \eqref{ScalarVectorTensor_FourierSpace_IofII} to the real space one of eq. \eqref{ScalarVectorTensor_RealSpace} then lets us read off:
\begin{align}
\label{ScalarVectorTensor_FourierSpace_ComponentsI}
\widetilde{\mathcal{E}}_{\text{I}\text{J}} 	= P_{\text{I}\text{J} \text{A}\text{B}} \widetilde{\mathcal{F}}_{\text{A}\text{B}} , 
\qquad\qquad
\widetilde{\mathcal{E}}_{\text{J}} 			= P_{\text{J}\text{K}} \frac{\widetilde{\mathcal{F}}_{\text{K}\text{L}} k_{\text{L}}}{i \vec{k}^2} , 
\qquad\qquad
\widetilde{\mathcal{E}}						= \delta^{\text{K}\text{L}} \widetilde{\mathcal{F}}_{\text{K}\text{L}}, \\
\label{ScalarVectorTensor_FourierSpace_ComponentsII}
\widetilde{\mathcal{W}}						=  -\frac{d-1}{d-2} \frac{1}{\vec{k}^4} \left\{ \widetilde{\mathcal{F}}_{\text{K}\text{L}} \cdot k_{\text{K}} k_{\text{L}} - \frac{\vec{k}^2}{d-1} \left( \widetilde{\mathcal{F}}_{\text{K}\text{L}} \delta^{\text{K}\text{L}} \right) \right\} .
\end{align}
This is an appropriate place to highlight, except for the trace term $\widetilde{\mathcal{E}}$, these scalar-vector and scalar-vector-tensor variables are non-local functions of the original tensor components because of the powers of momentum appearing in the denominator. For instance, by performing the inverse Fourier transform of eq. \eqref{ScalarVector_FourierSpace_IIofIII}, we may witness that the longitudinal piece $\mathcal{E}[\eta,\vec{x}]$ of $\mathcal{F}_\text{I}$ requires knowing the latter everywhere in space, as the former is computed from a convolution of the latter against the Euclidean Laplacian's Green's function:
\begin{align}
\mathcal{E}[\eta,\vec{x}] 
&= -\partial_\text{J} \int_{\mathbb{R}^{d-1}}\dd^{d-1}\vec{x}' \mathcal{G}^{\text{(E)}}[\vec{x}-\vec{x}'] \mathcal{F}_\text{J}[\eta,\vec{x}'] , \\
\mathcal{G}^{\text{(E)}}[\vec{x}-\vec{x}']
&\equiv \int_{\mathbb{R}^{d-1}} \frac{\dd^{d-1}\vec{k}}{(2\pi)^{d-1}} \frac{e^{i k_{\text{L}} (x^{\text{L}}-x'^{\text{L}})}}{\vec{k}^2} .
\end{align}
Similarly, the inverse Fourier transform of eq. \eqref{ScalarVectorTensor_FourierSpace_ComponentsII} yields the non-local ``tidal-shear" term $\mathcal{W}$ as the following convolution:
\begin{align}
\label{ScalarVectorTensor_ShearTerm}
\mathcal{W}[\eta,\vec{x}] &= \frac{d-1}{d-2} \left( \partial_{\text{I}} \partial_{\text{J}} - \frac{\delta_{\text{I}\text{J}}}{d-1} \vec{\nabla}^2 \right)  
\int_{\mathbb{R}^{d-1}}\dd^{d-1}\vec{x}' \left(\frac{1}{\vec{\nabla}^4}\right)[\vec{x}-\vec{x}'] \mathcal{F}_{\text{I}\text{J}}[\eta,\vec{x}'] , \\
\left(\frac{1}{\vec{\nabla}^4}\right)[\vec{x}-\vec{x}'] &\equiv \int_{\mathbb{R}^{d-1}} \frac{\dd^{d-1}\vec{k}}{(2\pi)^{d-1}} \frac{e^{i k_{\text{L}} (x^{\text{L}}-x'^{\text{L}})}}{\vec{k}^4} .
\end{align}
Just like the scalar-vector decomposition, it is useful to keep in mind for applications below, that these scalar-vector-tensor components are independent in the following sense. Any relation of the form
\begin{align}
\label{ScalarVectorTensor_v2_IofII}
\mathcal{U}_{\text{I}\text{J}} + \partial_{\{\text{I}} \mathcal{U}_{\text{J}\}} + \frac{\mathcal{U}}{d-1} \delta_{\text{I}\text{J}}
+ \left( \partial_{\text{I}} \partial_{\text{J}} - \frac{\delta_{\text{I}\text{J}}}{d-1} \vec{\nabla}^2 \right) \mathcal{V} = 0
\end{align}
with $\partial_\text{I} \mathcal{U}_\text{I} = \partial_\text{I} \mathcal{U}_{\text{I}\text{J}} = 0$ and $\mathcal{U}_{\text{I}\text{J}} = \mathcal{U}_{\text{J}\text{I}}$, implies that 
\begin{align}
\label{ScalarVectorTensor_v2_IIofII}
\mathcal{U}_{\text{I}\text{J}} = \mathcal{U}_{\text{J}} = \mathcal{U} = \mathcal{V} = 0 ;
\end{align}
as long as these fields admit spatial Fourier transforms that are well-behaved in the ``IR" limit, $\vec{k} \to \vec{0}$. 

Eq. \eqref{ScalarVectorTensor_v2_IIofII} follows from first taking the trace of eq. \eqref{ScalarVectorTensor_v2_IofII} to deduce $\mathcal{U}=0$; then, taking a single and double divergence in Fourier space,
\begin{align}
-\vec{k}^2 \widetilde{\mathcal{U}}_{\text{J}} - \frac{d-2}{d-1} i k_{\text{J}} \vec{k}^2 \widetilde{\mathcal{V}} 	&= 0, \\
\frac{d-2}{d-1} \vec{k}^4 \widetilde{\mathcal{V}} 																	&= 0 .
\end{align}
As long as $\vec{k}\neq\vec{0}$, $\widetilde{\mathcal{V}}=0$ from the second line; taking this into account on the first line then tells us $\widetilde{\mathcal{U}}_{\text{J}}=0$ too.

Consider an equation of the form $\mathcal{A}_{\text{I}\text{J}} = \mathcal{B}_{\text{I}\text{J}}$, where $\mathcal{A}_{\text{I}\text{J}}$ and $\mathcal{B}_{\text{I}\text{J}}$ are both symmetric $(d-1) \times (d-1)$ matrices of spacetime functions. The above result implies, for all finite wavelength modes, we may perform a scalar-vector-tensor decomposition of $\mathcal{A}_{\text{I}\text{J}}$ and $\mathcal{B}_{\text{I}\text{J}}$, and proceed to equate their ``$\mathcal{E}_{\text{I}\text{J}}$" transverse-traceless (TT), ``$\mathcal{E}_{\text{J}}$" divergence-free, ``$\mathcal{E}$" trace, and the ``$\mathcal{W}$" tidal-shear parts (cf. eq. \eqref{ScalarVectorTensor_RealSpace}); and obtain four independent/irreducible sets of equations from $\mathcal{A}_{\text{I}\text{J}} = \mathcal{B}_{\text{I}\text{J}}$. Similar remarks apply if we had instead $\mathcal{A}_{\text{I}} = \mathcal{B}_{\text{I}}$: we may equate their ``$\mathcal{E}$" gradient/longitudinal parts and ``$\mathcal{E}_\text{I}$" divergence-less ones (cf. eq. \eqref{ScalarVector_RealSpace}). This strategy will in fact be employed repeatedly in the decomposition of the metric and fluid variables in \S \eqref{Section_Bardeen} and that of the linearized Einstein's and perfect fluid equations in \S \eqref{Section_PT_EoMandSoln}.

\subsection{Gauge-Invariant Metric and Fluid Perturbation Variables in Cosmological Geometries}
\label{Section_Bardeen}
{\bf Metric Perturbations} \qquad Suppose we are provided with the transformation expressing some coordinate system $\{ x^\mu \}$ in terms of another $\{x'^\mu\}$; i.e., suppose $x^\mu = x^\mu[x']$ is given. The components of the spacetime metric $g_{\mu\nu}[x]$ in $\{x^\mu\}$ is then related to $g_{\mu'\nu'}[x']$ in $\{x'^\mu\}$ via the transformation
\begin{align}
\label{Bardeen_CoordinateTransformation_Metric}
g_{\mu'\nu'}\left[x'\right] 
= g_{\alpha\beta}\left[x[x']\right] \frac{\partial x^\alpha[x']}{\partial x'^\mu} \frac{\partial x^\beta[x']}{\partial x'^\nu} .
\end{align}
We are going to consider infinitesimal coordinate transformations of the $d$-dimensional perturbed cosmological metric in eq. \eqref{SpatiallFlatFLRW_PerturbedMetric}. By infinitesimal transformations, we mean one brought about by a vector $\xi^\mu$ through
\begin{align}
\label{Bardeen_CoordinateTransformation_Small}
x^\alpha \equiv x'^\alpha + \xi^\alpha[x'] , \qquad\qquad x' \equiv (\eta',\vec{x}') ,
\end{align}
where $\xi^\alpha$ is ``small" in the same sense that the metric perturbations we are about to consider are small \cite{Weinberg:2008zzc}. At first order in $\chi_{\mu\nu}$ and $\xi^\alpha$, according to eq. \eqref{Bardeen_CoordinateTransformation_Metric}, one would find that the components of eq. \eqref{SpatiallFlatFLRW_PerturbedMetric} in the $x'$-coordinate system are
\begin{align}
\label{Bardeen_MetricCoordinateTransformation_Small}
a\left[\eta[x']\right] &\left( \eta_{\alpha\beta} + \chi_{\alpha\beta}\left[ x[x'] \right] \right) 
\frac{\partial (x'^\alpha + \xi^\alpha[x'])}{\partial x'^\mu} \frac{\partial (x'^\beta + \xi^\beta[x'])}{\partial x'^\nu} \nonumber\\
&= a[\eta'] \left( \eta_{\mu\nu} + \chi_{\mu\nu}[x'] 
+ \partial_{ \{ \mu'} \xi_{\nu\}}[x'] + 2 \xi^0[x'] \frac{\dot{a}[\eta']}{a[\eta']} \eta_{\mu\nu} 
+ \mathcal{O}\left[ (\chi_{\alpha\beta})^2, (\xi_\alpha)^2, \chi_{\alpha\beta} \xi_\sigma \right]\right) ,
\end{align}
where we are moving indices with the flat metric, $\xi_\nu \equiv \eta_{\nu\sigma} \xi^\sigma$, and $\partial_{\mu'} \equiv \partial/\partial x'^\mu$.

Because coordinates in curved spacetimes are mere labels and have no intrinsic physical/geometric meaning until the metric itself is specified, it is customary to drop the primes on the `new' coordinate system on the right-hand-side of eq. \eqref{Bardeen_CoordinateTransformation_Small} and instead phrase the ensuing transformations as replacement rules. In the same vein, eq. \eqref{Bardeen_MetricCoordinateTransformation_Small} tells us the effect of the infinitesimal coordinate re-definition $x^\alpha \to x^\alpha + \xi^\alpha$ upon the metric can be, at first order, entirely attributed to the following transformation of the perturbations:
\begin{align}
\label{Bardeen_GaugeTransformationOfChi}
\chi_{\mu\nu} \to \chi_{\mu\nu} + \partial_{\{\mu} \xi_{\nu\}} + 2 \frac{\dot{a}}{a} \xi^0 \eta_{\mu\nu}, \qquad\qquad 
\left(\text{whereas }a[\eta] \to a[\eta]\right) .
\end{align}
If one has an expression strictly first order in $\chi_{\mu\nu}$ that arose from a generally covariant field theoretic calculation, to implement an infinitesimal coordinate transformation of the form in eq. \eqref{Bardeen_CoordinateTransformation_Small}, one only needs to replace every occurrence of $\chi_{\mu\nu}$ with the right hand side of eq. \eqref{Bardeen_GaugeTransformationOfChi}. In particular, the coordinates themselves do not need to be altered -- because any such change would only affect higher order perturbations.

More generally, for any rank$-(n \geq 1)$ perturbed tensor
\begin{align}
\mathcal{T}_{\mu_1 \dots \mu_n} \equiv \overline{\mathcal{T}}_{\mu_1 \dots \mu_n} + \delta \mathcal{T}_{\mu_1 \dots \mu_n} ,
\end{align}
we may follow the same reasoning above, and attribute all the infinitesimal coordinate transformations upon it to its perturbation $\delta \mathcal{T}_{\mu_1 \dots \mu_n}$; i.e., 
\begin{align}
\overline{\mathcal{T}}_{\mu_1 \dots \mu_n} 	&\to \overline{\mathcal{T}}_{\mu_1 \dots \mu_n} , \\
\label{GeneralGaugeTransformation}
\delta \mathcal{T}_{\mu_1 \dots \mu_n} 		&\to \delta \mathcal{T}_{\mu_1 \dots \mu_n}
+ \xi^\sigma \partial_\sigma \overline{\mathcal{T}}_{\mu_1 \dots \mu_n}
+ \sum_{\text{I}=1}^{n} \overline{\mathcal{T}}_{\mu_1\mu_2 \dots \mu_{\text{I}-1}\sigma\mu_{\text{I}+1} \dots \mu_{n-1}\mu_n} \partial_{\mu_{\text{I}}} \xi^\sigma ,
\end{align}
whenever we implement $x^\alpha \to x^\alpha + \xi^\alpha$. This immediately teaches us that, whenever the ``background" tensor $\overline{\mathcal{T}}_{\mu_1 \dots \mu_n}$ is zero, the first order perturbation of the same tensor is gauge invariant:
\begin{align}
\delta \mathcal{T}_{\mu_1 \dots \mu_n} \to \delta \mathcal{T}_{\mu_1 \dots \mu_n} 
\qquad\qquad \text{ if } \qquad\qquad
\overline{\mathcal{T}}_{\mu_1 \dots \mu_n} = 0 .
\end{align}
We have already invoked this fact above, when introducing the first order perturbation theory of the Einstein and fluid equations; and we will do so again when discussing the linearized Weyl tensor about a perturbed spatially flat FLRW spacetime in appendix \S \eqref{Section_GeodesicDistance} below.

Eq. \eqref{Bardeen_GaugeTransformationOfChi} is also teaching us that, to a significant extent, the values of $\chi_{\mu\nu}$ can be modified at will by choosing an appropriate vector $\xi^\alpha$, hence rendering the former's physical/geometric meaning obscure. This arbitrariness of $\chi_{\mu\nu}$ is the motivation for introducing first order gauge-invariant variables \cite{Bardeen:1980kt} -- we now turn our attention towards its $d$-dimensional generalization.

{\it Gauge-Invariant Perturbation Variables} \qquad The background cosmological metric (and the Minkowski limit, $a \to 1$) enjoys a global SO$_{d-1}$ invariance under $x^i \to \mathfrak{R}^i_{\phantom{i}j} x^j$, for all spacetime-independent matrices $\{ \mathfrak{R}^i_{\phantom{i}j} \}$ satisfying $\delta_{ab} \mathfrak{R}^a_{\phantom{a}i} \mathfrak{R}^b_{\phantom{b}j} = \delta_{ij}$. This is a guide to seeking the first order gauge-invariant metric perturbation variables and the PDEs that govern them. By performing an irreducible scalar-vector-tensor (SVT) decomposition of the metric perturbations as well as an analogous one for $\xi_\alpha$, we shall see that the SVT decomposition of the linearized Einstein-fluid equation system itself can be organized into a gauge-invariant form. 

The scalar-vector decomposition of the generator-of-coordinate-transformations is
\begin{align}
\label{Bardeen_SVDecomposition}
\xi_\mu = (\xi_0, \partial_i \ell + \ell_i), \qquad\qquad \partial_i \ell_i = 0 ;
\end{align}
whereas the SVT one of the metric perturbation $\chi_{\mu\nu}$ is
\begin{align}
\label{Bardeen_SVTDecomposition}
\chi_{00} \equiv E, \qquad \qquad \chi_{0i} \equiv \partial_i F + F_i, \nonumber\\
\chi_{ij} \equiv D_{ij} + \partial_{\{ i} D_{j\} } + \frac{D}{d-1} \delta_{ij} + \left( \partial_i \partial_j - \frac{\delta_{ij}}{d-1} \vec{\nabla}^2 \right) K ,
\end{align}
where these variables subject to the following constraints:
\begin{align}
\label{Bardeen_SVTConstraints}
\partial_i F_i = \delta^{ij} D_{ij} = \partial_i D_{ij} = \partial_i D_i = 0 .
\end{align}
Applying the irreducible decomposition of eq. \eqref{Bardeen_SVTDecomposition} to eq. \eqref{Bardeen_GaugeTransformationOfChi}, under $x^0 \to x^0 + \xi_0$ and $x^i \to x^i - (\partial_i \ell + \ell_i)$, we may gather 
\begin{align}
\label{Bardeen_CoordinateTransformation_MetricSVT_IofII}
E \to E + 2 \frac{\partial_0 (a \xi_0)}{a} , \qquad\qquad
F \to F + \dot{\ell} + \xi_0, \qquad\qquad 
F_i \to F_i + \dot{\ell}_i,  \\
D_j \to D_j + \ell_j, \qquad\qquad
D \to D+2\vec{\nabla}^2 \ell - 2(d-1)\frac{\dot{a}}{a} \xi_0, \qquad\qquad
K \to K + 2\ell , \nonumber
\end{align}
and the transverse-traceless graviton is gauge-invariant
\begin{align}
\label{Bardeen_CoordinateTransformation_MetricSVT_IIofII}
D_{ij} \equiv \chi_{ij}^{\text{TT}} \to D_{ij} .
\end{align}
A direct calculation will then verify that the following variables are first order gauge invariant ones. The $2$ scalar ones are
\begin{align}
\label{Bardeen_Psi}
\Psi &\equiv E - \frac{2}{a} \partial_0 \left\{ a \left( F - \frac{\dot{K}}{2} \right) \right\} , \\
\label{Bardeen_Phi}
\Phi &\equiv \frac{D - \vec{\nabla}^2 K}{d-1} + 2 \frac{\dot{a}}{a} \left( F - \frac{\dot{K}}{2} \right) ;
\end{align}
\footnote{We have re-normalized the definition of $\Psi$ in eq. \eqref{Bardeen_Psi} relative to that in \cite{Chu:2016qxp}; $2\Psi$ in \cite{Chu:2016qxp} is $\Psi$ here.}whereas the vector and tensor modes are, respectively,
\begin{align}
\label{Bardeen_VandDij}
V_i \equiv F_i - \dot{D}_i 
\qquad \qquad \text{ and }\qquad \qquad
D_{ij} \equiv \chi_{ij}^\text{TT} .
\end{align}
We will recall, as already remarked in \cite{Chu:2016qxp}, that the tensor mode $D_{ij}$, which obeys a minimally coupled massless wave equation, only exists when $d \geq 4$.

{\bf Scalar Fields For EFT of Fluids} \qquad We move on to understand how to parametrize, in a gauge-invariant manner, the first order fluid perturbations examined in this paper. In a $d$-dimensional spacetime there are $(d-1)$ scalar fields $\{ \varphi^\text{I} \vert \text{I} = 1,2,3,\dots,d-1 \}$ labeling the fluid elements in its co-moving frame. Because they are scalar fields, under a generic coordinate transformation $x^\alpha = x^\alpha[x']$,
\begin{align}
\label{Bardeen_CoordinateTransformation_Scalars}
\varphi^\text{I}[x'] = \varphi^\text{I}\left[ x[x'] \right] .
\end{align}
Earlier, we have shown that the $\overline{\varphi}^\text{I} = x^\text{I}$ solves the background fluid equations-of-motion and, with an appropriate choice of Lagrangian to govern their dynamics, produces a cosmology with a constant equation-of-state $w$. Just as for the gravitational side of the problem, we shall study the perturbations about this background $\overline{\varphi}^\text{I}$, namely $\varphi^\text{I} \equiv x^\text{I} + \Pi^\text{I}$ (cf. eq. \eqref{PerfectFluid_ExpansionFromEq}), and understand how the perturbations $\Pi^\text{I}$ change under an infinitesimal coordinate transformation, $x^\alpha \to x^\alpha + \xi^\alpha$. Eq. \eqref{Bardeen_CoordinateTransformation_Scalars} informs us that, to first order in $\xi^\alpha$, 
\begin{align}
x^\text{I} + \Pi^\text{I}[x] 
&\to x^\text{I}+\xi^\text{I} + \Pi^\text{I}[x+\xi] \nonumber\\
&= x^\text{I} + \Pi^\text{I}[x] + \xi^\text{I}[x] + \mathcal{O}\left[ \Pi^\text{I} \xi^\alpha, (\xi^\alpha)^2 \right] .
\end{align}
Building on the above discussion for metric perturbations -- for any strictly first order expression involving $\{ \Pi^\text{I} \}$ which descended from a generally covariant set of equations -- we see that implementing the infinitesimal coordinate transformation $x^\alpha \to x^\alpha + \xi^\alpha$ reduces to keeping the coordinates themselves fixed and making the replacement
\begin{align}
\label{Bardeen_GaugeTransformationOfPiI}
\Pi^\text{I}[x] \to \Pi^\text{I}[x] + \xi^\text{I}[x] .
\end{align}
If we further perform a scalar-vector decomposition of $\Pi^\text{I}$,
\begin{align}
\label{Bardeen_FluidSVDecomposition}
\Pi_\text{I} \equiv -\Pi^\text{I} = \partial_\text{I} P + P_\text{I}, \qquad\qquad \partial_\text{I} P_\text{I} = 0 ;
\end{align}
remembering $\xi_\alpha \equiv (\xi_0,\partial_i \ell + \ell_i)$, the irreducible pieces transform as
\begin{align}
P \to P + \ell \qquad\qquad \text{ and } \qquad\qquad P^\text{I} \to P^\text{I} + \ell^\text{I} .
\end{align}
A short calculation would verify that the following are first order gauge-invariant fluid perturbation variables:
\begin{align}
\label{Bardeen_GaugeInvariantFluidVariables}
\Xi^\text{I} \equiv P^\text{I} - D^\text{I} \qquad\qquad \text{ and } \qquad\qquad \Xi \equiv P - \frac{K}{2} ,
\end{align}
where the $D^\text{I}$ and $K$ are the metric perturbations in eq. \eqref{Bardeen_SVTDecomposition}.

\subsection{First Order Equations-Of-Motion and Solutions}
\label{Section_PT_EoMandSoln} 
Let us begin this section by collecting the various ingredients necessary to construct the first order gauge-invariant equations for the metric $(D_{ij}, V_i, \Phi, \Psi)$ and fluid $(\Xi_\text{I}, \Xi)$ perturbation variables.

{\bf SVT decomposition of astrophysical source} \qquad Firstly, we shall suppose that the astrophysical system at hand has an energy-momentum-shear-stress tensor $\,^{(\text{a})}T_{\mu\nu}$ whose scalar-vector-tensor decomposition is given by
\begin{align}
\label{Astro_SVT_IofII}
\,^{(\text{a})}T_{00} &= \rho, \qquad\qquad
\,^{(\text{a})}T_{0i} = \Sigma_i + \partial_i \Sigma, \\
\label{Astro_SVT_IIofII}
\,^{(\text{a})}T_{ij} &= \sigma_{ij} + \partial_{ \{i } \sigma_{ j \} } + \frac{\sigma}{d-1} \delta_{ij} 
+ \left( \partial_i \partial_j - \frac{\delta_{ij}}{d-1} \vec{\nabla}^2 \right) \Upsilon .
\end{align}
Because of the conservation of the stress tensor, these variables are not all independent. We have $\overline{\nabla}^\sigma \,^{\text{(a)}}T_{\sigma 0} = 0$ handing us
\begin{align}
\label{Astro_ConservationIofIII}
\frac{\partial_0 (a^{d-2} \rho)}{a^{d-2}} 	&= \vec{\nabla}^2 \Sigma + \frac{\dot{a}}{a} \left( \rho - \sigma \right) ;
\end{align}
while the longitudinal and transverse parts of $\overline{\nabla}^\sigma \,^{\text{(a)}}T_{\sigma i} = 0$ return
\begin{align}
\label{Astro_ConservationIIofIII}
\frac{\partial_0 (a^{d-2} \Sigma)}{a^{d-2}} &= \frac{\sigma + (d-2) \vec{\nabla}^2 \Upsilon}{d-1}, \\
\label{Astro_ConservationIIIofIII}
\frac{\partial_0 (a^{d-2} \Sigma_i)}{a^{d-2}} &= \vec{\nabla}^2 \sigma_i .
\end{align}
{\it Non-conservation of mass} \qquad In the cosmological geometry of eq. \eqref{SpatiallFlatFLRW_Metric}, the $d$-beins corresponding to the local Lorentz frame of a co-moving observer are given by $\varepsilon^\alpha_{\phantom{\alpha}\widehat{\mu}} = a^{-1} \delta^\alpha_\mu$.\footnote{Using these $d$-beins we record that the $\Pi_{ij}^{\text{(T)}}$ in eq. \eqref{GW_NullCone} should be related to $\sigma_{ij}$ in eq. \eqref{Astro_SVT_IIofII} through: $\,^{(\text{a})}T_{\widehat{i}\widehat{j}}^{\text{TT}} = \,^{(\text{a})}T_{\alpha\beta}^{\text{TT}} \varepsilon^\alpha_{\phantom{\alpha}\widehat{i}} \varepsilon^\beta_{\phantom{\beta}\widehat{j}} = a^{-2} \sigma_{ij} = - \Pi_{ij}^{\text{(T)}}$.} In such an orthonormal frame, the mass $\,^{(\text{a})}M$ of the system can be defined through the proper spatial volume integral of $\,^{(\text{a})}T_{\widehat{0}\widehat{0}} \equiv \varepsilon^\alpha_{\phantom{\alpha}\widehat{0}} \varepsilon^\beta_{\phantom{\beta}\widehat{0}} \,^{(\text{a})}T_{\alpha\beta} = a^{-2} \,^{(\text{a})}T_{00}$:
\begin{align}
\label{Astro_Mass}
\,^{(\text{a})}M[\eta] \equiv \int_{\mathbb{R}^{d-1}} \dd^{d-1}\vec{x}' a[\eta]^{d-3} \rho[\eta,\vec{x}'] .
\end{align}
Similarly, the pressure $\,^{\text{(a)}}P_i$ in the $i$th direction can be defined in terms of the proper spatial volume integral of $\,^{(\text{a})}T_{\widehat{i}\widehat{i}} = a^{-2} \,^{(\text{a})}T_{ii}$, with no summation over $i$ implied. This in turn leads us to
\begin{align}
\label{Astro_SumOfPressure}
\sum_{i=1}^{d-1} \,^{\text{(a)}}P_i[\eta] = \int_{\mathbb{R}^{d-1}} \dd^{d-1}\vec{x}' a[\eta]^{d-3} \sigma[\eta,\vec{x}'] .
\end{align}
By virtue of being in a cosmological environment, the mass $\,^{(\text{a})}M$ of the system is not, in fact, conserved. Taking a time derivative of eq. \eqref{Astro_Mass}, and making use of the conservation law in eq. \eqref{Astro_ConservationIofIII},
\begin{align}
\,^{(\text{a})}\dot{M} = \int_{\mathbb{R}^{d-1}} \dd^{d-1}\vec{x}' a^{d-3} \left( \vec{\nabla}^2 \Sigma - \frac{\dot{a}}{a} \sigma \right) .
\end{align}
Converting the Laplacian term to a surface integral at spatial infinity, we see that it is trivial. We may therefore gather, through the relation in eq. \eqref{Astro_SumOfPressure},
\begin{align}
\label{Astro_MassLossRate}
\,^{(\text{a})}\dot{M}[\eta] = - \frac{\dot{a}[\eta]}{a[\eta]} \sum_{i=1}^{d-1} \,^{\text{(a)}}P_i[\eta] .
\end{align}
{\bf First order perturbations of $\,^{\text{(f)}}T_{\mu\nu}$} \qquad We will also need the perfect fluid stress tensor in eq. 	\eqref{PerfectFluid_StressTensor} developed up to first order in $\{ \chi_{\mu\nu} \}$ and $\{ \Pi^\text{I} \}$. Useful intermediate results are
\begin{align}
\label{PerfectFluid_EntropyCurrentIofIII}
n^\mu &= \frac{(-)^d}{a^d} \left\{ 
 \delta^\mu_0 \left( -1 + \frac{\chi}{2} - \partial_{\text{I}} \Pi^{\text{I}} \right) + \delta^\mu_\text{I} \dot{\Pi}^\text{I} 
 + \dots \right\} , \qquad\qquad \chi \equiv \eta^{\sigma\rho} \chi_{\sigma\rho} , \\
 \label{PerfectFluid_EntropyCurrentIIofIII}
\frac{n_\mu n_\nu}{n^2} 
&= a^2 \left\{ \delta_\mu^0 \delta_\nu^0  (1 + \chi_{00}) + \delta_{\{\mu}^0 \delta_{\nu\}}^j \left( \dot{\Pi}^j + \chi_{0j} \right) 
 + \dots \right\} ,
\end{align}
and 
\begin{align}
 \label{PerfectFluid_EntropyCurrentIIIofIII}
\left(n^2\right)^{\frac{w+1}{2}}
= \frac{1}{ a^{(w+1)(d-1)} } \left\{ 1 + \frac{w+1}{2}\left(\delta^{ij} \chi_{ij} + 2 \partial_\text{I} \Pi^\text{I} \right) 
 + \dots \right\} .
\end{align}
The desired $\mathcal{O}[\chi_{\mu\nu},\Pi^{\text{I}}]$-accurate fluid stress tensor is
\begin{align}
\label{PerfectFluid_StressTensor_FirstOrder}
\,^{\text{(f)}}T_{\mu\nu}
= \frac{(d-2)(d-1)}{4 \pi \GN q_w^2 \eta^2} 
\Bigg\{ & \left(- w \eta_{\mu\nu} + (w+1) \delta_\mu^0 \delta_\nu^0\right) \left( 1 + \frac{w+1}{2}\left(\delta^{ij} \chi_{ij} + 2 \partial_\text{I} \Pi^\text{I} \right) + \dots \right) \\
&\qquad
- w \chi_{\mu\nu} + (w+1)\left( \delta_\mu^0 \delta_\nu^0  \chi_{00} + \delta_{\{\mu}^0 \delta_{\nu\}}^j \left( \dot{\Pi}^j + \chi_{0j} \right) + \dots \right) \Bigg\}  . \nonumber
\end{align}
In equations \eqref{PerfectFluid_EntropyCurrentIofIII} through \eqref{PerfectFluid_StressTensor_FirstOrder}, the ``$\dots$" represent terms that scale as $\mathcal{O}\left[(\chi_{\mu\nu})^2,(\Pi^{\text{I}})^2,\chi_{\mu\nu} \Pi^\text{I}\right]$ or higher.

{\bf Linearized Einstein's Equations} \qquad We are now ready to enumerate the EoM for the gauge-invariant metric perturbations $\Phi$, $\Psi$, $V_i$ and $D_{ij}$. We used {\sf xAct} \cite{xAct} to work out the linearized Einstein tensor $(G \vert 1)_{\mu\nu}$ and subtracted from it the first order piece of the fluid stress tensor in eq. \eqref{PerfectFluid_StressTensor_FirstOrder} to compute the left-hand-side of eq. \eqref{Setup_EinsteinEquations_FirstOrder}, $\Delta_{\mu\nu}$. By first decomposing $\chi_{\mu\nu}$ and $\Pi^\text{I}$ according to equations \eqref{Bardeen_SVTDecomposition} and \eqref{Bardeen_FluidSVDecomposition} respectively, the end result for $\Delta_{\mu\nu}$ is gauge-invariant if we group together the appropriate terms according to the definitions in equations \eqref{Bardeen_Psi}, \eqref{Bardeen_Phi}, \eqref{Bardeen_VandDij} and \eqref{Bardeen_GaugeInvariantFluidVariables}. As already remarked earlier, this is both a check of the calculation itself, as well as the starting point for the ensuing scalar-vector-tensor decomposition of the linearized Einstein's equation \eqref{Setup_EinsteinEquations_FirstOrder}.\footnote{Computationally speaking, note that $E$ only appears in $\Psi$ and $D$ only in $\Phi$. We may therefore use eq. \eqref{Bardeen_Psi} to replace all occurrences of $E$ with its equivalent expression in terms of $\Psi$, $F$ and $K$; eq. \eqref{Bardeen_Phi} to replace all $D$'s in terms of $\Phi$, $K$, and $F$; eq. \eqref{Bardeen_VandDij} to replace all $F_i$'s in terms of $V_i$ and $D_i$. One should find all the remaining $F$, $K$ and $D_i$ to cancel. It may also be useful, at this point, to refer to the decompositions of equations \eqref{ScalarVector_RealSpace} and \eqref{ScalarVectorTensor_RealSpace} and recall the discussion starting just before eq. \eqref{ScalarVectorTensor_v2_IofII} through the end of \S \eqref{Section_SVT}. Furthermore we highlight that, if one first computes the following EoM in Fourier-space, the formulas in equations \eqref{ScalarVector_FourierSpace_IIofIII}, \eqref{ScalarVector_FourierSpace_IIIofIII}, \eqref{ScalarVectorTensor_FourierSpace_ComponentsI} and \eqref{ScalarVectorTensor_FourierSpace_ComponentsII} may prove handy for extracting the relevant scalar-vector(-tensor) components.}

{\it Scalars} \qquad The scalar EoM for $\Phi$ and $\Psi$ arise from the $00$ component of eq. \eqref{Setup_EinsteinEquations_FirstOrder}, with $\,^{(\text{a})}T_{00} = \rho$ (cf. eq. \eqref{Astro_SVT_IofII}),
\begin{align}
\label{FirstOrderPTEqns_Scalar1of4}
\frac{1}{2} (d-2) \vec{\nabla}^2 \Phi 
- \frac{(d-2) (d-1)}{\eta q_w} \dot{\Phi}
- \frac{(d-2) (d-1)}{\eta^2 q_w^2} \left( (w+1) ((d-1) \Phi - 2 \vec{\nabla}^2 \Xi) + 2 \Psi \right)
= 8\pi\GN\rho ;
\end{align}
equating the gradient portions of the $0i$ components on both sides of eq. \eqref{Setup_EinsteinEquations_FirstOrder}, with $\,^{(\text{a})}T_{0i}$ given in eq. \eqref{Astro_SVT_IofII},
\begin{align}
\label{FirstOrderPTEqns_Scalar2of4}
\frac{1}{2} (d-2) \dot{\Phi} + \frac{(d-2)}{\eta q_w} \Psi + \frac{2 (d-2) (d-1) (w+1)}{\eta^2 q_w^2} \dot{\Xi}
= 8\pi\GN \Sigma ;
\end{align}
equating the spatial trace portions of the $ij$ components on both sides of eq. \eqref{Setup_EinsteinEquations_FirstOrder}, with $\,^{(\text{a})}T_{ij}$ given in eq. \eqref{Astro_SVT_IIofII},
\begin{align}
\label{FirstOrderPTEqns_Scalar3of4}
&\frac{1}{2} (d-2) \left( -\vec{\nabla}^2 \left\{ (d-3) \Phi - \Psi \right\} + (d-1) \ddot{\Phi} \right) 
+ \frac{(d-2) (d-1)}{\eta q_w} \left( (d-2) \dot{\Phi} + \dot{\Psi} \right) \\
&\qquad\qquad
- \frac{(d-2) (d-1)^2 w}{\eta^2 q_w^2} \left( (w+1) ((d-1) \Phi - 2 \vec{\nabla}^2 \Xi)+ 2 \Psi \right)
= 8\pi\GN \sigma ; \nonumber
\end{align}
as well as equating the tidal-shear part of the $ij$ components on both sides of eq. \eqref{Setup_EinsteinEquations_FirstOrder}, again with $\,^{(\text{a})}T_{ij}$ given in eq. \eqref{Astro_SVT_IIofII},
\begin{align}
\label{FirstOrderPTEqns_Scalar4of4}
\frac{1}{2} ((d-3) \Phi - \Psi) = 8\pi\GN \Upsilon .
\end{align}
Because $\Psi$ can be immediately obtained from $\Phi$ through eq. \eqref{FirstOrderPTEqns_Scalar4of4}, we will focus on solving $\Phi$ throughout the rest of this paper.

{\it Vector} \qquad The vector perturbations come from equating the divergence-free part of the $0i$ components on both sides of eq. \eqref{Setup_EinsteinEquations_FirstOrder}, with $\,^{(\text{a})}T_{0i}$ given in eq. \eqref{Astro_SVT_IofII},
\begin{align}
\label{FirstOrderPTEqns_Vector1of2}
\frac{1}{2}\vec{\nabla}^2 V_i - \frac{2 (d-2) (d-1) (w+1)}{\eta^2 q_w^2} (V_i - \dot{\Xi}_i) = 8\pi\GN \Sigma_i ;
\end{align}
and equating the divergence-less vector portion of the $ij$ components on both sides of eq. \eqref{Setup_EinsteinEquations_FirstOrder}, with $\,^{(\text{a})}T_{ij}$ given in eq. \eqref{Astro_SVT_IIofII},
\begin{align}
\label{FirstOrderPTEqns_Vector2of2}
\frac{1}{2} \dot{V}_i + \frac{d-2}{\eta q_w} V_i = \frac{\partial_0 \left( a^{d-2} V_i \right)}{2 a^{d-2}} = 8\pi\GN \sigma_i .
\end{align}
{\it Tensor} \qquad Finally, the tensor equation arise from equating the transverse-traceless parts of the $ij$ components on both sides of eq. \eqref{Setup_EinsteinEquations_FirstOrder}, with $\,^{(\text{a})}T_{ij}$ given in eq. \eqref{Astro_SVT_IIofII},
\begin{align}
\label{FirstOrderPTEqns_Tensor}
-\frac{1}{2} \left( \ddot{D}_{ij} + \frac{2(d-2)}{q_w \eta} \dot{D}_{ij} - \vec{\nabla}^2 D_{ij} \right) 
= - \frac{a^2}{2} \overline{\Box}^{\text{(S)}} D_{ij} = 8\pi\GN \sigma_{ij} .
\end{align}
The $\overline{\Box}^{\text{(S)}}$ is the scalar spacetime Laplacian; with the $\gb_{\mu\nu}$ in eq. \eqref{SpatiallFlatFLRW_Metric}, $\overline{\Box}^{\text{(S)}} D_{ij} = \partial_\mu(\sqrt{|\gb|} \gb^{\mu\nu} \partial_\nu D_{ij})/\sqrt{|\gb|}$. We also remind the reader that $q_w \equiv (d-3) + (d-1)w$ (cf. eq. \eqref{SpatiallFlatFLRW_Metric_w}).

{\bf Linearized Fluid EoM} \qquad Inserting into the fluid EoM in eq. \eqref{Setup_PerfectFluidEOM} the metric and fluid expansions of equations \eqref{SpatiallFlatFLRW_PerturbedMetric} and \eqref{PerfectFluid_ExpansionFromEq}, followed by their decompositions in equations \eqref{Bardeen_SVTDecomposition} and \eqref{Bardeen_FluidSVDecomposition}, a direct calculation then hand us the linearized fluid equations:
{\allowdisplaybreaks\begin{align}
\label{FirstOrderPTEqns_Fluid}
	&\rho_\text{f}''[a^{2(1-d)}/2] a^{2(1-d)} \left\{ \partial_{\text{I}} \left( \partial_\text{K} \Pi^\text{K} + \frac{\delta^{mn} \chi_{mn}}{2} \right) - (1-d) \frac{\dot{a}}{a} \left( \chi_{0\text{I}} + \dot{\Pi}^\text{I} \right) \right\}  \\
	&+ \rho_\text{f}'[a^{2(1-d)}/2] \left\{ \partial_{\text{I}} \left( \partial_{\text{K}} \Pi^{\text{K}} + \frac{\delta^{mn} \chi_{mn}}{2} \right) 
	- (1-d) \frac{\dot{a}}{a} \left( \chi_{0\text{I}} + \dot{\Pi}^\text{I} \right) 
	+ \frac{1}{2} \partial_{\text{I}} \chi_{00} - \frac{1}{a} \partial_0 \left\{ a \left( \chi_{0\text{I}} + \dot{\Pi}^\text{I} \right) \right\} \right\} = 0 . \nonumber
	\end{align}}
We then proceed to group terms according to the definitions laid out in equations \eqref{Bardeen_Psi}, \eqref{Bardeen_Phi}, \eqref{Bardeen_VandDij} and \eqref{Bardeen_GaugeInvariantFluidVariables}. For all finite wavelength modes, the ``gradient" part of eq. \eqref{FirstOrderPTEqns_Fluid} provides us with an equation for $\Xi$:
\begin{align}
\label{FirstOrderPTEqns_FluidScalar_rhof}
\rho_\text{f}''[a^{2(1-d)}/2] a^{2(1-d)} & \left\{ -\vec{\nabla}^2 \Xi - (d-1) \frac{\dot{a}}{a} \dot{\Xi} + \frac{d-1}{2} \Phi \right\}  \nonumber\\
&\qquad\qquad
+ \rho_\text{f}'[a^{2(1-d)}/2] \left\{ 
-\vec{\nabla}^2 \Xi - (d-2) \frac{\dot{a}}{a} \dot{\Xi} + \ddot{\Xi}
+ \frac{d-1}{2} \Phi + \frac{1}{2}\Psi \right\} = 0 .
\end{align}
The divergence-free portion of eq. \eqref{FirstOrderPTEqns_Fluid} yields the equation for $\Xi_\text{I}$:
\begin{align}
\label{FirstOrderPTEqns_FluidVector_rhof}
\rho_\text{f}''[a^{2(1-d)}/2] a^{2(1-d)} & \left\{ 
(d-1) \frac{\dot{a}}{a} \left( V_\text{I} - \dot{\Xi}_\text{I} \right) \right\} \nonumber\\
&\qquad\qquad
+ \rho_\text{f}'[a^{2(1-d)}/2] \left\{
(d-2) \frac{\dot{a}}{a} \left( V_\text{I} - \dot{\Xi}_\text{I} \right) 
- \left( \dot{V}_\text{I} - \ddot{\Xi}_\text{I} \right)
\right\} = 0 .
\end{align}
Equations \eqref{FirstOrderPTEqns_FluidScalar_rhof} and \eqref{FirstOrderPTEqns_FluidVector_rhof} hold for any $\rho_\text{f}$ evaluated on the background fluid solution $\overline{\varphi}^\text{I} = x^\text{I}$. If we specialize to the case of interest, namely with $\rho_\text{f}$ given in eq. \eqref{PerfectFluid_PowerLawLagrangian}, we find equations \eqref{FirstOrderPTEqns_FluidScalar_rhof} and \eqref{FirstOrderPTEqns_FluidVector_rhof} becoming respectively
\begin{align}
\label{FirstOrderPTEqns_FluidScalar}
(w+1) \left\{ \ddot{\Xi} - \frac{2((d-1)w-1)}{q_w \cdot \eta} \dot{\Xi} - w \vec{\nabla}^2 \Xi + \frac{(d-1) w}{2} \Phi + \frac{1}{2}\Psi \right\} = 0 
\end{align}
and
\begin{align}
\label{FirstOrderPTEqns_FluidVector}
(w+1) a^{(d-1)w-1} \partial_0 \left( \frac{V_i - \dot{\Xi}_i}{ a^{(d-1)w-1} } \right) = 0 .
\end{align}
As discussed in \S \eqref{Section_Setup}, the $w=-1$ case corresponds to Einstein's equations with a positive cosmological constant. The background metric solution is de Sitter spacetime and, in particular, there is really no background cosmological fluid driving its evolution -- this is reflected by the fact that, when we put $w=-1$ in equations \eqref{FirstOrderPTEqns_FluidScalar} and \eqref{FirstOrderPTEqns_FluidVector}, these fluid EoM become a pair of irrelevant $0=0$ identities.

With the EoM laid out, we will turn to describing their solutions for the remaining of this section.

{\bf Tensor mode solutions} \qquad The tensor mode wave equation \eqref{FirstOrderPTEqns_Tensor} can be re-expressed as 
\begin{align}
\label{FirstOrderPTEqns_Tensor_PDE}
\left( \partial^2 + \frac{(d-2)((d-1)w-1)}{q_w^2 \eta^2} \right) \left( a^{\frac{d-2}{2}} D_{ij} \right) = -16\pi\GN a^{\frac{d-2}{2}} \sigma_{ij} ,
\end{align}
with $\partial^2 \equiv \eta^{\mu\nu} \partial_\mu \partial_\nu$. The solution to the inverse of the wave operator taking the form $\partial^2 - \kappa(\kappa+1)/\eta^2$ in eq. \eqref{FirstOrderPTEqns_Tensor_PDE}, and throughout the rest of this paper, is reviewed in appendix \eqref{Section_PDESolution}; we will simply report the results in the main text. Here, the retarded solution is
\begin{align}
\label{FirstOrderPTEqns_Tensor_Soln}
a[\eta]^{\frac{d-2}{2}} D_{ij}[\eta,\vec{x}]
&= -16\pi\GN \int_{\mathbb{R}^{d-1}} \dd^{d-1}\vec{x}' \int_{\eta_\circ}^\eta \dd\eta' a[\eta']^{\frac{d-2}{2}} \mathcal{G}^{\text{(T)}}[\eta,\eta';\sbar]  \sigma_{ij}[\eta',\vec{x}'] ,
\end{align}
where the symmetric Green's functions are
\begin{align}
\label{FirstOrderPTEqns_Tensor_GSoln_Evend}
\mathcal{G}^{\text{(T)}}_{\text{even $d \geq 4$}}[\eta,\eta';\sbar] 
&=\left( \frac{1}{2\pi} \frac{\partial}{\partial \sbar} \right)^{\frac{d-2}{2}} \left(\frac{\Theta[\sbar]}{2} P_{-\frac{d-2}{q_w}}[1+s] \right) , \\
\label{FirstOrderPTEqns_Tensor_GSoln_Oddd}
\mathcal{G}^{\text{(T)}}_{\text{odd $d \geq 5$}}[\eta,\eta';\sbar] 
&=\left( \frac{1}{2\pi} \frac{\partial}{\partial \sbar} \right)^{\frac{d-3}{2}} \left(\frac{\Theta[\sbar]}{4\pi} \frac{ \left(s + \sqrt{s(s+2)} + 1\right)^{\frac{(w-1)(d-1)}{q_w}} + 1}{\sqrt{\sbar(s+2)} \left(s + \sqrt{s(s+2)} + 1\right)^{\frac{(w-1)(d-1)}{2q_w}}}
 \right) , \\
\sbar &\equiv \frac{1}{2}\left((\eta-\eta')^2-(\vec{x}-\vec{x}')^2\right) , \qquad\qquad
s \equiv \frac{\sbar}{\eta\eta'} .
\end{align}
The $P_\nu$ occurring in eq. \eqref{FirstOrderPTEqns_Tensor_GSoln_Evend} is the Legendre function, and will show up repeatedly in the even dimensional solutions in this paper. 

Let us highlight the 4D case $D_{ij}^\text{(4D)} = D_{ij}^{(\text{4D direct})} + D_{ij}^{(\text{4D tail})}$. Its light cone part is
\begin{align}
\label{FirstOrderPTEqns_Tensor_Soln_4Ddirect}
D_{ij}^{(\text{4D direct})}[\eta,\vec{x}]
&= -4 \GN \Theta_p \int_{\mathbb{R}^{3}} \dd^{3}\vec{x}' \frac{a[\eta_r]}{a[\eta]} \frac{\sigma_{ij}[\eta_r,\vec{x}']}{|\vec{x}-\vec{x}'|} ,
\end{align}
with retarded time $\eta_r \equiv \eta-|\vec{x}-\vec{x}'|$; and the symbol $\Theta_p$, as explained in appendix \eqref{Section_PDESolution}, reminds us that the solution is non-zero -- whenever $q_w > 0$ -- only within the particle horizon of the GW source. The tail portion is
\begin{align}
\label{FirstOrderPTEqns_Tensor_Soln_4Dtail}
D_{ij}^{(\text{4D tail})}[\eta,\vec{x}]
	&= -4 \GN \int_{\mathbb{R}^{3}} \dd^{3}\vec{x}' \int_{\eta_\circ}^{\eta_r} \dd\eta' \frac{a[\eta']}{a[\eta] \cdot \eta\eta'} 
	P'_{-\frac{2}{1+3w}}\left[ 1 + \frac{(\eta-\eta')^2 - (\vec{x}-\vec{x}')^2}{2\eta\eta'} \right]
	\sigma_{ij}[\eta',\vec{x}'] .
\end{align}
{\it Null waves} \qquad Before before moving on to the vector and scalar solutions, let us compare the cosmological solution of eq. \eqref{FirstOrderPTEqns_Tensor_Soln} and its Minkowski version, i.e., the TT part of eq. \eqref{LinearizedGravity_Minkowski_EinsteinEquation_deDonderContinuousSource_Evend} below. The scale factors in the former, multiplying $D_{ij}$ on the left-hand-side tells us, to leading order, the portion of tensor waves propagating strictly on the null cone -- including their associated memories -- is expected to redshift as $1/a[\eta]^{(d-2)/2}$ relative to their Minkowskian cousins, at least in all relevant even dimensions $d=4,6,8,\dots$.

{\bf Vector solutions} \qquad According to eq. \eqref{FirstOrderPTEqns_FluidVector}, if there were no perturbations to begin with in the asymptotic past, then $V_i - \dot{\Xi}_i = 0$ for all time. To see this, we integrate once to obtain
\begin{align}
\label{FirstOrderPTEqns_Vector0of2}
V_i[\eta,\vec{x}] - \dot{\Xi}_i[\eta,\vec{x}] 
= \left(\frac{a[\eta]}{a[\eta']}\right)^{(d-1)w-1} (V_i[\eta',\vec{x}] - \dot{\Xi}_i[\eta',\vec{x}]) ,
\end{align}
for some fixed initial time $\eta'$. If this initial time $\eta'$ is taken to be in the far past, so that we may assume no metric perturbations has been generated, this means $V_i[\eta',\vec{x}] = \dot{\Xi}_i[\eta',\vec{x}] = 0$ and therefore $V_i - \dot{\Xi}_i = 0$. Using this in eq. \eqref{FirstOrderPTEqns_Vector1of2} now hands us
\begin{align}
\label{FirstOrderPTEqns_Vector1of2_v2}
\frac{1}{2}\vec{\nabla}^2 V_i = 8\pi\GN \Sigma_i ,
\end{align}
which may be readily solved via the Green's function of the Laplacian in $(d-1)$-space,
\begin{align}
\label{FirstOrderPTEqns_Vector_SolnIofII}
V_i[\eta,\vec{x}] = -16\pi\GN \int_{\mathbb{R}^{d-1}} \dd^{d-1}\vec{x}' \frac{\Gamma\left[\frac{d-3}{2}\right] \cdot \Sigma_i[\eta,\vec{x}']}{4\pi^{ \frac{d-1}{2} } |\vec{x}-\vec{x}'|^{d-3}} .
\end{align}
Our assumption that $V_i$ tends to zero in the distant past therefore amounts to assuming that $\Sigma_i$ itself is negligible in the distant past. As a check of the consistency of eq. \eqref{FirstOrderPTEqns_Vector1of2_v2}, we may apply the operator $\partial_0 (a^{d-2} \cdot )/a^{d-2}$ on both sides, followed by using eq. \eqref{FirstOrderPTEqns_Vector2of2},
\begin{align}
8\pi\GN \vec{\nabla}^2 \sigma_i = 8\pi\GN \frac{\partial_0 (a^{d-2} \Sigma_i)}{a^{d-2}} ,
\end{align}
which is simply the conservation law of eq. \eqref{Astro_ConservationIIIofIII}.

Alternatively, we may also integrate eq. \eqref{FirstOrderPTEqns_Vector2of2} once to obtain $\partial_{\{i } V_{ j\} }$ in terms of the $\partial_{\{i } \sigma_{ j\} }$ component of $\,^{(\text{a})}T_{\mu\nu}$.
\begin{align}
\label{FirstOrderPTEqns_Vector_SolnIIofII}
\partial_{\{i } V_{ j\} }[\eta,\vec{x}]
= 16\pi\GN \int_{\eta'}^{\eta} \left(\frac{a[\eta'']}{a[\eta]}\right)^{d-2} \partial_{\{i } \sigma_{ j\} }[\eta'',\vec{x}] \dd\eta''
+ \left(\frac{a[\eta']}{a[\eta]}\right)^{d-2} \partial_{\{i } V_{ j\} }[\eta',\vec{x}].
\end{align} 
{\bf Scalar solution in de Sitter ($w=-1$)} \qquad In de Sitter spacetime, or equivalently when $w=-1$, the fluid equations become irrelevant. We may now set $w=-1$ in equations \eqref{FirstOrderPTEqns_Scalar1of4} and \eqref{FirstOrderPTEqns_Scalar2of4}, and eliminate $\Psi$ to arrive at
\begin{align}
\frac{d-2}{2} \vec{\nabla}^2 \Phi = 8\pi\GN \left( \rho - \frac{d-1}{\eta} \Sigma \right) .
\end{align}
The solution is
\begin{align}
\label{FirstOrderPTEqns_Scalar_dSSoln}
\Phi[\eta,\vec{x}] = -\frac{16\pi\GN}{d-2} \int_{\mathbb{R}^{d-1}} \dd^{d-1}\vec{x}' \frac{\Gamma\left[\frac{d-3}{2}\right]}{4\pi^{ \frac{d-1}{2} } |\vec{x}-\vec{x}'|^{d-3}} \left( \rho[\eta,\vec{x}'] + H a[\eta] (d-1) \Sigma[\eta,\vec{x}'] \right) ,
\end{align}
where we have introduced a constant Hubble parameter $H > 0$ and expressed the de Sitter scale factor as $a[\eta] \equiv -1/(H\eta)$.

{\bf Scalar solution in matter dominated universe ($w=0$)} \qquad By putting $w=0$ in eq. \eqref{FirstOrderPTEqns_Scalar3of4}, and invoking eq. \eqref{FirstOrderPTEqns_Scalar4of4} to eliminate $\Psi$, we immediately obtain an ordinary differential equation in time for $\Phi$, sourced by $\sigma$ and $\Upsilon$:
\begin{align}
\label{FirstOrderPTEqns_Scalar_MatterPDE}
\ddot{\Phi} + \frac{2(2 d-5)}{(d-3) \eta} \dot{\Phi} 
= \frac{\partial_0 \left( a^{2d-5} \dot{\Phi} \right)}{a^{2d-5}}
= 16\pi\GN \left( \frac{\partial_0 \left( a^{d-2} \Sigma \right)}{(d-2) \cdot a^{d-2}} + \frac{2 \dot{\Upsilon}}{(d-3) \eta} \right) .
\end{align}
If we assume the $\Phi$ and $\dot{\Phi}$ were zero in the distant past, then we may integrate twice to obtain its solution:
\begin{align}
\label{FirstOrderPTEqns_Scalar_MatterSoln}
\Phi[\eta,\vec{x}]
= 16\pi\GN \int_{\eta_\circ}^{\eta} \frac{\dd\eta'''}{a[\eta''']^{2d-5}} \int_{\eta_\circ}^{\eta'''} \dd\eta'' a[\eta'']^{2d-5} \left( \frac{\partial_0 \left( a[\eta'']^{d-2} \Sigma[\eta'',\vec{x}] \right)}{(d-2) \cdot a[\eta'']^{d-2}} 
+ \frac{2 \dot{\Upsilon}[\eta'',\vec{x}]}{(d-3) \eta''} \right) .
\end{align}
{\bf Scalar solution for $0 < w \leq 1$} \qquad The reason why the $w=-1$ and $w=0$ cases need to be treated separately is because the character of the differential equations involved becomes different at these 2 discrete points on the EoS parameter line. For instance, setting $w=0$ in eq. \eqref{FirstOrderPTEqns_FluidScalar} sets to zero its Laplacian term. As such, when $w=-1$ or $w=0$, the following manipulations will not be available to reduce the scalar equations to a single wave equation for $\Phi$. 

We first use the fluid eq. \eqref{FirstOrderPTEqns_FluidScalar} to replace the $\vec{\nabla}^2 \Xi$ in eq. \eqref{FirstOrderPTEqns_Scalar1of4} with an expression involving $\ddot{\Xi}$, $\dot{\Xi}$, $\Phi$ and $\Psi$. Then, to get rid of the remaining $\ddot{\Xi}$ and $\dot{\Xi}$ in eq. \eqref{FirstOrderPTEqns_Scalar1of4}, we first solve $\dot{\Xi}$ in terms of $\dot{\Phi}$, $\Psi$ and $\Sigma$ using eq. \eqref{FirstOrderPTEqns_Scalar2of4}; by taking one time derivative of this latter result, $\ddot{\Xi}$ can also be replaced in terms of $\dot{\Phi}$, $\ddot{\Phi}$, $\Psi$, $\dot{\Psi}$, $\Sigma$ and $\dot{\Sigma}$. 
Finally, we replace all $\Psi$ in terms of $\Phi$ using eq. \eqref{FirstOrderPTEqns_Scalar4of4}. After all these steps, one should arrive at
\begin{align}
\label{FirstOrderPTEqns_Scalar1of4_v2}
\ddot{\Phi} + \frac{2 ((d-1)w + 2d-5)}{\eta q_w} \dot{\Phi} - w \vec{\nabla}^2\Phi
&= 16\pi\GN \left( \frac{\partial_0 \left(a^{d-2} \Sigma\right)}{(d-2) a^{d-2}}
- \frac{w}{d-2} \rho
+ \frac{2 \dot{\Upsilon}}{\eta q_w} \right) .
\end{align}
\footnote{As a check of our calculations here, note that setting $d=4$ on the left hand side of eq. \eqref{FirstOrderPTEqns_Scalar1of4_v2} recovers that of eq. (7.57) in Mukhanov's textbook \cite{Mukhanov:2005sc}. (Mukhanov sets $\Psi=\Phi$.)}The left hand side of eq. \eqref{FirstOrderPTEqns_Scalar1of4_v2} is proportional to the scalar spacetime Laplacian $\,^{(\Phi)}\Box$ acting on $\Phi$, but in the geometry
\begin{align}
\label{FirstOrderPTEqns_ScalarEffectiveMetric}
\,^{(\Phi)}g_{\mu\nu} \dd x^\mu \dd x^\nu
\equiv \left(\frac{\eta}{\eta_0}\right)^{ 2 \left(\frac{2d-5 + (d-1)w}{(d-2)q_w}\right) } 
\left( (\dd\eta)^2 - \frac{1}{w} \dd\vec{x}\cdot\dd\vec{x} \right) ;
\end{align}
in detail,
\begin{align}
\left(\frac{\eta}{\eta_0}\right)^{ 4 \left(\frac{2d-5 + (d-1)w}{(d-2)q_w}\right) }  
\,^{(\Phi)}g^{\mu\nu} \nabla_\mu \nabla_\nu \Phi = \ddot{\Phi} + \frac{2 ((d-1)w + 2d-5)}{\eta q_w} \dot{\Phi} - w \vec{\nabla}^2\Phi .
\end{align}
{\it Physical range of EoS} \qquad For $w > 0$, the speed of the $\Phi$-wavefront is $\sqrt{w}$.\footnote{In cosmology-speak $c_s^2 \equiv w$ is commonly called the ``sound speed squared".} Therefore, in order for the scalar-hydrodynamical system to not admit superluminal signals, its EoS must be constrained to be $w \leq 1$. We now re-write the scalar mode wave equation \eqref{FirstOrderPTEqns_Scalar1of4_v2} as
\begin{align}
\frac{1}{a^{\frac{1}{2}(2d-5+w(d-1))}} &\left( \,^{(w)}\partial^2 - \frac{(d-2)(2d-5+w(d-1))}{q_w^2 \eta^2} \right)\left( a^{\frac{1}{2}(2d-5+w(d-1))} \Phi \right) \\
&=  16\pi\GN \left( \frac{\partial_0 \left(a^{d-2} \Sigma\right)}{(d-2) \cdot a^{d-2}} - w \frac{\rho}{d-2} 
+ \frac{2 \dot{\Upsilon}}{\eta q_w} \right) , \nonumber
\end{align}
with $\,^{(w)}\partial^2 \equiv \partial_0^2 - w \vec{\nabla}^2$. The $\Phi$ solution is
\begin{align}
\label{FirstOrderPTEqns_PhiSolution_w>0}
a[\eta]^{ \frac{1}{2}\left( 2d-5+(d-1)w \right) }\Phi[\eta,\vec{x}]
&= \frac{16\pi\GN}{w^{\frac{d-1}{2}}} \int_{\mathbb{R}^{d-1}} \dd^{d-1}\vec{x}' \int_{\eta_\circ}^\eta\dd\eta' a[\eta']^{ \frac{1}{2}\left( 2d-5+(d-1)w \right) } \mathcal{G}^{(\Phi)}[\eta,\eta';\sbar_w] \\
&\qquad\qquad\times
\left( \frac{\partial_0 \left(a[\eta']^{d-2} \Sigma[\eta',\vec{x}']\right)}{(d-2) \cdot a[\eta']^{d-2}} - w \frac{\rho[\eta',\vec{x}']}{d-2} + \frac{2 \dot{\Upsilon}[\eta',\vec{x}']}{q_w \eta'} \right) , \nonumber
\end{align} 
where the symmetric Green's functions are
\begin{align}
\label{FirstOrderPTEqns_PhiG_w>0_Evend}
\mathcal{G}^{(\Phi)}_{\text{even $d \geq 4$}}[\eta,\eta';\sbar_w]
&= \left( \frac{1}{2\pi} \frac{\partial}{\partial\sbar_w} \right)^{ \frac{d-2}{2} } \left( \frac{\Theta[\sbar_w]}{2} P_{ -\frac{2d-5 + (d-1)w}{q_w} }\left[1+s_w\right] \right)\\
\label{FirstOrderPTEqns_PhiG_w>0_Oddd}
\mathcal{G}^{(\Phi)}_{\text{odd $d \geq 5$}}[\eta,\eta';\sbar_w]
&= \left( \frac{1}{2\pi} \frac{\partial}{\partial\sbar_w} \right)^{\frac{d-3}{2}} \left(\frac{\Theta[\sbar_w]}{4\pi} \frac{ \left(s_w + \sqrt{s_w(s_w+2)} + 1\right)^{\frac{3d-7 + (d-1)w}{q_w}} + 1}{\sqrt{\sbar_w(s_w+2)} \left(s_w + \sqrt{s_w(s_w+2)} + 1\right)^{\frac{3d-7 + (d-1)w}{2q_w}}}
\right) , \\
\sbar_w &\equiv \frac{1}{2}\left( (\eta-\eta')^2 - \frac{(\vec{x}-\vec{x}')^2}{w} \right), \qquad\qquad s_w \equiv \frac{\sbar_w}{\eta\eta'} .
\end{align}
The presence of $1/w^{\frac{d-1}{2}}$ in eq. \eqref{FirstOrderPTEqns_PhiSolution_w>0} reminds us that the $w=0$ case -- considered separately above -- is disconnected from its $w>0$ counterpart. We also remark that, unlike the tensor mode above, the scalar mode radiation can be sourced by 3 irreducible components $(\Sigma,\rho,\Upsilon)$ of the energy-momentum-shear-stress of the astrophysical system. We again highlight the 4D case by writing down its retarded Green's function:
\begin{align}
\label{FirstOrderPTEqns_PhiG_w>0_4D}
\mathcal{G}^{(\Phi\vert+)}_{\text{4D},0 < w \leq 1}[\eta,\eta';|\vec{x}-\vec{x}'|]
&= \frac{1}{4\pi} \Bigg( \sqrt{w} \frac{\delta[\eta-\eta'-|\vec{x}-\vec{x}'|/\sqrt{w}]}{|\vec{x}-\vec{x}'|} \\
&\qquad\qquad
+ \frac{\Theta[\eta-\eta'-|\vec{x}-\vec{x}'|/\sqrt{w}]}{\eta\eta'} P'_{ -\frac{3(1+w)}{1+3w} }\left[1+\frac{(\eta-\eta')^2-(\vec{x}-\vec{x}')^2/w}{2\eta\eta'} \right] \Bigg) \nonumber .
\end{align}
As already pointed out by Tolish and Wald \cite{Tolish:2016ggo}, it is possible to generate Cherenkov $\Phi$-radiation if there are astrophysical sources moving faster than $\sqrt{w} < 1$ with respect to the cosmic rest frame.

{\it Negative EoS} \qquad When $w < 0$, the scalar mode equation \eqref{FirstOrderPTEqns_Scalar1of4_v2} becomes a Poisson-type one because the coefficient of the spatial Laplacian term is now positive. If one wishes to quantize the $\Phi$ fluctuations -- as is the case during the primordial epoch of the universe, when the initial seeds of inhomogeneities were believed to be laid down -- this would lead to pathologies because half of the $\Phi$'s mode functions, which solve the homogeneous equation
\begin{align}
\ddot{\Phi} + \frac{2 ((d-1)w + 2d-5)}{\eta q_w} \dot{\Phi} + |w| \vec{\nabla}^2\Phi = 0 ,
\end{align}
would grow unbounded in amplitude with increasing time. Specifically, in Fourier space, where $\Phi \to \widetilde{\Phi}[\eta] e^{i\vec{k}\cdot\vec{x}}$, one would find that $\widetilde{\Phi}[\eta]$ is a linear combination involving Hankel functions: $\eta^{(1-\kappa)/2} H^{(1)}_{(\kappa-1)/2}[-i|w|^{\frac{1}{2}}k\eta]$ and $\eta^{(1-\kappa)/2} H^{(2)}_{(\kappa-1)/2}[-i|w|^{\frac{1}{2}}k\eta]$, with $\kappa \equiv 2 ((d-1)w + 2d-5)/q_w$.\footnote{A simpler example would be to quantize the field $\Phi$ obeying $\vec{\nabla}^2 \Phi = 0$, in some $(d-1)$-Euclidean space, but insist that $x^1 \equiv t$ is `time'. By performing a Fourier transform with respect to the rest of the Cartesian coordinates $\vec{x}_\perp$, namely $\Phi[t,\vec{x}_\perp] \to \widetilde{\Phi}[t] e^{i\vec{k}_\perp \cdot \vec{x}_\perp}$, one would find that $\widetilde{\Phi}[t]$ is a linear combination of $\exp[\pm |\vec{k}_\perp|t]$.} At the purely classical level, even though we believe the relevant $w < 0$ Green's functions for eq. \eqref{FirstOrderPTEqns_Scalar1of4_v2} exists, the latter's Poisson-type nature means the ensuing solutions would depend on the sources $(\Sigma,\rho,\Upsilon)$ in the future of the observer's spacetime event. Altogether, these reasons compel us to restrict our attention to the EoS range $0 < w \leq 1$.

{\bf Fluid Solutions} \qquad When $w=-1$, or equivalently, about de Sitter spacetime, the fluid equations are irrelevant. 

When $w=0$, eq. \eqref{FirstOrderPTEqns_FluidScalar} allows us to integrate twice with respect to time and solve the longitudinal $\Xi$ portion of the fluid perturbation in terms of the scalar metric perturbation $\Psi$ and its value $\Xi[\eta',\vec{x}]$ at some initial time $\eta'$.
\begin{align}
\Xi[\eta,\vec{x}] 
= -\frac{1}{2} \int_{\eta'}^{\eta} \frac{\dd\eta'''}{a[\eta''']} \int_{\eta'}^{\eta'''} a[\eta'']\Psi[\eta'',\vec{x}] 
+ \Xi[\eta',\vec{x}].
\end{align}
For $0 < w \leq 1$, the longitudinal $\Xi$ portion of the fluid obeys a wave equation sourced by the scalar metric perturbations $\Phi$ and $\Psi$. We re-express eq. \eqref{FirstOrderPTEqns_FluidScalar} as
\begin{align}
\frac{1}{a^{ \frac{1}{2}\left( 1-(d-1)w \right) }} 
\left( \,^{(w)}\partial^2 - \frac{((d - 1) w-1) (d-4 + 2 (d-1)w)}{q_w^2 \eta^2} \right) 
&\left( a^{ \frac{1}{2}\left( 1-(d-1)w \right) } \Xi \right) \nonumber\\
&= -\frac{d-1}{2} w \Phi - \frac{1}{2}\Psi,
\end{align}
where $\,^{(w)}\partial^2 \equiv \partial_0^2 - w \vec{\nabla}^2$. We will not consider the ranges $w<0$ and $w>1$ for the same reasons when discussing the wave solutions for $\Phi$ above. The retarded solutions are
\begin{align}
a[\eta]^{ \frac{1}{2}\left( 1-(d-1)w \right) } \Xi[\eta ,\vec{x}]
= - \int_{\mathbb{R}^{d-1}} \dd^{d-1}\vec{x}' \int_{\eta_\circ}^{\eta} \dd\eta' & a[\eta']^{ \frac{1}{2}\left( 1-(d-1)w \right) } \mathcal{G}^{(\Xi)}[\eta,\eta';\sbar] \nonumber\\
&\times \left( \frac{1}{2} (d-1) w \Phi[\eta',\vec{x}'] + \frac{1}{2}\Psi[\eta',\vec{x}'] \right) ,
\end{align}
with the following symmetric Green's functions:
\begin{align}
\mathcal{G}^{(\Xi)}_{\text{even $d \geq 4$}}[\eta,\eta';\sbar] 
&= \left( \frac{1}{2\pi} \frac{\partial}{\partial\sbar_w} \right)^{ \frac{d-2}{2} } \left( \frac{\Theta[\sbar_w]}{2} P_{\frac{(d-1)w-1}{q_w}}[1+s_w] \right) , \\
\mathcal{G}^{(\Xi)}_{\text{odd $d \geq 5$}}[\eta,\eta';\sbar] 
&= \left( \frac{1}{2\pi} \frac{\partial}{\partial\sbar_w} \right)^{ \frac{d-3}{2} } 
\left( \frac{\Theta[\sbar_w]}{4\pi} \frac{1 + \left( s_w + \sqrt{s_w(s_w+2)} + 1 \right)^{ \frac{(d-5) + 3 (d-1) w}{q_w} }}{\sqrt{\sbar_w(s_w+2)} \left( s_w + \sqrt{s_w(s_w+2)} + 1 \right)^{ \frac{(d-5) + 3 (d-1) w}{2q_w} }} \right) , \\
\sbar_w &\equiv \frac{(\eta-\eta')^2 - (\vec{x}-\vec{x}')^2/w}{2}, \qquad\qquad
s_w \equiv \frac{\sbar_w}{\eta\eta'}, \qquad\qquad
0 < w \leq 1 .
\end{align}
Finally, for the solution of the transverse portion $\Xi_i$ of the fluid perturbation -- we recall in the discussion above, that we wish to implement the initial conditions $V_i[\eta_\circ,\vec{x}] = \Xi_i[\eta_\circ,\vec{x}] = 0$. This implies from eq. \eqref{FirstOrderPTEqns_Vector0of2} that $\dot{\Xi}_i = V_i$ for all times. Therefore, for any $w$, the transverse $\Xi_i$ portion of the fluid is
\begin{align}
\Xi_i[\eta,\vec{x}] = \int_{\eta'}^{\eta} V_i[\eta',\vec{x}] \dd\eta' + \Xi_i[\eta',\vec{x}] ,
\end{align}
where $\Xi_i[\eta',\vec{x}]$ is its value at some arbitrary initial time $\eta'$.

\subsection{Synchronous Gauge: Gravitational Waves and Their Memories}
\label{Section_PT_GWsAndMemory}
Gravitational memory is usually defined as the permanent fractional displacement experienced by a pair of co-moving test masses after the passage of a primary GW train emitted by a distant astrophysical source. We will begin with the observation that the synchronous gauge is the appropriate choice of coordinates to perform such a GW memory calculation. This is because, the geodesic equations obeyed by co-moving observers are satisfied automatically, and what remains is to compute the metric perturbations that occur in the fractional displacement formula of eq. \eqref{FractionalDistortion_AtConstantCosmicTime_FirstOrderHuman}. We will do so by converting the first order gauge-invariant results of the previous section to the synchronous gauge, and proceed to work out the GW signatures -- including potential memory effects -- from generic isolated astrophysical sources.

{\bf Synchronous gauge} \qquad The synchronous gauge refers to the choice of coordinates such that all metric perturbations occur solely within the spatial metric -- namely,
\begin{align}
\label{SpatiallFlatFLRW_PerturbedMetric_SynchronousGauge}
\dd s^2 = a[\eta]^2 \left( (\dd\eta)^2 - \left(\delta_{ij} - \chi_{ij}\right) \dd x^i \dd x^j \right) .
\end{align}
As can be verified by a direct calculation, the timelike trajectory with zero spatial velocity
\begin{align}
\label{SpatiallFlatFLRW_PerturbedMetric_SynchronousGauge_ComovingTrajectory}
Z^\mu = \left( \eta, \vec{Z}_0 \right) ,
\end{align}
with $\vec{Z}_0$ constant (i.e., it is co-moving) -- satisfies the geodesic equation 
\begin{align}
\frac{\dd^2 Z^\alpha}{\dd \tau^2} + \Gamma^\alpha_{\phantom{\alpha}\mu\nu} \frac{\dd Z^\mu}{\dd \tau} \frac{\dd Z^\nu}{\dd \tau} = 0 
\end{align}
to all orders in $\chi_{ij}$, with proper time $\tau$ defined through $\dd\tau = a[\eta]\dd\eta$. In other words: in the synchronous gauge metric, if a geodesic worldline has zero spatial proper velocity at a given instant in time, it is co-moving for all time. This allows us to identify, from eq. \eqref{SpatiallFlatFLRW_PerturbedMetric_SynchronousGauge}, that the induced metric on constant--time surfaces measures geodesic spatial distances between any pair of such co-moving geodesic test masses:
\begin{align}
\label{SpatiallFlatFLRW_PerturbedSpatialMetric_SynchronousGauge}
\dd \vec{L}^2 = a[\eta]^2 \left( \delta_{ij} - \chi_{ij} \right) \dd x^i \dd x^j  .
\end{align}
As we shall show in appendix \eqref{Section_GeodesicDistance}, assuming the hypothetical GW experiment does not take place over cosmological timescales, this leads us to the fractional displacement formula of eq. \eqref{FractionalDistortion_AtConstantCosmicTime_FirstOrderHuman}. Note that, by normalizing the tangent vector $\dd Z^\mu/\dd \tau$ of the trajectory in eq. \eqref{SpatiallFlatFLRW_PerturbedMetric_SynchronousGauge_ComovingTrajectory} to unity, namely
\begin{align}
\mathfrak{U}^\mu \equiv a^{-1} \delta^\mu_0 ,
\end{align}
we may re-express the synchronous gauge metric of eq. \eqref{SpatiallFlatFLRW_PerturbedMetric_SynchronousGauge} into the following form
\begin{align}
\label{SpatiallFlatFLRW_PerturbedMetric_SynchronousGauge_CovariantSplit}
g_{\mu\nu} = \mathfrak{U}_\mu \mathfrak{U}_\nu - \mathfrak{H}_{\mu\nu} ,
\end{align}
where $\mathfrak{H}_{\mu\nu} \dd x^\mu \dd x^\nu \equiv -\dd \vec{L}^2$ (cf. eq. \eqref{SpatiallFlatFLRW_PerturbedSpatialMetric_SynchronousGauge}) and $\mathfrak{U}^\mu \mathfrak{H}_{\mu\nu} = 0$. This puts into a generally covariant form the statement that, as long as the synchronous gauge coordinate system itself remains valid, the timelike geodesic observers of eq. \eqref{SpatiallFlatFLRW_PerturbedMetric_SynchronousGauge_ComovingTrajectory} have worldlines that always remain orthogonal to the constant conformal time (i.e., $\dd \eta = 0$) surfaces, since $U_\mu \dd x^\mu = a[\eta] \dd\eta$. 

Now, for eq. \eqref{FractionalDistortion_AtConstantCosmicTime_FirstOrderHuman} to be useful, we need to understand how to extract the synchronous gauge metric perturbations from the solutions to the first order gauge invariant metric perturbation variables we obtained in the previous section. By referring to eq. \eqref{Bardeen_SVTDecomposition}, we see synchronous gauge means 
\begin{align}
\label{SynchronousGauge_EandF}
E = F = F_i = 0 .
\end{align}
Starting with eq. \eqref{Bardeen_Psi}, $K$ can be determined from $\Psi$ by integrating $a^{-1} \partial_0 (a \dot{K}) = \Psi$:
\begin{align}
K[\eta,\vec{x}] &= \int_{\eta'}^{\eta}\frac{\dd\eta'''}{a[\eta''']} \int_{\eta'}^{\eta'''} a[\eta''] \Psi[\eta'',\vec{x}] \dd\eta'' 
+ \dot{K}[\eta',\vec{x}] \int_{\eta'}^{\eta} \frac{a[\eta']}{a[\eta'']} \dd\eta'' 
+ K[\eta',\vec{x}] ,
\end{align}
for some arbitrary but fixed initial time $\eta'$. Employing this solution for $K$, eq. \eqref{Bardeen_Phi} can be used to solve for the trace of the spatial metric perturbation:
\begin{align}
\frac{D[\eta,\vec{x}]}{d-1} 
= \Phi[\eta,\vec{x}] + \frac{\vec{\nabla}^2 K[\eta,\vec{x}]}{d-1}
+ \frac{\dot{a}[\eta]}{a[\eta]} \left( \int_{\eta'}^{\eta} \frac{a[\eta'']}{a[\eta]} \Psi[\eta'',\vec{x}] \dd\eta'' + \frac{a[\eta']}{a[\eta]} \dot{K}[\eta',\vec{x}] \right)  .
\end{align}
This implies the ``spatial-trace plus tidal-shear" part of the spatial metric perturbation in eq. \eqref{Bardeen_SVTDecomposition} is
\begin{align}
\label{SynchronousGaugeMetric_StepI}
\delta_{ij} \frac{D}{d-1} &+ \left( \partial_i \partial_j - \delta_{ij} \frac{\vec{\nabla}^2}{d-1} \right) K
= \delta_{ij} \left( \Phi[\eta,\vec{x}] 
+ \frac{\dot{a}[\eta]}{a[\eta]} \left( \int_{\eta'}^{\eta} \frac{a[\eta'']}{a[\eta]} \Psi[\eta'',\vec{x}] \dd\eta'' + \frac{a[\eta']}{a[\eta]} \dot{K}[\eta',\vec{x}] \right) \right) \nonumber \\
&+ \partial_i \partial_j \left\{ 
\int_{\eta'}^{\eta}\frac{\dd\eta'''}{a[\eta''']} \int_{\eta'}^{\eta'''} \dd\eta'' a[\eta''] \Psi[\eta'',\vec{x}]
+ \dot{K}[\eta',\vec{x}] \int_{\eta'}^{\eta} \frac{a[\eta']}{a[\eta'']} \dd\eta'' 
+ K[\eta',\vec{x}] 
\right\} .
\end{align}
Finally, eq. \eqref{Bardeen_VandDij} tells us the portion of the spatial metric involving the divergence-free vector is
\begin{align}
\label{SynchronousGaugeMetric_StepII}
\partial_{ \{ i } D_{ j \} }[\eta,\vec{x}]
&= - \int_{\eta'}^{\eta} \partial_{ \{ i } V_{ j \} }[\eta'',\vec{x}] \dd\eta'' 
		+ \partial_{ \{ i } D_{ j \} }[\eta',\vec{x}] .
\end{align}
We may now gather, by inserting equations \eqref{SynchronousGaugeMetric_StepI} and \eqref{SynchronousGaugeMetric_StepII} into eq. \eqref{Bardeen_SVTDecomposition}: given $D_{ij}$, $V_i$, $\Phi$ and $\Psi$, the synchronous gauge metric perturbation is 
\begin{align}
\label{SpatiallFlatFLRW_PerturbedSpatialMetric_SynchronousGaugeFromGaugeInvariant}
\chi_{ij}^{\text{(Syn.)}}[\eta \geq \eta',\vec{x}]
&= D_{ij}[\eta,\vec{x}] - \int_{\eta'}^{\eta} \partial_{ \{ i } V_{ j \} }[\eta'',\vec{x}] \dd\eta'' 
+ \partial_{ \{ i } D_{ j \} }[\eta',\vec{x}] \\
&+ \delta_{ij} \left\{ \Phi[\eta,\vec{x}] 
+ \frac{\dot{a}[\eta]}{a[\eta]} \left( \int_{\eta'}^{\eta} \frac{a[\eta'']}{a[\eta]} \Psi[\eta'',\vec{x}] \dd\eta'' + \frac{a[\eta']}{a[\eta]} \dot{K}[\eta',\vec{x}] \right) \right\} \nonumber \\
&+ \int_{\eta'}^{\eta}\frac{\dd\eta'''}{a[\eta''']} \int_{\eta'}^{\eta'''} \dd\eta'' a[\eta''] \partial_i \partial_j \Psi[\eta'',\vec{x}] 
+ \partial_i \partial_j \dot{K}[\eta',\vec{x}] \int_{\eta'}^{\eta} \frac{a[\eta']}{a[\eta'']} \dd\eta'' 
+ \partial_i \partial_j K[\eta',\vec{x}] . \nonumber
\end{align}
\footnote{That we have been able to build a purely-spatial metric perturbation in eq. \eqref{SpatiallFlatFLRW_PerturbedSpatialMetric_SynchronousGaugeFromGaugeInvariant}, given arbitrary gauge-invariant perturbations $(\Phi,\Psi,V_i,D_{ij})$, constitutes a constructive proof of the existence of the synchronous gauge itself.}The presence of the arbitrary initial metric perturbation $K[\eta',\vec{x}]$ and its velocity $\dot{K}[\eta',\vec{x}]$ indicates the synchronous gauge does not completely eliminate the freedom to alter the coordinate system. However, to recover a manifestly spatially flat FLRW cosmology when there are no GW sources -- i.e., when all the $(\Phi,\Psi,V_i,D_{ij})$ are zero -- we shall take the liberty to choose
\begin{align}
\label{SynchronousGauge_InitialZeroes}
K[\eta',\vec{x}] = \dot{K}[\eta',\vec{x}] = 0 .
\end{align}
Furthermore, since it was arbitrary up till this point, let us now define $\eta'$ to be the beginning of our hypothetical GW experiment, which we further suppose measures the $(\delta L/L_0)[\eta]$ in eq. \eqref{FractionalDistortion_AtConstantCosmicTime_FirstOrderHuman}. At this point, with eq. \eqref{SynchronousGauge_InitialZeroes} applied to eq. \eqref{SpatiallFlatFLRW_PerturbedSpatialMetric_SynchronousGaugeFromGaugeInvariant}, the total change in the spatial metric $\Delta \chi_{ij}^{\text{(Syn.)}}[\eta \geq \eta', \vec{x}] \equiv \chi_{ij}^{\text{(Syn.)}}[\eta,\vec{x}] - \chi_{ij}^{\text{(Syn.)}}[\eta',\vec{x}]$, required to calculate eq. \eqref{FractionalDistortion_AtConstantCosmicTime_FirstOrderHuman}, takes the expression
\begin{align}
\label{SpatiallFlatFLRW_PerturbedSpatialMetric_SynchronousGaugeFromGaugeInvariant_Delta}
\Delta \chi_{ij}^{\text{(Syn.)}}[\eta \geq \eta',\vec{x}]
&= \Delta D_{ij}[\eta,\vec{x}] - \int_{\eta'}^{\eta} \partial_{ \{ i } V_{ j \} }[\eta'',\vec{x}] \dd\eta'' \nonumber\\
&+ \delta_{ij} \left\{ \Delta \Phi[\eta,\vec{x}] 
+ \frac{\dot{a}[\eta]}{a[\eta]} \int_{\eta'}^{\eta} \frac{a[\eta'']}{a[\eta]} \Psi[\eta'',\vec{x}] \dd\eta'' \right\} \\
&+ \int_{\eta'}^{\eta}\frac{\dd\eta'''}{a[\eta''']} \int_{\eta'}^{\eta'''} \dd\eta'' a[\eta''] \partial_i \partial_j \left( (d-3)\Phi[\eta'',\vec{x}] - 16\pi\GN \Upsilon[\eta'',\vec{x}] \right) , \nonumber
\end{align}
where, through eq. \eqref{FirstOrderPTEqns_Scalar4of4}, we have replaced $\Psi$ in the double integral term with the appropriate linear combination of $\Phi$ and $\Upsilon$; and further defined
\begin{align}
\Delta D_{ij}[\eta \geq \eta',\vec{x}] 	&\equiv D_{ij}[\eta,\vec{x}] - D_{ij}[\eta',\vec{x}] , \\
\Delta \Phi[\eta \geq \eta',\vec{x}] 	&\equiv \Phi[\eta,\vec{x}] - \Phi[\eta',\vec{x}] .
\end{align}
In eq. \eqref{FirstOrderPTEqns_Vector_SolnIofII}, if we re-write $\Sigma_i$ in terms of the matter stress tensor $\,^{(\text{a})}T_{0i}$, and employ the result in the $\partial_{\{ i} V_{ j\} }$ term of eq. \eqref{SpatiallFlatFLRW_PerturbedSpatialMetric_SynchronousGaugeFromGaugeInvariant_Delta}, we may estimate -- by counting the number of spatial derivatives -- that this vector term likely scales as (timescale of GW experiment)/(observer-source spatial distance) relative to the dominant piece of $\Delta D_{ij}$. Assuming that the GW detector lies at sufficiently large distances away from the astrophysical source will then allow us to drop this $\partial_{ \{ i } V_{ j \} }$ term in eq. \eqref{SpatiallFlatFLRW_PerturbedSpatialMetric_SynchronousGaugeFromGaugeInvariant_Delta}. Next, if we infer from eq. \eqref{FirstOrderPTEqns_Scalar4of4} that $\Phi \sim \Psi$, then the single-integral term on the second line in eq. \eqref{SpatiallFlatFLRW_PerturbedSpatialMetric_SynchronousGaugeFromGaugeInvariant_Delta} should scale relative to $\Delta \Phi$ as, very roughly, (timescale of GW experiment)/(age of universe) -- due to the presence of the $\dot{a}/a$ pre-factor. 
In the last line of eq. \eqref{SpatiallFlatFLRW_PerturbedSpatialMetric_SynchronousGaugeFromGaugeInvariant_Delta}, let us assume that the astrophysical source is sufficiently localized so that $\GN \partial_i \partial_j \Upsilon \sim (\Phi,\GN \Upsilon)/$(observer-source spatial distance)$^2$; moreover, over the timescales of human experiments the $a[\eta'']$ and $a[\eta''']$ terms roughly cancel each other. Altogether, for distant astrophysical sources and for human GW experiments, eq. \eqref{SpatiallFlatFLRW_PerturbedSpatialMetric_SynchronousGaugeFromGaugeInvariant_Delta} is reduced to
\begin{align}
\label{SpatiallFlatFLRW_PerturbedSpatialMetric_SynchronousGaugeFromGaugeInvariant_Delta_v2}
\Delta \chi_{ij}^{\text{(Syn.)}}[\eta \geq \eta',\vec{x}]
&\approx \Delta D_{ij}[\eta,\vec{x}] 
+ \delta_{ij} \Delta \Phi[\eta,\vec{x}] 
+ (d-3) \int_{\eta'}^{\eta} \dd\eta''' \int_{\eta'}^{\eta'''} \dd\eta'' \partial_i \partial_j \Phi[\eta'',\vec{x}] .
\end{align}
{\bf GWs and their memories in de Sitter ($w=-1$)} \qquad As we will now proceed to further argue, for the de Sitter $(w=-1)$ and matter dominated $(w=0)$ cases, the $\Phi$ contribution in eq. \eqref{SpatiallFlatFLRW_PerturbedSpatialMetric_SynchronousGaugeFromGaugeInvariant_Delta_v2} to eq. \eqref{FractionalDistortion_AtConstantCosmicTime_FirstOrderHuman} may in fact be discarded.
\begin{align}
\label{FractionalDistortion_AtConstantCosmicTime_FirstOrderHuman_dSMatter}
\left(\frac{\delta L}{L_0}\right)[\eta > \eta';\text{de Sitter, Matter}] 
= - \frac{\widehat{n}^i \widehat{n}^j}{2} \int_{0}^{1} \Delta D_{ij}[\vec{Y}_0 + \lambda(\vec{Z}_0-\vec{Y}_0)] \dd \lambda 
\end{align}
The Poisson-type solution for $\Phi$ in eq. \eqref{FirstOrderPTEqns_Scalar_dSSoln} suggests it may not fluctuate as rapidly as the wave solutions of its tensor mode counterpart in eq. \eqref{FirstOrderPTEqns_Tensor_Soln}, and therefore $\Delta \Phi$ is sub-dominant relative to $\Delta D_{ij}$. To check this intuition we apply the operator $\partial_0(a^{d-2} \cdot )/a^{d-2}$ on both sides of eq. \eqref{FirstOrderPTEqns_Scalar_dSSoln}, and use the conservation laws in equations \eqref{Astro_ConservationIofIII} and \eqref{Astro_ConservationIIofIII} to reveal
\begin{align}
\label{FirstOrderPTEqns_Scalar_dSSoln_TimeDerivative}
\frac{\partial_0 \left( a^{d-2} \Phi \right)}{a^{d-2}}
&= 16\pi\GN \left(\frac{\Sigma}{d-2} + H a \cdot \Upsilon\right) - H a \cdot \Phi .
\end{align}
Far away from the source, we first recall the relation $\Sigma = \vec{\nabla}^{-2} \partial_l \,^{(\text{a})}T_{0l}$; which then allows us to argue that $\GN \Sigma \sim (r_s/r)^{d-3} \times (v/r)$; where $r$ is the observer-source spatial distance, and if $r_s$ is the Schwarzschild radius and $v$ is the typical speed(s) of the internal constituents of the source, we have $r_s^{d-3} \cdot v \sim \GN \int \dd^{d-1}\vec{x}' T_{0i} \sim \GN \times \text{(mass of GW source)} \times v$. By comparing the $H$-independent terms of equations \eqref{FirstOrderPTEqns_Scalar_dSSoln} and \eqref{FirstOrderPTEqns_Scalar_dSSoln_TimeDerivative}, we thus estimate that $\dot{\Phi}/\Phi$ should scale as $v/r$, plus additional terms that are suppressed by $H \sim 1/$(age of universe). These considerations lead us to interpret eq. \eqref{FirstOrderPTEqns_Scalar_dSSoln_TimeDerivative} to mean that $\Phi$ varies appreciably only in the near zone or over cosmological timescales, and may be dropped in the formula of eq. \eqref{SpatiallFlatFLRW_PerturbedSpatialMetric_SynchronousGaugeFromGaugeInvariant_Delta_v2}.

{\it Tail-induced memory} \qquad We have already observed in \cite{Chu:2016qxp} that the tail of the TT tensor mode solution in eq. \eqref{FirstOrderPTEqns_Tensor_Soln}, for all even dimensional de Sitter spacetimes higher than two, is a spacetime constant. Through eq. \eqref{FractionalDistortion_AtConstantCosmicTime_FirstOrderHuman_dSMatter}, this in turn leads a spacetime constant permanent fractional distortion of the space around the GW detector. Let $H>0$ be the constant Hubble parameter for de Sitter spacetime, so that $a[\eta] = -1/(H\eta)$. Then, eq. \eqref{FractionalDistortion_AtConstantCosmicTime_FirstOrderHuman_dSMatter} receives the following contribution over the duration $\eta \in [\eta_\text{i}, \eta_\text{f}]$ of active GW generation:
\begin{align}
\Delta D_{ij}^{\text{(tail)}}[\eta,\vec{x}]
&= -16\pi \GN \frac{H^{d-2}}{(2\pi)^{\frac{d-2}{2}}} \frac{(d-2)!}{2^{\frac{d}{2}} \left(\frac{d-2}{2}\right)!} 
\int_{\mathbb{R}^{d-1}} \dd^{d-1}\vec{x}' \int_{\eta_\text{i}}^{\eta_\text{f}} \dd\eta' a[\eta']^{d-2} \sigma_{ij}[\eta',\vec{x}'] .
\end{align}
This has no analog in even-dimensional $d \geq 4$ flat spacetime because the memory effect there falls off as $1/\text{(observer-source spatial distance)}^{d-3}$.

{\bf GWs and their memories in matter dominated universes ($w=0$)} \qquad Over timescales of human GW experiments, the cosmic scale factors do not evolve significantly and, similar in spirit to the $w=-1$ case, we may assert that $\Delta \Phi \sim \GN \int \dd\eta''' \Sigma \sim v \cdot (\text{timescale of GW experiment}/r) \times (r_s/r)^{d-3}$ plus the cosmic-time (i.e., $1/\eta''' \sim 1/$(age of universe)) suppressed $\Upsilon$ term, where $r \equiv$ observer-source spatial distance. This suggests the $\Phi$-terms are suppressed by $(\text{timescale of GW experiment}/r)$ relative to the dominant $1/r^{d-3}$ behavior of $\Delta D_{ij}$, and we thus continue to neglect the former in eq. \eqref{SpatiallFlatFLRW_PerturbedSpatialMetric_SynchronousGaugeFromGaugeInvariant_Delta_v2} when $w=0$.

{\it Tail-induced memory} \qquad We have already observed in \cite{Chu:2015yua} and \cite{Chu:2016qxp} that the TT tensor mode solution in eq. \eqref{FirstOrderPTEqns_Tensor_Soln}, for $(3+1)$-dimensional matter dominated universes, has a tail that decays with increasing time but is otherwise space-independent. Over timescales of human GW experiments, this leads to a tail-induced memory effect very similar to the de Sitter case just mentioned above, with no analog in flat spacetime. The contribution to eq. \eqref{FractionalDistortion_AtConstantCosmicTime_FirstOrderHuman_dSMatter} is
\begin{align}
\Delta D_{ij}^{(\text{4D, tail})}[\eta_r > \eta_{\text{f}},\vec{x}]
&= -4 \GN \int_{\mathbb{R}^{3}} \dd^{3}\vec{x}' \int_{\eta_\text{i}}^{\eta_\text{f}} \dd\eta' \frac{a[\eta']}{a[\eta] \cdot \eta\eta'} 
\sigma_{ij}[\eta',\vec{x}'] .
\end{align}
{\bf GWs and their memories for $0 < w \leq 1$} \qquad We now turn to the scenario where both $\Phi$ and $D_{ij}$ admit wave solutions. Let us first split the tensor and scalar modes into their null/acoustic cone (``direct") and tail terms:
\begin{align}
D_{ij} 	&= D_{ij}^{(\text{direct})} + D_{ij}^{\text{(tail)}} , \\
\Phi 	&= \Phi^{(\text{direct})} + \Phi^{\text{(tail)}} .
\end{align}
Following that, we may also split eq. \eqref{SpatiallFlatFLRW_PerturbedSpatialMetric_SynchronousGaugeFromGaugeInvariant_Delta_v2} into its null/acoustic cone and tail parts:
\begin{align}
\label{SpatiallFlatFLRW_PerturbedSpatialMetric_SynchronousGaugeFromGaugeInvariant_Delta_DirectVsTail}
\Delta \chi_{ij}^{\text{(Syn.)}}
= \Delta \chi_{ij}^{\text{(Syn.$\vert$direct)}} + \Delta \chi_{ij}^{\text{(Syn.$\vert$tail)}} ,
\end{align}
where
\begin{align}
\label{SpatiallFlatFLRW_PerturbedSpatialMetric_SynchronousGaugeFromGaugeInvariant_Delta_Direct}
\Delta \chi_{ij}^{\text{(Syn.$\vert$direct)}}[\eta \geq \eta',\vec{x}]
&\approx \Delta D_{ij}^{\text{(direct)}}[\eta,\vec{x}] 
+ \delta_{ij} \Delta \Phi^{\text{(direct)}}[\eta,\vec{x}] \\
&\qquad\qquad\qquad
+ (d-3) \int_{\eta'}^{\eta} \dd\eta''' \int_{\eta'}^{\eta'''} \dd\eta'' \partial_i \partial_j \Phi^{\text{(direct)}}[\eta'',\vec{x}] , \nonumber\\
\label{SpatiallFlatFLRW_PerturbedSpatialMetric_SynchronousGaugeFromGaugeInvariant_Delta_Tail}
\Delta \chi_{ij}^{\text{(Syn.$\vert$tail)}}[\eta \geq \eta',\vec{x}]
&\approx \Delta D_{ij}^{\text{(tail)}}[\eta,\vec{x}] 
+ \delta_{ij} \Delta \Phi^{\text{(tail)}}[\eta,\vec{x}] \\
&\qquad\qquad\qquad
+ (d-3) \int_{\eta'}^{\eta} \dd\eta''' \int_{\eta'}^{\eta'''} \dd\eta'' \partial_i \partial_j \Phi^{\text{(tail)}}[\eta'',\vec{x}] . \nonumber
\end{align}
In what follows, we shall repeatedly take the far zone limit, where the observer is assumed to be located at a distance from the astrophysical body very large compared to the latter's characteristic size. At leading order, if $\vec{x}$ and $\vec{x}'$ respectively refer to the observer and source spatial locations, this amounts to the replacement $|\vec{x} - \vec{x}'| \to |\vec{x}|$ as long as the source is sufficiently spatially localized. (This also means we have defined $\vec{x}=\vec{x}'=\vec{0}$ to lie within the astrophysical system.)

{\it Tails} \qquad We begin by seeking, for all $d=4,6,8,\dots$, whether there is a space-independent tail induced GW memory effect -- similar to the de Sitter and matter dominated cases just discussed -- in cosmologies where the dominant pressure-to-energy density ratio is positive and non-trivial, i.e., $0 < w \leq 1$. 

Now, the tail of the tensor mode Green's function is (cf. eq. \eqref{FirstOrderPTEqns_Tensor_GSoln_Evend})
\begin{align}
\label{TensorMemory_GTail}
\mathcal{G}^{\text{(T$\vert$tail)}}_{\text{even $d \geq 4$}}[\eta,\eta';\sbar] 
= \frac{\Theta[\sbar]}{2} \left( \frac{1}{2\pi\eta\eta'} \right)^{\frac{d-2}{2}} 
	\left(\frac{\partial}{\partial s}\right)^{\frac{d-2}{2}} \,_2F_1\left[ \frac{d-2}{q_w}, 1-\frac{d-2}{q_w};1;-\frac{s}{2}\right] ; 
\end{align}
and that of the $\Phi$ Green's function is (cf. eq. \eqref{FirstOrderPTEqns_PhiG_w>0_Evend})
\begin{align}
\label{ScalarMemory_GTail}
\mathcal{G}^{(\Phi\vert\text{tail})}_{\text{even $d \geq 4$}}[\eta,\eta';\sbar_w]
= \frac{\Theta[\sbar_w]}{2} \left( \frac{1}{2\pi \eta\eta'}\right)^{ \frac{d-2}{2} } & \left( \frac{\partial}{\partial s_w} \right)^{ \frac{d-2}{2} } \\
& \times \,_2F_1\left[ \frac{2d-5 + (d-1)w}{q_w}, 1-\frac{2d-5 + (d-1)w}{q_w};1;-\frac{s_w}{2}\right] . \nonumber
\end{align}
We have converted $\partial_{\sbar}$'s to $\partial_s$'s, and used the hypergeometric-function equivalent expression for the Legendre function, namely $P_{\nu}[1+s]=\,_2F_1[-\nu,1+\nu;1;-s/2]$ (see eq. 8.702 of \cite{G&S}). The reason for doing so is the following. To obtain a space-independent answer from the $(d-2)/2$ derivatives of $\,_2F_1$ in equations \eqref{TensorMemory_GTail} and \eqref{ScalarMemory_GTail}, the $\,_2F_1$ needs to be a polynomial in $-s/2$ (or $-s_w/2$) of degree $(d-2)/2$. To have a trivial tail, we would need the $\,_2F_1$ to be a polynomial of degree $0$ through $-1 + (d-2)/2$. In fact, $\,_2F_1[\alpha,\beta;\gamma;z]$ is a polynomial in $z$ of non-negative integer degree $\ell$ whenever either $\alpha = -\ell$ or $\beta = -\ell$. To obtain a space-independent tail for the tensor waves, therefore,
\begin{align}
\frac{d-2}{q_w} + \frac{d-2}{2} = 0 \qquad\qquad \text{ or } \qquad\qquad
1-\frac{d-2}{q_w} + \frac{d-2}{2} = 0 ;
\end{align}
and to achieve the same for the scalar waves,
\begin{align}
\frac{2d-5 + (d-1)w}{q_w} + \frac{d-2}{2} = 0 \qquad\qquad \text{ or } \qquad\qquad
1-\frac{2d-5 + (d-1)w}{q_w} + \frac{d-2}{2} = 0 .
\end{align}
Some algebra would inform us that, no such tensor mode solution exists for all relevant even dimensions; whereas no such scalar solution can be found for even $d \geq 6$. 

However, in $(3+1)$-dimensional radiation dominated universes -- i.e., with $w=1/3$ -- cosmological scalar GW tails do not decay with increasing distance from the astrophysical source along the acoustic cone $\eta - \sqrt{3}|\vec{x}| = $ constant:
\begin{align}
\label{ScalarMemory_Tail}
\Phi^{\text{(tail)}}_{\text{4D}, w=1/3}[\eta,\vec{x}]
= \frac{4 \GN}{\eta_0^2 a[\eta]^3} \cdot 3^{\frac{3}{2}} \int_{\mathbb{R}^{3}} & \dd^{3}\vec{x}' \int_{0}^{\eta - \sqrt{3}|\vec{x} - \vec{x}'| - 0^+} \dd\eta' a
\left( \frac{\partial_0 \left(a^{2} \Sigma\right)}{2 a^{2}} - \frac{\rho}{6} 
+ \frac{2 \dot{\Upsilon}}{\eta' q_w} \right) .
\end{align}
As a reminder, the scale factor of a radiation dominated 4D spatially flat universe is $a_{w=1/3}[\eta] = \eta/\eta_0$.

In the far zone and assuming the characteristic timescale of the astrophysical source is much shorter than that of cosmic expansion, acting $\partial_i \partial_j$ on eq. \eqref{ScalarMemory_Tail} yields at leading order
\begin{align}
\label{ScalarMemory_Tail_partialij}
\partial_i \partial_j \Phi^{\text{(tail)}}_{\text{4D}, w=1/3}[\eta,\vec{x}]
&\approx 3 \widehat{r}_i \widehat{r}_j \partial_\eta^2 \Phi^{\text{(tail)}}_{\text{4D}, w=1/3}[\eta,\vec{x}] ,
\qquad\qquad
\widehat{r}_i \equiv \frac{x^i}{|\vec{x}|} ,
\end{align}
plus terms that are suppressed relative to the one on the right hand side by 1/(age of the universe) or by 1/(observer-source spatial distance). In a 4D radiation dominated universe, the contribution of the tail effect in eq. \eqref{SpatiallFlatFLRW_PerturbedSpatialMetric_SynchronousGaugeFromGaugeInvariant_Delta_Tail} to gravitational memory in eq. \eqref{FractionalDistortion_AtConstantCosmicTime_FirstOrderHuman} is therefore
\begin{align}
\label{SpatiallFlatFLRW_PerturbedSpatialMetric_SynchronousGaugeFromGaugeInvariant_Delta_Tail_Radiation}
\Delta \chi_{ij}^{\text{(Syn.$\vert$4D tail, $0 < w \leq 1$)}}[\eta \geq \eta',\vec{x}]
&\approx \Delta D_{ij}^{\text{(4D tail, $0 < w \leq 1$)}}[\eta,\vec{x}] \\
&+ \left( \delta_{ij} + 3 \widehat{r}_i \widehat{r}_j \right) \Delta \Phi^{\text{(tail)}}_{\text{4D}, w=1/3}[\eta,\vec{x}] 
- 3 \widehat{r}_i \widehat{r}_j (\eta-\eta') \partial_{\eta'} \Phi^{\text{(tail)}}_{\text{4D}, w=1/3}[\eta',\vec{x}] . \nonumber
\end{align}
Suppose the sources were actively producing $\Phi$-waves over the time interval $\eta' \in [\eta_\text{i},\eta_\text{f}]$; and suppose the GW experiment began way before this active period (i.e., $\eta' \ll \eta_\text{i} - |\vec{x}|$) so that $\partial_{\eta'} \Phi^{\text{(tail)}}[\eta',\vec{x}] = 0$. Then, the associated scalar contribution to the memory effect in eq. \eqref{SpatiallFlatFLRW_PerturbedSpatialMetric_SynchronousGaugeFromGaugeInvariant_Delta_Tail_Radiation} -- after the primary wave train has passed the observer at $(\eta,\vec{x})$ -- can be entirely attributed to the space-independent expression
\begin{align}
\Delta \Phi^{\text{(tail)}}_{\text{4D}, w=1/3}\left[ \eta \gtrsim \eta_{\text{f}} - |\vec{x}| \right]
= \frac{4 \GN}{\eta_0^2 a[\eta]^3} \cdot 3^{\frac{3}{2}} \Theta_p & \int_{\mathbb{R}^{3}} \dd^{3}\vec{x}' \int_{\eta_\text{i}}^{\eta_\text{f}} \dd\eta' a 
\left( \frac{\partial_0 \left(a^{2} \Sigma\right)}{2 a^{2}} - \frac{\rho}{6} + \frac{2 \dot{\Upsilon}}{q_w \cdot \eta'} \right) .
\end{align}
\footnote{This memory effect will be zero if there is some way for this cumulative $\Phi$-signal to destructively interfere over the time interval $\eta' \in [\eta_\text{i},\eta_\text{f}]$. Such destructive interference does occur for the tensor mode in even dimensional de Sitter higher than two \cite{Chu:2016qxp} whenever both the pressure and mass quadrupole moments of the astrophysical system take the same values at $\eta = \eta_\text{i}$ and at $\eta = \eta_\text{f}$; i.e., whenever they revert to their original values.}Just like the tensor GWs in 4D matter dominated universes, these scalar GWs would likely yield a tail induced GW memory effect that is roughly spacetime constant over timescales of human experiments.

Finally, we find there are no even dimensional $D_{ij}$--GW tails when $w=1/(d-1)$; in 4D this corresponds to the radiation dominated universe, where we have already noted the absence of tensor tails in \cite{Chu:2015yua}. In the same vein, there are no $\Phi$--GW tails when $w=1/(d-1)$ for even $d \geq 6$.

{\it 4D Acoustic Cone} \qquad For $0 < w \leq 1$, i.e., without specializing to radiation domination, the scalar wavefront in the same 4D cosmology -- which propagate strictly at speed $\sqrt{w}$ -- take the form
\begin{align}
\label{ScalarMemory_NullCone}
a[\eta]^{ \frac{3}{2}\left( 1+w \right) }\Phi^{\text{(direct)}}_{\text{4D},0 < w \leq 1}[\eta,\vec{x}]
= \frac{4\GN}{w} \Theta_p \int_{\mathbb{R}^{3}} \dd^{3}\vec{x}' \frac{a[\eta_r]^{ \frac{3}{2}\left( 1+w \right) }}{|\vec{x}-\vec{x}'|} 
\left( \frac{\partial_{0} \left(a^{2} \Sigma\right)}{2 a^{2}} - w \frac{\rho}{2} + \frac{2 \dot{\Upsilon}}{\eta_r q_w} \right) , 
\end{align}
where all times within the spatial integral are evaluated at retarded time $\eta_r \equiv \eta - |\vec{x}-\vec{x}'|/\sqrt{w}$. In the far zone limit, acting $\partial_i \partial_j$ on this eq. \eqref{ScalarMemory_NullCone} yields
\begin{align}
\label{partialijPhiDirect}
\partial_i \partial_j \Phi^{\text{(4D direct)}}
\approx \frac{\widehat{r}_i \widehat{r}_j}{w} \partial_0^2 \Phi^{\text{(4D direct)}}, 
\qquad\qquad \widehat{r}_i \equiv \frac{x^i}{|\vec{x}|} .
\end{align}
plus corrections that scale as 1/(observer-source spatial distance) relative to this leading order expression. Eq. \eqref{partialijPhiDirect} implies eq. \eqref{SpatiallFlatFLRW_PerturbedSpatialMetric_SynchronousGaugeFromGaugeInvariant_Delta_Direct} becomes
\begin{align}
\label{SpatiallFlatFLRW_PerturbedSpatialMetric_SynchronousGaugeFromGaugeInvariant_Delta_4DDirect}
\Delta \chi_{ij}^{\text{(Syn.$\vert$4D direct)}}[\eta \geq \eta',\vec{x}]	
&\approx \Delta D_{ij}^{\text{(4D direct)}}[\eta,\vec{x}] \\
&+ \left( \delta_{ij} + \frac{\widehat{r}_i \widehat{r}_j}{w} \right)\Delta \Phi^{\text{(4D direct)}}[\eta,\vec{x}] 
- \frac{\widehat{r}_i \widehat{r}_j}{w} (\eta-\eta') \partial_{\eta'} \Phi^{\text{(4D direct)}}[\eta',\vec{x}] . \nonumber
\end{align}
Suppose the GW experiment was operational way before the active period of scalar GW production, so that $\partial_{\eta'} \Phi^{\text{(4D direct)}}[\eta',\vec{x}] = 0$. As already alluded to for the tensor GWs, this portion of the scalar GWs that travels strictly on the acoustic-cone should produce a non-zero memory effect whenever the matter components involved have different configurations well after and well before the duration of dominant GW production, whose peak cosmic time we will denote as $\eta_\star$. Specifically, the scalar GW memory effect at the far zone detector location, after the primary GW train has gone by, contributes to eq. \eqref{SpatiallFlatFLRW_PerturbedSpatialMetric_SynchronousGaugeFromGaugeInvariant_Delta_4DDirect} in the following manner:
\begin{align}
\label{ScalarMemory_NullCone_Delta}
\Delta \Phi^{\text{(direct)}}_{\text{4D},0 < w \leq 1}[\eta,\vec{x}]
&\approx \frac{4\GN}{w} \Theta_p \Delta \int_{\mathbb{R}^{3}} \dd^{3}\vec{x}' \frac{a^{ \frac{3}{2}\left( 1+w \right) }}{|\vec{x}-\vec{x}'|} 
\left( \frac{\partial_{0} \left(a^{2} \Sigma\right)}{2 a^{2}} - w \frac{\rho}{2} + \frac{2 \dot{\Upsilon}}{\eta q_w} \right) ,
\end{align}
where we have defined
{\allowdisplaybreaks\begin{align}
\Delta \int_{\mathbb{R}^{3}} \dd^{3}\vec{x}' \frac{a[\eta]^{ \frac{3}{2}\left( 1+w \right) }}{|\vec{x}-\vec{x}'|} 
&\left( \frac{\partial_{0} \left(a^{2} \Sigma\right)}{2 a^{2}} - w \frac{\rho}{2} + \frac{2 \dot{\Upsilon}}{\eta q_w} \right) \nonumber\\
&\equiv
\lim_{\eta_r \gg \eta_\star} 
\int_{\mathbb{R}^{3}} \dd^{3}\vec{x}' \frac{a[\eta_r]^{ \frac{3}{2}\left( 1+w \right) }}{|\vec{x}-\vec{x}'|} 
\left( \frac{\partial_{0} \left(a^{2} \Sigma\right)}{2 a^{2}} - w \frac{\rho}{2} + \frac{2 \dot{\Upsilon}}{\eta_r q_w} \right) \\
&\qquad\qquad
- \lim_{\eta_r \ll \eta_\star}
\int_{\mathbb{R}^{3}} \dd^{3}\vec{x}' \frac{a[\eta_r]^{ \frac{3}{2}\left( 1+w \right) }}{|\vec{x}-\vec{x}'|} 
\left( \frac{\partial_{0} \left(a^{2} \Sigma\right)}{2 a^{2}} - w \frac{\rho}{2} + \frac{2 \dot{\Upsilon}}{\eta_r q_w} \right) . \nonumber
\end{align}}
This is an appropriate place to reiterate the estimates made in \cite{Chu:2015yua}, that cosmological GW tails are suppressed relative to their null/acoustic cone counterparts whenever (I) the observer-source spatial distance is small compared to the characteristic size of the observable universe; and (II) the duration of the GW source is short relative to the age of the universe.\footnote{Although the estimates in \cite{Chu:2015yua} were performed for the tensor mode $D_{ij}$ only, they should carry over to the scalar ones $(\Phi,\Psi)$ without appreciable differences because the only relevant objects are time integrals, scale factors and their derivatives.} Therefore, the acoustic cone scalar-GW memory effect encoded within eq. \eqref{ScalarMemory_NullCone_Delta} is expected to dominate over the corresponding tail contribution for most physical cases of interest.

Because the $\Sigma$ and $\Upsilon$ terms of eq. \eqref{ScalarMemory_NullCone_Delta} are non-local functions of the astrophysical stress tensor $\,^{(\text{a})}T_{\mu\nu}$, this poses technical challenges in finding full-fledged explicit examples of scalar memory in cosmology -- we hope to return to these issue in the future.\footnote{In a previous version of this paper, by invoking the conservation eq. \eqref{Astro_ConservationIIofIII} to replace the $\Sigma$ term of eq. \eqref{ScalarMemory_NullCone_Delta} with ones involving $\sigma$ and $\Upsilon$, we had mistakenly thought that the $\sigma$ and $\rho$ terms would lead explicitly to a memory effect for the Tolish-Wald scattering setup. However, momentum conservation likely prohibits it \cite{PrivateCommunicationTW}. We have thus reinstated the $\Sigma$ term throughout our equations.}

We remind the reader, the 4D null-cone $D_{ij}$-memory was not discussed in this section, because we already did so around eq. \eqref{GWMemory_NullCone}.

{\bf Summary} \qquad We summarize the 4D wave properties of the first order gauge-invariant metric perturbation variables in the following table. GW tensor $D_{ij}$ null-cone and scalar $(\Phi,\Psi)$ acoustic-cone propagation will produce memory whenever the relevant components of the astrophysical source (cf. equations \eqref{Astro_SVT_IofII} and \eqref{Astro_SVT_IIofII}) -- i.e., $\sigma_{ij}$ for the tensor, and $(\Sigma, \rho, \Upsilon)$ for the scalar -- do not have the same configuration in the asymptotic future as they did in the asymptotic past. Whenever there are space-independent tails, we believe it is likely there will be an associated GW memory effect, unless destructive interference occurs. We have also provided physical arguments to persuade the reader, it is only the gauge-invariant variables admitting wave solutions that contribute to the fractional distortion $\delta L/L_0$ (cf. eq. \eqref{FractionalDistortion_AtConstantCosmicTime_FirstOrderHuman}) measured by our GW experiment.
\begin{center}
\begin{tabular}{| l | c | c | c |}
\hline
Equation-of-state $w$				& Wave Equations? 		& Space-Independent Tails 		& No tails	\\ \hline
de Sitter ($w=-1$) 					& $D_{ij}$				& $D_{ij}$ (spacetime constant)	& 			\\
Matter domination ($w=0$)			& $D_{ij}$				& $D_{ij}$						&			\\
Radiation domination ($w=1/3$)		& $D_{ij},\Phi,\Psi$	& $\Phi,\Psi$					& $D_{ij}$	\\
$0 < w \leq 1$, $w \neq 1/3$		& $D_{ij},\Phi,\Psi$	&								&			\\
\hline
\end{tabular}
\end{center}
In higher dimensions, $d > 4$, the results are summarized as follows. The rightmost 2 columns apply for even dimensions only.
\begin{center}
\begin{tabular}{| l | c | c | c |}
\hline
Equation-of-state $w$				& Wave Equations? 		& Space-Independent Tails 				& No tails 				\\ \hline
de Sitter ($w=-1$) 					& $D_{ij}$				& $D_{ij}$ (spacetime constant)			& 						\\
Matter domination ($w=0$) 			& $D_{ij}$				& 										& 						\\
$w=1/(d-1)$							& $D_{ij},\Phi,\Psi$	& 										& $D_{ij}, \Phi, \Psi$	\\
$0 < w \leq 1$, $w \neq 1/(d-1)$	& $D_{ij},\Phi,\Psi$	&										&						\\
\hline
\end{tabular}
\end{center}

\section{Hamiltonian/Temporal Gauge and Electric Memory}
\label{Section_GaugeInvariantElectromagnetism}
In \cite{Bieri:2013hqa}, Bieri and Garfinkle proposed an electric analog to the GW memory effect. Here, we will examine it in the cosmological context. According to the Lorentz force law, in the geometry of eq. \eqref{SpatiallFlatFLRW_Metric}, the proper spatial acceleration of an electric charge $q$ of mass $m$ is governed by
\begin{align}
\label{LorentzForceLaw}
\frac{\dd^2 Y^i}{\dd \tau^2} + \Gamma^i_{\phantom{i}\alpha\beta} \frac{\dd Y^\alpha}{\dd \tau} \frac{\dd Y^\beta}{\dd \tau}
= \frac{q}{m} a^{-2} \left( F^{i}_{\phantom{i}0} \frac{\dd Y^0}{\dd \tau} + F^{i}_{\phantom{i}j} \frac{\dd Y^j}{\dd \tau} \right) ,
\end{align}
where $\tau$ denotes proper time, and we have neglected all self-force terms. The non-zero Christoffel symbol terms of the left hand side of eq. \eqref{LorentzForceLaw} are $\Gamma^i_{\phantom{i}0j} (\dd Y^0/\dd \tau) (\dd Y^j/\dd \tau)$. We shall, however, actually neglect them in what follows because they are all proportional to $\dot{a}/a$, which means as long as our electric memory experiment does not take place over cosmological timescales, they must play a highly sub-dominant role. 

The Faraday tensor $F_{\mu\nu} \equiv \partial_{[\mu} A_{\nu]}$ contains the electric 
\begin{align}
F_{0i} = -F_{i0} = F^{i}_{\phantom{i}0} = \dot{A}_i - \partial_i A_0 
\end{align}
and the magnetic fields
\begin{align}
F_{ij} = -F^{i}_{\phantom{i}j} = \partial_i A_j - \partial_j A_i .
\end{align}
Whenever the electric term in eq. \eqref{LorentzForceLaw} dominates over the magnetic one -- as is often the case in non-relativistic systems\footnote{This dropping of the magnetic contribution to the Lorentz force law is why we prefer to call the ensuing memory effect an `electric' as opposed to an `electromagnetic' one.} -- and if we further choose the Hamiltonian/temporal gauge \cite{Jackson:2002rj}
\begin{align}
\label{HamiltonianGauge}
A_0 = A^0 = 0 ;
\end{align}
then integrating both sides of eq. \eqref{LorentzForceLaw} with respect to proper time, from some initial cosmic time $\eta'$ to the present $\eta$, gives us
\begin{align}
\label{LorentzForceLaw_Integrated_Step1}
\left[\frac{\dd Y^i}{\dd \tau''}\right]_{\tau''=\tau_\text{i}}^{\tau''=\tau}
= \frac{q}{m} \int_{\eta'}^{\eta} \frac{\dd\eta''}{a[\eta'']^2} \left( \partial_0 A_i + \dots \right) ,
\end{align}
where $\tau_\text{i}$ and $\tau$ are the proper times corresponding respectively to $\eta'$ and $\eta$. As already alluded to, if we watch our test charge move about in a cosmological spacetime, but only over timescales of human experiments, $a[\eta'']$ will not change very much. We shall choose to evaluate it at $\eta$ and pull it out of the integral in eq. \eqref{LorentzForceLaw_Integrated_Step1}. This tells us, within the Hamiltonian/temporal gauge of eq. \eqref{HamiltonianGauge}, that the velocity kick received by our test electric charge due the passage of some electromagnetic signal, is proportional to the change in the spatial part of the vector potential:
\begin{align}
\label{LorentzForceLaw_Integrated}
\left[\frac{\dd Y^i}{\dd \tau}\right]_{\eta'}^{\eta}
\approx \frac{q}{m a[\eta]^2} \left[ A_i[\eta'',\vec{Y}[\eta'']] \right]_{\eta''=\eta'}^{\eta''=\eta} .
\end{align}
This is the Bieri-Garfinkle electric memory effect generalized to generic spatially flat FLRW cosmologies: it is the permanent change in the proper spatial velocity of test charges,\footnote{For a complementary viewpoint on the electromagnetic memory effect, see Susskind \cite{Susskind:2015hpa}.} and is to be contrasted against the gravitational case involving permanent fractional displacement of pairs of test masses.

We will now proceed to calculate this electric memory generated by some externally prescribed current $\mathcal{J}^\nu$, by first solving Maxwell's equations: 
\begin{align}
\label{Maxwell}
\nabla_\mu F^{\mu\nu} = \frac{\partial_\mu \left( a^{d-4} F^{\mu\nu} \right)}{a^d} = \mathcal{J}^\nu .
\end{align}
We will do so within a gauge-invariant formalism, analogous to our analysis for the linearized Einstein-fluid equations above. Following that, we will translate our results to the Hamiltonian/temporal gauge, and extract the change in the vector potential. For completeness, in appendix \eqref{Section_EMandLinearizedYM}, we will also solve the same equations in a generalized Lorenz gauge.

{\bf Gauge-invariant formalism} \qquad The vector potential $A_\mu$ leaves the Faraday tensor $F_{\mu\nu}$ invariant under the gauge transformation
\begin{align}
\label{Maxwell_GaugeTransformation}
A_\mu \to A_\mu - \partial_\mu \mathcal{C} ,
\end{align}
for any arbitrary function $\mathcal{C}$. Let us perform a scalar-vector decomposition 
\begin{align}
A_i \equiv \alpha_i + \partial_i \alpha, \qquad\qquad \partial_i \alpha_i = 0 ,
\end{align}
and note that under the gauge transformation in eq. \eqref{Maxwell_GaugeTransformation},
\begin{align}
A_0 \to A_0 - \dot{\mathcal{C}} , \qquad\qquad
\alpha \to \alpha - \mathcal{C} ;
\end{align}
while the transverse portion of the spatial vector potential is already gauge-invariant.
\begin{align}
\label{Maxwell_GaugeInvariant_Vector}
\alpha_i \equiv A_i^\text{T} \to \alpha_i .
\end{align}
Therefore, we may form the following gauge-invariant scalar
\begin{align} 
\label{Maxwell_GaugeInvariant_Scalar}
\Phi &\equiv A_0 - \dot{\alpha} .
\end{align}
In terms of these variables, the Faraday tensor reads
\begin{align}
\label{Maxwell_FaradayTensor}
F_{0i} = \dot{\alpha}_i - \partial_i \Phi,
\qquad\qquad
F_{ij} = \partial_{[i} \alpha_{j]} .
\end{align}
The scalar-vector decomposition of the electromagnetic current is
\begin{align}
\mathcal{J}^0 &= \mathcal{J}_0 \equiv \rho , \\
\mathcal{J}_i &= -\mathcal{J}^i = \Gamma_i + \partial_i \Gamma, \qquad\qquad \partial_i \Gamma_i = 0 .
\end{align}
Similar in spirit to the conservation of the matter stress tensor, the divergence-free property of this electromagnetic current, $\nabla_\mu \mathcal{J}^\mu = 0$, tells us the electric charge density $\rho$ and the gradient portion of the spatial current are related:
\begin{align}
\label{Maxwell_CurrentConservation}
\frac{\partial_0 (a^d \rho)}{a^d} = \vec{\nabla}^2 \Gamma .
\end{align}
Maxwell's eq. \eqref{Maxwell} with the gauge-invariant variables employed through eq. \eqref{Maxwell_FaradayTensor} now take the following form. The zeroth component is
\begin{align}
\label{Maxwell_GaugeInvariant_IofIII}
-\vec{\nabla}^2 \Phi &= a^4 \rho ;
\end{align}
while the gradient and the divergence-free portions of the spatial components are, respectively,
\begin{align}
\label{Maxwell_GaugeInvariant_IIofIII}
\frac{\partial_0 \left( a^{d-4} \dot{\alpha}_i \right)}{a^{d-4}} - \vec{\nabla}^2 \alpha_i &= a^4 \Gamma_i , \\
\label{Maxwell_GaugeInvariant_IIIIofIII}
\frac{\partial_0 \left( a^{d-4} \Phi \right)}{a^{d-4}} &= -a^4 \Gamma  .
\end{align}
Eq. \eqref{Maxwell_GaugeInvariant_IIIIofIII} is actually redundant; it can be obtained by acting on both sides of eq. \eqref{Maxwell_GaugeInvariant_IofIII} with $\partial_0 (a^{d-4} \cdot )/a^{d-4}$, followed by utilizing eq. \eqref{Maxwell_CurrentConservation}.

{\bf Solutions} \qquad The gauge-invariant scalar of eq. \eqref{Maxwell_GaugeInvariant_IofIII} can be solved readily using the Green's function of the Laplacian in $(d-1)$-Euclidean space.
\begin{align}
\label{Maxwell_PotentialSolution}
\Phi[\eta,\vec{x}] = a[\eta]^4 \int_{\mathbb{R}^{d-1}} \dd^{d-1}\vec{x}' \frac{\Gamma\left[ \frac{d-3}{2} \right] \cdot \rho[\eta,\vec{x}']}{4\pi^{\frac{d-1}{2}} |\vec{x}-\vec{x}'|^{d-3}} .
\end{align}
The gauge-invariant vector mode of eq. \eqref{Maxwell_GaugeInvariant_IIofIII} can be solved by first re-expressing its wave equation as
\begin{align}
\left\{ \partial^2 - \frac{d-4}{2} \left( \frac{\ddot{a}}{a} + \frac{d-6}{2} \left(\frac{\dot{a}}{a}\right)^2 \right) \right\} \left( a^{ \frac{d-4}{2} } \alpha_i \right) 
&= a^{\frac{d+4}{2}} \Gamma_i .
\end{align}
If we specialize to the constant$-w$ cosmologies we have been considering in this paper, i.e., equations \eqref{SpatiallFlatFLRW_Metric} and \eqref{SpatiallFlatFLRW_Metric_w}, we find this is reduced to
\begin{align}
\left\{ \partial^2 + (d-4) \frac{1 + (d-1) w}{q_w^2} \frac{1}{\eta^2}\right\} \left( a^{ \frac{d-4}{2} } \alpha_i \right) 
&= a^{ \frac{d+4}{2} } \Gamma_i .
\end{align}
The retarded solution is
\begin{align}
\label{Maxwell_RadiationSolution}
a[\eta]^{\frac{d-4}{2}} \alpha_i[\eta,\vec{x}]
&= \int_{\mathbb{R}^{d-1}} \dd^{d-1}\vec{x}' \int_{\eta_\circ}^{\eta} \dd\eta' a[\eta']^{ \frac{d+4}{2} } \mathcal{G}^{(\gamma)}[\eta,\eta';\sbar]
\Gamma_i\left[ \eta',\vec{x}'\right] ,
\end{align}
where the symmetric Green's functions are
\begin{align}
\label{Maxwell_RadiationG_EvenD}
\mathcal{G}^{(\gamma)}_{\text{even $d\geq 4$}}[\eta,\eta';\sbar] 
&= \left(\frac{1}{2\pi} \frac{\partial}{\partial\sbar}\right)^{ \frac{d-2}{2} } \left(\frac{\Theta[\sbar]}{2} P_{-\frac{d-4}{q_w}}\left[ 1+s \right]\right) , \\
\label{Maxwell_RadiationG_OddD}
\mathcal{G}^{(\gamma)}_{\text{odd $d\geq 5$}}[\eta,\eta';\sbar] 
&= \left(\frac{1}{2\pi} \frac{\partial}{\partial\sbar}\right)^{ \frac{d-3}{2} } \left(\frac{\Theta[\sbar]}{4\pi} \frac{ \left(s+\sqrt{s(s+2)}+1\right)^{ \frac{5-d + (d-1)w}{q_w} } + 1 }{\sqrt{\sbar(s+2)} \left(s+\sqrt{s(s+2)}+1\right)^{ \frac{5-d + (d-1)w}{2q_w} }}\right) , \\
\sbar &= \frac{1}{2}\left( (\eta-\eta')^2 - (\vec{x}-\vec{x}')^2 \right), \qquad\qquad s \equiv \frac{\sbar}{\eta\eta'} .
\end{align}
{\it (3+1)-dimensions} \qquad Let us note that, apart from a re-scaling $\mathcal{J}^\nu \to a^4 \mathcal{J}^\nu$, the $d=4$ solutions are identical to the electromagnetic ones in 4D Minkowski spacetime; this is due to the conformal symmetry of the system at hand, and could already be seen by setting $d=4$ in eq. \eqref{Maxwell}.
\begin{align}
\Phi_4[\eta,\vec{x}] &= a[\eta]^4 \int_{\mathbb{R}^{3}} \dd^{3}\vec{x}' \frac{\rho[\eta,\vec{x}']}{4\pi |\vec{x}-\vec{x}'|} , \\
\alpha_{i_4}[\eta,\vec{x}] &= \Theta_p \int_{\mathbb{R}^{3}} \dd^{3}\vec{x}' a[\eta_r]^{4} \frac{\Gamma_i\left[\eta_r, \vec{x}'\right]}{4\pi |\vec{x}-\vec{x}'|} ,
\end{align}
where $\eta_r \equiv \eta-|\vec{x}-\vec{x}'|$ is the retarded time.

{\bf Cosmological electric memory} \qquad Imposing the Hamiltonian/temporal gauge in eq. \eqref{HamiltonianGauge} on the scalar gauge-invariant variable in eq. \eqref{Maxwell_GaugeInvariant_Scalar},
\begin{align}
\partial_i \alpha[\eta \geq \eta',\vec{x}] = -\int_{\eta'}^{\eta} \partial_i \Phi[\eta'',\vec{x}] \dd\eta'' + \partial_i \alpha[\eta',\vec{x}] ;
\end{align}
which in turn means the change in the vector potential that gives rise to the electric memory effect of eq. \eqref{LorentzForceLaw_Integrated} receives contributions from the gauge-invariant vector and scalar in the following way:
\begin{align}
\Delta A_i[\eta,\vec{x}]
&\equiv A_i[\eta,\vec{x}] - A_i[\eta',\vec{x}] \nonumber\\
\label{Maxwell_Memory_v1}
&= \Delta \alpha_i[\eta,\vec{x}] - \int_{\eta'}^{\eta} \partial_i \Phi[\eta'',\vec{x}] \dd\eta'' ,
\end{align}
where
\begin{align}
\Delta \alpha_i[\eta,\vec{x}] \equiv \alpha_i[\eta,\vec{x}] - \alpha_i[\eta',\vec{x}] .
\end{align}
By integrating the spatial derivative of eq. \eqref{Maxwell_GaugeInvariant_IIIIofIII} with respect to time,
\begin{align}
\label{Maxwell_Memory_CoulombContribution}
\partial_i \Phi[\eta,\vec{x}] 
&= -\frac{1}{a[\eta]^{d-4}} \int_{\eta'}^{\eta} a[\eta'']^d \partial_i \Gamma[\eta'',\vec{x}] \dd\eta'' 
+ \left(\frac{a[\eta']}{a[\eta]}\right)^{d-4} \partial_i \Phi[\eta',\vec{x}] .
\end{align}
Assuming the electromagnetic current is spatially localized, so its irreducible component $\partial_i \Gamma$ can be assumed to be zero at the test charge location, we conclude that $\partial_i \Phi[\eta,\vec{x}]$ far away from the source $\mathcal{J}^\nu$ depends on time only through the overall prefactor $1/a[\eta]^{d-4}$ on the rightmost term of eq. \eqref{Maxwell_Memory_CoulombContribution}. Referring to the $\Phi$ solution in eq. \eqref{Maxwell_PotentialSolution}, we see that eq. \eqref{Maxwell_Memory_CoulombContribution} provides a Coulomb-potential contribution to the electric memory formula of eq. \eqref{Maxwell_Memory_v1} that scales as $\Phi \times $(timescale of experiment)/(observer-source spatial distance); whereas the $\alpha_i$ is the radiative contribution to electric memory.

{\it Space-Independent Electromagnetic Tails} \qquad In even dimensions $d \geq 4$, by converting the $\alpha_i$-tail function $P_{-\frac{d-4}{q_w}}\left[ 1+s \right]$ in eq. \eqref{Maxwell_RadiationG_EvenD} into $\,_2F_1$'s, and following similar reasoning carried out for the gravitational case in the previous section -- we find that there are non-trivial space-independent electromagnetic tails for $d > 4$ whenever
\begin{align}
\label{EM_SpaceIndependentTails_w}
w = -1 + \frac{4}{(d-1)(d-2)} \qquad\qquad \text{ or } \qquad\qquad w = -1 + \frac{4}{d} \frac{d-2}{d-1} .
\end{align}
Explicitly, the wave tails of the gauge-invariant vector read
\begin{align}
a[\eta]^{\frac{d-4}{2}} \alpha_i^{\text{(tail)}}[\eta,\vec{x}]
&= \frac{(d-2)!}{\left(\frac{d-2}{2}\right)!} \int_{\mathbb{R}^{d-1}} \dd^{d-1}\vec{x}' \int_{\eta_\circ}^{\eta_r - 0^+} \dd\eta' 
\frac{a[\eta']^{ \frac{d+4}{2} }}{2(4\pi \eta\eta')^{\frac{d-2}{2}}}  \Gamma_i\left[ \eta',\vec{x}'\right] ,
\end{align}
which indicates -- if $\eta' \in [\eta_\text{i}, \eta_\text{f}]$ corresponds to the active period of electromagnetic wave production -- the associated electric memory registered at the observer, after the primary wave train has passed, is
\begin{align}
\label{EM_ElectricMemory_Tail}
\Delta \alpha_i^{\text{(tail)}}[\eta,\vec{x}]
&\approx a[\eta]^{-\frac{d-4}{2}} \frac{(d-2)!}{\left(\frac{d-2}{2}\right)!} \int_{\mathbb{R}^{d-1}} \dd^{d-1}\vec{x}' \int_{\eta_\text{i}}^{\eta_\text{f}} \dd\eta' 
\frac{a[\eta']^{ \frac{d+4}{2} }}{2(4\pi \eta\eta')^{\frac{d-2}{2}}}  \Gamma_i\left[ \eta',\vec{x}'\right] .
\end{align}
{\it Zero-tail conditions} \qquad It turns out there are no electromagnetic tails in even $d > 4$ when 
\begin{align}
w = -\frac{(d-3) + (d-4) \ell^{-1}}{d-1} \qquad\qquad \text{ or } \qquad\qquad
w = -\frac{(d-3) - (d-4) (\ell+1)^{-1}}{d-1} ,
\end{align}
where $\ell$ runs from $0$ through $-1 + (d-2)/2$.\footnote{In the gravitational case, we restricted our attention to $w = -1$ and $0 \leq w \leq 1$ because of potential pathologies arising in first order perturbation theory when $w < -1$, $-1 < w < 0$ or $w > 1$. Here in this section, because we do not need to consider gravitational perturbations, we choose to remain agnostic about the underlying matter responsible for the constant$-w$ background cosmology.}

\section{Summary and Future Directions}
\label{Section_Summary}
In the background spatially flat FLRW cosmologies of dimensions $d \geq 4$ described by equations \eqref{SpatiallFlatFLRW_Metric} and \eqref{SpatiallFlatFLRW_Metric_w}, we have solved the metric perturbations $\chi_{\mu\nu}$ in eq. \eqref{SpatiallFlatFLRW_PerturbedMetric} due to an isolated astrophysical system whose scalar-vector-tensor decomposition we defined in equations \eqref{Astro_SVT_IofII} and \eqref{Astro_SVT_IIofII}. We have done so using the gauge-invariant metric variables in equations \eqref{Bardeen_Psi}, \eqref{Bardeen_Phi}, and \eqref{Bardeen_VandDij}, which in turn are based on the scalar-vector-tensor decomposition of $\chi_{\mu\nu}$ in eq. \eqref{Bardeen_SVTDecomposition}. Except in the de Sitter case (where $w=-1$), it is also necessary to simultaneously consider the perturbations of the background perfect fluid that drives the evolution of the cosmology itself; we defined its deviation from equilibrium in eq. \eqref{PerfectFluid_ExpansionFromEq} and further re-phrased it in gauge-invariant terms through a scalar-vector decomposition in equations \eqref{Bardeen_FluidSVDecomposition} and \eqref{Bardeen_GaugeInvariantFluidVariables}. The linearized Einstein's equations governing the scalar perturbations can be found in equations \eqref{FirstOrderPTEqns_Scalar1of4} through \eqref{FirstOrderPTEqns_Scalar4of4}; the vector ones in equations \eqref{FirstOrderPTEqns_Vector1of2} and \eqref{FirstOrderPTEqns_Vector2of2}; and the sole tensor field in eq. \eqref{FirstOrderPTEqns_Tensor}. The fluid EoM involving the scalars and vectors are, respectively, equations \eqref{FirstOrderPTEqns_FluidScalar} and \eqref{FirstOrderPTEqns_FluidVector}.

We find that, to avoid superluminality, acausal dependence on the future configuration of the GW source(s) and potential quantum instabilities of the gauge-invariant metric scalar fields, the EoS of the cosmic fluid should obey $0 \leq w \leq 1$ or $w=-1$. (The latter case is de Sitter spacetime, where there really is no background cosmological fluid.) With this in mind, one can find the metric tensor perturbation wave-solution in eq. \eqref{FirstOrderPTEqns_Tensor_Soln}; and the metric vector perturbation solution in eq. \eqref{FirstOrderPTEqns_Vector_SolnIofII} expressed in a Coulomb-type form, or in eq. \eqref{FirstOrderPTEqns_Vector_SolnIIofII} as a time-integral involving a certain shear part of the matter source. The scalar metric perturbations in de Sitter spacetime yields a Coulomb-type solution in eq. \eqref{FirstOrderPTEqns_Scalar_dSSoln}. In matter dominated universes, $w=0$, they can be expressed as a double time integral involving the GW source in eq. \eqref{FirstOrderPTEqns_Scalar_MatterSoln}. Finally, it is when $0 < w \leq 1$ that they admit wave solutions like their tensor counterpart -- see eq. \eqref{FirstOrderPTEqns_PhiSolution_w>0}. Unlike the tensor mode whose wavefront travel at unit speed, however, these scalar ones do so at speed $0 < \sqrt{w} \leq 1$, which means Cherenkov radiation is possible. 

\begin{figure}
\begin{center}
\includegraphics[width=4in]{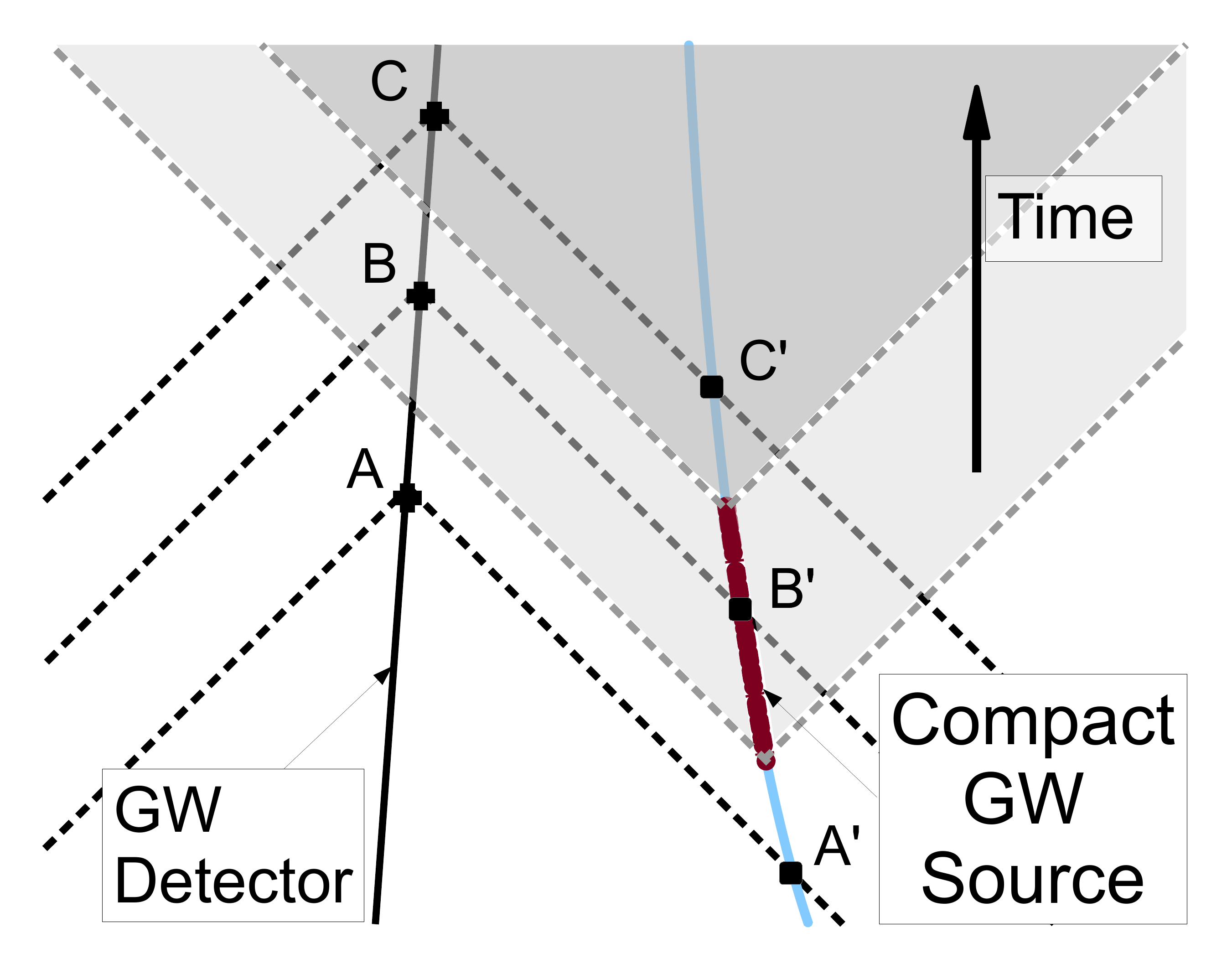}
\caption{(Figure borrowed from \cite{Chu:2015yua} and modified slightly.) This is a spacetime diagram depicting a hypothetical GW event involving an isolated astrophysical system (right world line). The dashed-dotted segment of the right world line denotes the full duration during which GWs are produced, corresponding to $\eta_\text{i} \leq \eta \leq \eta_\text{f}$ in the main text. The GWs are heard by a distant detector (left world line). The black dashed lines emanating from the worldline of the GW detector are either the past acoustic or null cones of events $A$, $B$ and $C$. The bottom pair of light gray dashed lines emanating from the right world line is either the forward acoustic or null cone of the starting point of the GW-generation process; the top pair is that of the ending point. {\it Acoustic/Null cone GWs:} We have highlighted, in equations \eqref{GWMemory_NullCone} and \eqref{ScalarMemory_NullCone_Delta}, that the portion of GW memory transmitted on the acoustic/null cone arises because the GW source configuration does not return to its original state after the main GW event. Specifically, if the $\sigma_{ij}$ part of the astrophysical system at $C'$ is not the same as that it were at $A'$, the GW detector operational from $A$ through $C$ would experience a non-trivial tensor null-cone memory effect. Similarly, over human GW experiment timescales, if $(\dot{\Sigma},\rho,\dot{\Upsilon})$ have changed significantly from $A'$ to $C'$, there will be a non-zero scalar acoustic-cone memory effect.}
\label{GWFigure}
\end{center}
\end{figure}

In Fig. \eqref{GWFigure}, we summarize the causal structure of these GW solutions, including their potential memories. The gauge-invariant tensor $D_{ij}$ admits wave solutions whose wavefronts move at unit speed. The light gray shaded region consists of $D_{ij}$-waves propagating both on and inside the null cone; and darker gray region is filled with only its wave tails. There are no tensor tails in 4D radiation dominated universes; while in dS$_{4+2n}$ as well as 4D matter dominated ones the tensor tail amplitude does not change with increasing spatial distance from the GW source. This latter feature may lead to a novel contribution to the memory effect. On the other hand, within the EoS interval $0 < w \leq 1$, the wave solutions of the gauge-invariant scalar modes $\Phi$ and $\Psi$ have wavefronts propagating at speed $\sqrt{w}$. The light gray region would be filled with scalar waves traveling both on and within their acoustic cones; and the darker gray region with only their wave tails, whose detailed properties depend in principle on the entire history of the source (i.e., the dashed-dotted segment). In a 4D radiation dominated universe, the scalar GW tail becomes independent of spatial coordinates along the acoustic cone $\eta-\sqrt{3}|\vec{x}|=$ constant; and might thus bring about memory. 

In \S \eqref{Section_GaugeInvariantElectromagnetism}, we grouped the Maxwell U$_1$ gauge potential into gauge-invariant variables in equations \eqref{Maxwell_GaugeInvariant_Vector} and \eqref{Maxwell_GaugeInvariant_Scalar}. Electromagnetism in the cosmologies of equations \eqref{SpatiallFlatFLRW_Metric} and \eqref{SpatiallFlatFLRW_Metric_w} is then encapsulated in equations \eqref{Maxwell_GaugeInvariant_IofIII} through \eqref{Maxwell_GaugeInvariant_IIIIofIII} whereas their solutions are to be found in equations \eqref{Maxwell_PotentialSolution} and \eqref{Maxwell_RadiationSolution}. We then identified a potential electric memory effect in eq. \eqref{EM_ElectricMemory_Tail} due to the space-independent nature of the part of the gauge-invariant transverse vector that travels inside the light cone at specific values of $w$ given by eq. \eqref{EM_SpaceIndependentTails_w}.

In this work, we have re-confirmed that, conformal re-scaling, dimension-reduction, Nariai's ansatz \cite{Nariai} (and the Euclidean Green's function of the $(d-1)$-spatial Laplacian) lead us to analytic solutions to the full set of partial differential equations arising from General Relativity linearized about the cosmological spacetimes of equations \eqref{SpatiallFlatFLRW_Metric} and \eqref{SpatiallFlatFLRW_Metric_w} for all $d \geq 4$. This extends the results of \cite{Chu:2016qxp}, which did so for the specific case of de Sitter spacetime $(w=-1)$. To investigate the cosmological generalization of Christodoulou's nonlinear GW memory effect \cite{Christodoulou:1991cr} and to extend the de Sitter quadrupole radiation formula derived by Ashtekhar et al. \cite{Ashtekar:2015lxa} to a more general EoS $w$, we would need to go beyond the linear order in perturbation theory. It is not clear to us that the gauge-invariant formalism used in this paper is the most computationally efficient means of doing so -- the gauge-fixed one of \cite{Iliopoulos:1998wq}, \cite{Ashtekar:2015lxa}, and \cite{Chu:2016qxp} might be worth re-visiting for this purpose. Moreover, while we have tried to employ physically motivated arguments to persuade the reader that the fractional distortion measured by a GW detector only receives contributions from the gauge-invariant tensor mode when $w=-1$ or $w=0$; and from both the tensor and scalar modes when $0 < w \leq 1$ -- in this age of ``precision cosmology" it may be physically important to further sharpen these arguments to either corroborate their robustness or to find cases where they fail. Finally, with the position-space Green's functions of both the graviton and photon fully worked out in constant$-w$ cosmologies in all relevant dimensions, we may now embark on a program to understand quantitatively the associated tail-induced self-force problem first uncovered by DeWitt and Brehme \cite{DeWitt:1960fc}.

When $0 < w \leq 1$, we have revealed through equations \eqref{Astro_MassLossRate} and \eqref{FirstOrderPTEqns_PhiSolution_w>0} that astrophysical systems could potentially lose some of their mass through scalar gravitational radiation. (A scalar field mechanism for such mass loss has been previously noted by Burko, Harte and Poisson \cite{Burko:2002ge}.) It would be instructive to find concrete setups where this phenomenon can be illustrated in detail. One may also wonder if this phenomenon has any impact on the evolution of the Large Scale Structure of our universe. Finding a specific example to study its GW emission could also aid in understanding the procedure alluded to in the introduction of this paper -- how the matching of the cosmological waveforms in the near/intermediate zone to the far zone of the Minkowskian one is actually carried out. In particular, we note that, while the TT tensor metric perturbation is always the only gauge-invariant variable that admits wave solutions in Minkowski spacetime\footnote{This is demonstrated in appendix \S \eqref{Section_LinearizedEinstein_GaugeInvariant}.}, in $(0 < w \leq 1)$-cosmologies both the gauge-invariant scalar and tensor modes do so. How does this mismatch in the number of radiative degrees-of-freedom play into the GW matching procedure, and in turn, affect the physical properties of the GWs themselves?

\section{Acknowledgments}

I wish to acknowledge Nishant Agarwal, Thorsten Battefeld, and Audrey Mithani for useful discussions on cosmological perturbation theory; Pasha Bolokhov for probing questions; as well as Alexander Tolish and Robert Wald for detailed exchanges regarding their work in \cite{Tolish:2016ggo}. I am grateful to Simon Caron-Huot for answering my question, ``What's the analog of a point mass in Yang-Mills theory?" (see \cite{Wong:1970fu}); and to Gary Gibbons for highlighting the important property that Kerr-Schild metrics linearize Einstein's equations. I have also benefited greatly from the numerous conversations I had with Zvi Bern and his group on the topic of ``GR $=$ YM$^2$," during my week-long Spring 2014 visit to UCLA. During the final stages of this work, I spent many hours at the Tavern on the Hill -- and I wish to express my appreciation towards its excellent hospitable staff. Finally, the bulk of this work would not be possible without the software {\sf Mathematica} \cite{Mathematica} and the tensor package {\sf xAct} \cite{xAct}.

\appendix

\section{Linearized Einstein's Equations on (Minkowski)$_{d \geq 4}$}
\label{Section_LinearizedEinstein}
In this section we will review both the gauge-invariant and de Donder gauge paths to obtain solutions of Einstein's equations $G_{\mu\nu} = 8\pi \GN T_{\mu\nu}$, with the cosmological constant set to zero, linearized about a Minkowski spacetime background. That is, we shall take the full spacetime geometry to be
\begin{align}
\label{LinearizedGravity_Metric}
g_{\mu\nu} = \eta_{\mu\nu} + \chi_{\mu\nu}
\end{align}
and expand the Einstein to linear order in $\chi_{\mu\nu}$, namely
\begin{align}
G_{\mu\nu} 	&= (G \vert 1)_{\mu\nu} + \mathcal{O}[(\chi_{\alpha\beta})^2] , \\
\label{LinearizedGravity_Minkowski_EinsteinTensor_Unbarred}
(G \vert 1)_{\mu\nu} 
			&= \frac{1}{2} \left( 
- \partial^2 \chi_{\mu\nu} + \eta_{\mu\nu} \left( \partial^2 \chi - \partial^\sigma \partial^\rho \chi_{\sigma\rho} \right)
+ \partial_\sigma \partial_{ \{ \mu } \chi^\sigma_{\phantom{\sigma} \nu \}} - \partial_\mu \partial_\nu \chi
\right) .
\end{align}
Throughout this section, all indices will be moved with the flat metric $\eta_{\mu\nu}$. If we defined the barred graviton as
\begin{align}
\overline{\chi}_{\mu\nu} \equiv \chi_{\mu\nu} - \frac{\eta_{\mu\nu}}{2} \eta^{\rho\sigma} \chi_{\rho\sigma} ,
\end{align}
the linearized Einstein tensor becomes
\begin{align}
\label{LinearizedGravity_Minkowski_EinsteinTensor}
(G \vert 1)_{\mu\nu} &=
-\frac{1}{2} \left( \partial^2 \overline{\chi}_{\mu\nu} - \partial_{\{\mu} \partial^\alpha \overline{\chi}_{\nu\} \alpha} + \eta_{\mu\nu} \partial_\alpha \partial_\beta \overline{\chi}^{\alpha\beta} \right) + \mathcal{O}[(\overline{\chi}_{\sigma\rho})^2] .
\end{align}
Evaluating the energy-momentum-shear-stress tensor of matter $T_{\mu\nu}$ about the flat background, $\overline{T}_{\mu\nu} \equiv T_{\mu\nu}[g_{\mu\nu} = \eta_{\mu\nu}]$, we thus wish to solve
\begin{align}
\label{LinearizedGravity_Minkowski_EinsteinEquation}
(G \vert 1)_{\mu\nu} = 8 \pi \GN \overline{T}_{\mu\nu} .
\end{align}

\subsection{Gauge-Invariant Formalism} 
\label{Section_LinearizedEinstein_GaugeInvariant}
We begin with the gauge-invariant formalism to solve the linearized Einstein's eq. \eqref{LinearizedGravity_Minkowski_EinsteinEquation}.\footnote{Much of this subsection is the $(d \geq 4)$-dimensional generalization of Flanagan and Hughes \cite{Flanagan:2005yc} \S 2.2.} This will be the $a \to 1$ limit of what we have achieved in the main text, but we will attempt to keep the discussion fairly self-contained. 

The infinitesimal coordinate transformation
\begin{align}
x^\alpha \to x^\alpha + \xi^\alpha 
\end{align}
induces a change in the graviton field $\chi_{\mu\nu}$ of eq. \eqref{LinearizedGravity_Metric} as
\begin{align}
\label{LinearizedGravity_GaugeTransformedGraviton}
\chi_{\mu\nu} \to \chi_{\mu\nu} + \partial_{ \{ \mu } \xi_{ \nu\} } .
\end{align}
{\bf Gauge-invariant variables} \qquad To obtain a gauge-invariant formulation for the dynamics of $\chi_{\mu\nu}$, we start with the following irreducible scalar-vector-tensor decomposition of the metric perturbation $\chi_{\mu\nu}$ as well as an analogous one for $\xi$. That is, we do
\begin{align}
\label{LinearizedGravity_GaugeTransformationDecomposed}
\xi_\mu = (\xi_0, \partial_i \ell + \ell_i), \qquad\qquad \partial_i \ell_i = 0 .
\end{align}
and
\begin{align}
\label{LinearizedGravity_SVTDecomposition}
\chi_{00} \equiv E, \qquad \qquad \chi_{0i} \equiv \partial_i F + F_i, \nonumber\\
\chi_{ij} \equiv D_{ij} + \partial_{\{ i} D_{j\} } + \frac{D}{d-1} \delta_{ij} + \left( \partial_i \partial_j - \frac{\delta_{ij}}{d-1} \vec{\nabla}^2 \right) K ,
\end{align}
where these variables obey the following constraints:
\begin{align}
\label{LinearizedGravity_Constraints}
\partial_i F_i = \delta^{ij} D_{ij} = \partial_i D_{ij} = \delta^{ij} \partial_i D_j = 0 .
\end{align}
The gauge transformation of the graviton in eq. \eqref{LinearizedGravity_GaugeTransformedGraviton}, in terms of the variables in equations \eqref{LinearizedGravity_GaugeTransformationDecomposed} and \eqref{LinearizedGravity_SVTDecomposition}, reads
\begin{align}
E \to E + 2 \dot{\xi}_0 , \qquad\qquad
F \to F + \dot{\ell} + \xi_0, \qquad\qquad 
F_i \to F_i + \dot{\ell}_i,  \\
D_j \to D_j + \ell_j, \qquad\qquad
D \to D+2\vec{\nabla}^2 \ell , \qquad\qquad
K \to K + 2\ell ;
\end{align}
while the transverse-traceless graviton is already gauge-invariant
\begin{align}
D_{ij} \to D_{ij} .
\end{align}
One may verify via a direct calculation that the following are gauge-invariant variables formed from the graviton components. The $2$ scalar ones are
\begin{align}
\label{LinearizedGravity_GaugeInvariant_MetricPerturbations_IofII}
\overline{\Psi} \equiv E - 2 \dot{F} + \ddot{K} , \qquad\qquad
\overline{\Phi} \equiv \frac{D - \vec{\nabla}^2 K}{d-1} ;
\end{align}
whereas the vector and tensor modes are, respectively,
\begin{align}
\label{LinearizedGravity_GaugeInvariant_MetricPerturbations_IIofII}
\overline{V}_i \equiv F_i - \dot{D}_i
\qquad \qquad \text{ and }\qquad \qquad
\overline{D}_{ij} \equiv D_{ij} \equiv \chi_{ij}^\text{TT} .
\end{align}
(As already explained in \cite{Chu:2016qxp}, there are TT tensor modes only when $d \geq 4$.)

We will need to perform a similar scalar-vector-tensor decomposition for the stress-energy tensor on the right hand side of eq. \eqref{LinearizedGravity_Minkowski_EinsteinEquation}.
\begin{align}
\label{LinearizedGravity_SVTDecomposition_Matter}
\overline{T}_{00} \equiv \rho, \qquad\qquad 
\overline{T}_{0i} \equiv \partial_i \Sigma + \Sigma_i, \nonumber\\
\overline{T}_{ij} \equiv \sigma_{ij} + \partial_{\{ i} \sigma_{j\} } + \frac{\sigma}{d-1} \delta_{ij} + \left( \partial_i \partial_j - \frac{\delta_{ij}}{d-1} \vec{\nabla}^2 \right) \Upsilon ,
\end{align}
where these variables obey the following constraints:
\begin{align}
\label{LinearizedGravity_Constraints_Matter}
\partial_i \Sigma_i = \delta^{ij} \sigma_{ij} = \partial_i \sigma_{ij} = \partial_i \Sigma_i = 0 .
\end{align}
Furthermore, these matter variables are not all independent because consistency of Einstein's equations require the stress-energy tensor to be divergence-less, $\partial^\mu \overline{T}_{\mu\nu} = 0$. A direct calculation would translate the latter to
\begin{align}
\label{LinearizedGravity_MatterConservation}
\dot{\rho} = \vec{\nabla}^2 \Sigma, \qquad\qquad
\dot{\Sigma}_i = \vec{\nabla}^2 \sigma_i, \qquad\qquad
\dot{\Sigma} = \frac{\sigma + (d-2) \vec{\nabla}^2 \Upsilon}{d-1} .
\end{align}
The interpretation is that, once $\rho$, $\Sigma_i$, $\sigma$ and $\sigma_{ij}$ are specified, the $\Sigma$, $\sigma_i$ and $\Upsilon$ can be determined by an appropriate convolution involving the Euclidean Green's function $\vec{\nabla}^{-2}$.

{\bf Linearized Einstein's Equations} \qquad The Riemann and Einstein tensor of the weak field metric in eq. \eqref{LinearizedGravity_Metric} must be gauge-invariant at first order in the perturbation $\chi_{\mu\nu}$, because they are zero when $\chi_{\mu\nu} \to 0$. Employing the variables of eq. \eqref{LinearizedGravity_SVTDecomposition} in the linearized Einstein tensor $(G \vert 1)_{\mu\nu}$ (see eq. \eqref{LinearizedGravity_Minkowski_EinsteinTensor_Unbarred}), and grouping terms in the spirit of the scalar-vector-tensor decomposition in eq. \eqref{LinearizedGravity_SVTDecomposition} -- a brute force calculation would then yield $(G \vert 1)_{\mu\nu}$ in terms of gauge-invariant metric perturbations defined in equations \eqref{LinearizedGravity_GaugeInvariant_MetricPerturbations_IofII} and \eqref{LinearizedGravity_GaugeInvariant_MetricPerturbations_IIofII}:
\begin{align}
(G \vert 1)_{00} &= \frac{d-2}{2} \vec{\nabla}^2 \overline{\Phi}, \qquad\qquad
(G \vert 1)_{0i} = \frac{1}{2} \left( \vec{\nabla}^2 \overline{V}_i+ (d-2) \partial_i \dot{\overline{\Phi}} \right) , \\
(G \vert 1)_{ij} &= \frac{1}{2} \Bigg( 
- \partial^2 \overline{D}_{ij} + \partial_{\{i} \dot{\overline{V}}_{j\}}
- \frac{\delta_{ij}}{d-1} (d-2) \left( \vec{\nabla}^2 \left\{ (d-3) \overline{\Phi} - \overline{\Psi} \right\} - (d-1) \ddot{\overline{\Phi}} \right) \\
&\qquad\qquad\qquad\qquad
+ \left(\partial_i \partial_j - \frac{\delta_{ij}}{d-1} \vec{\nabla}^2\right) \left( (d-3) \overline{\Phi} - \overline{\Psi} \right)
\Bigg) . \nonumber
\end{align}
Imposing Einstein's equations $(G \vert 1)_{\mu\nu} = 8\pi\GN \overline{T}_{\mu\nu}$, with the SVT decomposition of matter in eq. \eqref{LinearizedGravity_SVTDecomposition_Matter}, returns the equations
\begin{align}
\label{LinearizedGravity_SVTDecomposition_EinsteinIofIII}
(d-2) \vec{\nabla}^2 \overline{\Phi} = 16\pi\GN \rho, \qquad\qquad
(d-2) \dot{\overline{\Phi}} = 16\pi\GN \Sigma, \qquad\qquad
(d-3) \overline{\Phi} - \overline{\Psi} = 16\pi\GN \Upsilon, \\
\label{LinearizedGravity_SVTDecomposition_EinsteinIIofIII}
(d-2) \left( \vec{\nabla}^2 \left\{ (d-3) \overline{\Phi} - \overline{\Psi} \right\} - (d-1) \ddot{\overline{\Phi}} \right) = -16\pi\GN \sigma , \\
\label{LinearizedGravity_SVTDecomposition_EinsteinIIIofIII}
\partial^2 \overline{D}_{ij} = -16\pi\GN \sigma_{ij}, \qquad\qquad
\vec{\nabla}^2 \overline{V}_i = 16\pi\GN \Sigma_i, \qquad\qquad
\dot{\overline{V}}_i = 16\pi\GN \sigma_i .
\end{align}
Because the Einstein tensor itself obeys the Bianchi identity, one may expect not all these equations are independent. We will now demonstrate that they can be reduced to the set:
\begin{align}
\label{LinearizedGravity_SVTDecomposition_Einstein_Scalars}
(d-2) \vec{\nabla}^2 \overline{\Phi} = 16\pi\GN \rho, \qquad\qquad
(d-3) \overline{\Phi} - \overline{\Psi} = 16\pi\GN \Upsilon, \\
\label{LinearizedGravity_SVTDecomposition_Einstein_VectorTensor}
\vec{\nabla}^2 \overline{V}_i = 16\pi\GN \Sigma_i, \qquad\qquad
\partial^2 \overline{D}_{ij} = -16\pi\GN \sigma_{ij} .
\end{align}
We may start by differentiating the first equation from the left in eq. \eqref{LinearizedGravity_SVTDecomposition_Einstein_Scalars} once with respect to time, and employ $\dot{\rho} = \vec{\nabla}^2 \Sigma$ from eq. \eqref{LinearizedGravity_MatterConservation} to eliminate $(d-2) \dot{\overline{\Phi}} = 16\pi\GN \Sigma$ in eq. \eqref{LinearizedGravity_SVTDecomposition_EinsteinIofIII}. Similarly, taking a time derivative of the first equation from the left in eq. \eqref{LinearizedGravity_SVTDecomposition_Einstein_VectorTensor} followed by replacing $\dot{\Sigma}_i \to \vec{\nabla}^2 \sigma_i$ from \eqref{LinearizedGravity_MatterConservation} gets rid of $\dot{\overline{V}}_i = 16\pi\GN \sigma_i$ in eq. \eqref{LinearizedGravity_SVTDecomposition_EinsteinIIIofIII}. Finally, replacing the $(d-3) \overline{\Phi} - \overline{\Psi}$ in eq. \eqref{LinearizedGravity_SVTDecomposition_EinsteinIIofIII} with $16\pi\GN \Upsilon$ (eq. \eqref{LinearizedGravity_SVTDecomposition_Einstein_Scalars}) and using the rightmost equation of eq. \eqref{LinearizedGravity_MatterConservation}, yields $(d-2) \ddot{\overline{\Phi}} = 16\pi\GN \dot{\Sigma}$, which is the time derivative of the second equation in eq. \eqref{LinearizedGravity_SVTDecomposition_EinsteinIofIII}.

In equations \eqref{LinearizedGravity_SVTDecomposition_Einstein_Scalars} and \eqref{LinearizedGravity_SVTDecomposition_Einstein_VectorTensor}, the scalar and vector perturbations obey Poisson's equations; only the tensor perturbation $\overline{D}_{ij}$ admits wave solutions:
\begin{align} 
\label{LinearizedGravity_Minkowski_EinsteinSolution_Scalar}
\overline{\Phi}[t,\vec{x}] 	
&= -\frac{16\pi\GN}{d-2} \int_{\mathbb{R}^{d-1}} \dd^{d-1}\vec{x}' \frac{\Gamma\left[ \frac{d-3}{2} \right] \cdot \rho[t,\vec{x}']}{4 \pi^{\frac{d-1}{2}} \vert \vec{x}-\vec{x}' \vert^{d-3}},  \\
\label{LinearizedGravity_Minkowski_EinsteinSolution_Vector}
\overline{V}_i[t,\vec{x}] 
&= -16\pi\GN \int_{\mathbb{R}^{d-1}} \dd^{d-1}\vec{x}' \frac{\Gamma\left[ \frac{d-3}{2} \right] \cdot \Sigma_i[t,\vec{x}']}{4 \pi^{\frac{d-1}{2}} \vert \vec{x}-\vec{x}' \vert^{d-3}}, \\
\label{LinearizedGravity_Minkowski_EinsteinSolution_Tensor}
\overline{D}_{ij}[t,\vec{x}] 
&= -16\pi\GN \int_{-\infty}^{t} \dd t' \int_{\mathbb{R}^{d-1}} \dd^{d-1} \vec{x}' \overline{\mathcal{G}}_d[\sbar] \sigma_{ij}[t',\vec{x}'] .
\end{align}
where the symmetric Green's functions $\overline{\mathcal{G}}$ in eq. \eqref{LinearizedGravity_Minkowski_EinsteinSolution_Tensor} are
\begin{align}
\label{MasslessScalarG_Evend}
\overline{\mathcal{G}}_{\text{even }d \geq 4}[\sbar] 
&= \frac{1}{(2\pi)^{\frac{d-4}{2}}} \left(\frac{\partial}{\partial \sbar}\right)^{ \frac{d-4}{2} } \left( \frac{\delta[\sbar]}{4\pi} \right) , \\
\label{MasslessScalarG_Oddd}
\overline{\mathcal{G}}_{\text{odd }d \geq 5}[\sbar] 
&= \frac{1}{(2\pi)^{\frac{d-3}{2}}} \left(\frac{\partial}{\partial \sbar}\right)^{ \frac{d-3}{2} } \left( \frac{\Theta[\sbar]}{2\pi \sqrt{2\sbar}} \right) , \\
\sbar &\equiv \frac{1}{2} \{ (t-t')^2 - (\vec{x}-\vec{x}')^2 \} .
\end{align}
They obey
\begin{align}
\partial_x^2 \overline{\mathcal{G}}_d[\sbar] = \partial_{x'}^2 \overline{\mathcal{G}}_d[\sbar] = 2 \delta^{(d)}[x-x'] .
\end{align}
{\it Coulomb vs. Waves} \qquad It is instructive to compare the time derivative of $\overline{\Phi}$ in eq. \eqref{LinearizedGravity_Minkowski_EinsteinSolution_Scalar} with that of $\overline{D}_{ij}$ in eq. \eqref{LinearizedGravity_Minkowski_EinsteinSolution_Tensor}. For the former, let us employ $\dot{\rho} = \vec{\nabla}^2 \Sigma$ in eq. \eqref{LinearizedGravity_MatterConservation}:
\begin{align}
\label{LinearizedGravity_Minkowski_EinsteinSolution_ScalarDot}
\dot{\overline{\Phi}}[t,\vec{x}] 	
&= -\frac{16\pi\GN}{d-2} \partial_i \int_{\mathbb{R}^{d-1}} \dd^{d-1}\vec{x}' \frac{\Gamma\left[ \frac{d-3}{2} \right] \cdot \partial_{i'} \Sigma[t,\vec{x}']}{4 \pi^{\frac{d-1}{2}} \vert \vec{x}-\vec{x}' \vert^{d-3}}, 
\end{align}
where $\partial_i \equiv \partial/\partial x^i$, $\partial_{i'} \equiv \partial/\partial x'^i$; we have integrated-by-parts one of the spatial derivatives and used the translation-symmetry of the $|\vec{x}-\vec{x}'|$ to pull it out of the integral. We see that, in the far zone -- where the observer is much farther from the source than the latter's characteristic size -- each time derivative of $\overline{\Psi}$ scales as $1/$(observer-source spatial distance). Note too that the total mass $M$ of the system
\begin{align}
\label{LinearizedGravity_Minkowski_TotalMass}
M \equiv \int_{\mathbb{R}^{d-1}} \dd^{d-1}\vec{x}' \rho[t,\vec{x}'] 
\end{align}
is conserved in General Relativity linearized about Minkowski spacetime. This is because, if we differentiate both sides of eq. \eqref{LinearizedGravity_Minkowski_TotalMass} with respect to time, and make the replacement $\dot{\rho} \to \vec{\nabla}^2 \Sigma$ (eq. \eqref{LinearizedGravity_MatterConservation}) within the integrand on the right-hand-side, the resulting volume integral can be converted to a surface one involving the flux of the $\partial_i \Sigma$ portion of $\overline{T}_{0i}$, which is zero if the surface is at spatial infinity. This in turn tells us the far zone scalar profile in eq. \eqref{LinearizedGravity_Minkowski_EinsteinSolution_Scalar}, i.e., the portion that scales as $1/$(observer-source spatial distance)$^{d-3}$, is strictly time independent.

For the time derivative of the wave solution in eq. \eqref{LinearizedGravity_Minkowski_EinsteinSolution_Tensor}, because they have the same form, we may refer to those in equation pairs \eqref{LinearizedGravity_Minkowski_EinsteinEquation_deDonderContinuousSource_Evend}--\eqref{LinearizedGravity_Minkowski_EinsteinEquation_deDonderContinuousSource_Evend_grad} and \eqref{LinearizedGravity_Minkowski_EinsteinEquation_deDonderContinuousSource_Oddd}--\eqref{LinearizedGravity_Minkowski_EinsteinEquation_deDonderContinuousSource_Oddd_grad} below to deduce, each time derivative of $\overline{D}_{ij}$ scales as $1/$(characteristic timescale of $\sigma_{ij}$). As expected, in the far zone, $\overline{D}_{ij}$ varies as rapidly as the GW source itself.

Altogether, these considerations suggest that the Coulomb-type solution of $\overline{\Phi}$ is not radiative -- it does not carry energy-momentum to infinity -- but the $\overline{D}_{ij}$ wave solution in eq. \eqref{LinearizedGravity_Minkowski_EinsteinSolution_Tensor} does.\footnote{Our analysis here is, of course, the ``poor person's" route to distinguishing radiative from non-radiative solutions. Strictly speaking, one would have to expand Einstein's equations to second order in metric perturbations, in order to determine which first-order gauge-invariant variables appear in the $0i$-components of the graviton pseudo stress tensor.} As we will further argue below, this means it is the tensor mode alone that contributes appreciably to the tidal squeezing and stretching of the space around a distant GW detector.

{\bf GW Detector In Flat Spacetime} \qquad In a background Minkowski spacetime, the fractional distortion $\delta L/L_0$ measured by a GW detector due to a astrophysical system located at a large distance, would need to be derived in a similar fashion as we did the cosmological equations \eqref{FractionalDistortion_AtConstantCosmicTime_FirstOrderHuman} and \eqref{FractionalDistortion_AtConstantCosmicTime_Deltachiij}. In particular, we would need the synchronous gauge metric perturbations produced by this GW source. If we are able to compute the gauge-invariant metric perturbations through equations \eqref{LinearizedGravity_Minkowski_EinsteinSolution_Scalar}, \eqref{LinearizedGravity_Minkowski_EinsteinSolution_Vector}, \eqref{LinearizedGravity_Minkowski_EinsteinSolution_Tensor}, as well as $(d-3) \overline{\Phi} - \overline{\Psi} = 16\pi\GN \Upsilon$ in eq. \eqref{LinearizedGravity_SVTDecomposition_Einstein_Scalars}; then eq. \eqref{SpatiallFlatFLRW_PerturbedSpatialMetric_SynchronousGaugeFromGaugeInvariant_Delta} would supply the necessary result if we set $a \to 1$, followed by replacing $\eta \to t$ and all unbarred metric variables with barred ones. Stepping through similar estimates performed in the discussion right after eq. \eqref{SpatiallFlatFLRW_PerturbedSpatialMetric_SynchronousGaugeFromGaugeInvariant_Delta}, one would infer -- due to Coulomb-type solutions of $(\overline{\Phi},\overline{\Psi},\overline{V}_i)$ -- the integral terms involving $\partial_{\{i} \overline{V}_{j\}}$ and $\partial_i \partial_j \overline{\Psi}$ would very likely scale respectively as (timescale of GW experiment)/(observer-source spatial distance) and ((timescale of GW experiment)/(observer-source spatial distance))$^2$ relative to the $\Delta \overline{\Phi}$ term. Furthermore, from the ``{\it Coulomb vs. Waves}" discussion above, one would expect $\Delta \overline{\Phi}$ itself to be highly sub-dominant to $\Delta \overline{D}_{ij}$ because the characteristic timescale governing changes in amplitude is the observer-source spatial distance for $\overline{\Phi}$ while it is the GW source(s)' intrinsic frequencies for $\overline{D}_{ij}$.

To sum: the fractional distortion measured by a GW detector in a flat spacetime is similar to its counterpart (eq. \eqref{FractionalDistortion_AtConstantCosmicTime_FirstOrderHuman_dSMatter}) in a matter dominated or de Sitter cosmology -- it receives contributions primarily from the tensor mode alone.
\begin{align}
\label{FractionalDistortion_AtConstantMinkowskiTime_FirstOrderHuman}
\left(\frac{\delta L}{L_0}\right)[\eta > \eta';\text{Minkowski}] 
= - \frac{\widehat{n}^i \widehat{n}^j}{2} \int_{0}^{1} \Delta \overline{D}_{ij}[\vec{Y}_0 + \lambda(\vec{Z}_0-\vec{Y}_0)] \dd \lambda 
\end{align}

\subsection{de Donder gauge: Li\'{e}nard--Wiechert Graviton Fields and GW Memory}
\label{Section_LinearizedEinstein_deDonder}
We now turn to solving eq. \eqref{LinearizedGravity_Minkowski_EinsteinEquation} by imposing the de Donder gauge
\begin{align}
\label{LinearizedGravity_deDonderGauge}
\partial^\sigma \overline{\chi}_{\sigma \mu} = 0 .
\end{align}
(This is analogous to imposing the Lorenz gauge condition $\partial^\sigma A_\sigma = 0$ in electromagnetism, where $A_\sigma$ is the $U_1$ gauge potential.) Eq. \eqref{LinearizedGravity_deDonderGauge} applied to eq. \eqref{LinearizedGravity_Minkowski_EinsteinTensor} will reduce the linearized Einstein's eq. \eqref{LinearizedGravity_Minkowski_EinsteinEquation} to
\begin{align}
\label{LinearizedGravity_Minkowski_EinsteinEquation_deDonder}
\partial^2 \overline{\chi}_{\mu\nu} = -16 \pi \GN \overline{T}_{\mu\nu} .
\end{align}
This converts the problem at hand to a $d \times d$ matrix of minimally coupled massless scalar wave equations. The retarded solutions are
\begin{align}
\label{LinearizedGravity_Minkowski_EinsteinEquation_deDonderSolution}
\overline{\chi}_{\mu\nu}[t,\vec{x}] 
= -16 \pi \GN \int_{-\infty}^t \dd t' \int_{\mathbb{R}^{d-1}} \dd^{d-1}\vec{x}' \overline{\mathcal{G}}_d[\sbar] \overline{T}_{\mu\nu}[t',\vec{x}'] ,
\end{align}
where the symmetric Green's functions $\{\overline{\mathcal{G}}_d[\sbar]\}$ can be found in equations \eqref{MasslessScalarG_Evend} and \eqref{MasslessScalarG_Oddd}.

{\bf Li\'{e}nard--Wiechert graviton fields} \qquad As a simple application of these formulas, we shall work out the (Li\'{e}nard--Wiechert) graviton field sourced by a point mass in motion in all dimensions higher than $3$. We will witness that the linear graviton field of a point mass propagates strictly on the null cone in even dimensions $d \geq 4$ and exclusively within the null cone (i.e., it is pure tail) in odd dimensions $d \geq 5$. Following that, we generalize to all even dimensions $d \geq 4$ the known 4-dimensional GW memory engendered by 2 (or more) compact bodies scattering off each other due to their mutual gravity.

The stress energy tensor of a point mass $M$, with timelike trajectory $Y^\mu[\tau]$ and proper time $\tau$, moving about in some geometry $g_{\mu\nu}$ -- is given by
\begin{align}
\label{StressTensor_PointMass}
T^{\mu\nu}[x] = 
M \int_{-\infty}^{+\infty} \dd \tau \dot{Y}^\mu[\tau] \dot{Y}^\nu[\tau] \frac{\delta^{(d)}[x-Y[\tau]]}{\sqrt{|g[x]|}}, 
\qquad\qquad
\dot{Y}^\mu \equiv \frac{\dd Y^\mu[\tau]}{\dd \tau} .
\end{align}
\footnote{We will define proper time $\tau$ in relation to coordinate time $t$ through the differential equation $\dd \tau = \sqrt{g_{\alpha\beta} (\dd Y^\alpha/\dd t) (\dd Y^\beta/\dd t)} \dd t$; in particular, $\sqrt{g_{\alpha\beta} (\dd Y^\alpha/\dd \tau) (\dd Y^\beta/\dd \tau)}=1$.}To obtain the linear Li\'{e}nard--Wiechert graviton field $\overline{\chi}^{\mu\nu}$ in Minkowski spacetime, we first set $g_{\mu\nu} = \eta_{\mu\nu}$, so that $\sqrt{|g[x]|}=1$. After integrating over spacetime $(t',\vec{x}')$ in eq. \eqref{LinearizedGravity_Minkowski_EinsteinEquation_deDonderSolution}, there is a sole remaining integral, with respect to $\tau$. We then notice that all the $\partial_{\sbar}$ derivatives in equations \eqref{MasslessScalarG_Evend} and \eqref{MasslessScalarG_Oddd} can be expressed as
\begin{align}
\label{Minkowski_sbardot}
\frac{\partial}{\partial \sbar} = \frac{1}{\dot{\sbar}} \frac{\dd}{\dd \tau} ,
\qquad\qquad
\dot{\sbar} = \dot{Y}[\tau] \cdot (Y[\tau]-x) .
\end{align}
(The ``$\cdot$" is the Minkowski dot product; for example, $X \cdot Y = \eta_{\mu\nu} X^\mu Y^\nu$.) At this juncture,
\begin{align}
\label{LinearizedGravity_Minkowski_EinsteinEquation_deDonderLW_Evend_1}
\text{Even $d \geq 4$ : }
\overline{\chi}^{\mu\nu}[t,\vec{x}] 
&= -16 \pi \GN M 
\int_{t'=-\infty}^{t'=t} \dd \tau \frac{\dot{Y}^\mu[\tau] \dot{Y}^\nu[\tau]}{(2\pi)^{\frac{d-4}{2}}} 
\left(\frac{1}{\dot{\sbar}} \frac{\dd}{\dd \tau}\right)^{ \frac{d-4}{2} } \left( \frac{\delta[\sbar]}{4\pi} \right), \\
\label{LinearizedGravity_Minkowski_EinsteinEquation_deDonderLW_Oddd_1}
\text{Odd $d \geq 5$ : }
\overline{\chi}^{\mu\nu}[t,\vec{x}] 
&= -16 \pi \GN M 
\int_{t'=-\infty}^{t'=t} \dd \tau \frac{\dot{Y}^\mu[\tau] \dot{Y}^\nu[\tau]}{(2\pi)^{\frac{d-3}{2}}} 
\left(\frac{1}{\dot{\sbar}} \frac{\dd}{\dd \tau}\right)^{ \frac{d-3}{2} } \left( \frac{\Theta[\sbar]}{2\pi \sqrt{2\sbar}} \right) .
\end{align}
To compute the energy-momentum-stress-shear of these GWs, we would need their gradients.
\begin{align}
	\label{LinearizedGravity_Minkowski_EinsteinEquation_deDonderLW_Evend_1grad}
	\text{Even $d \geq 4$ : }
	\partial_\alpha \overline{\chi}^{\mu\nu}[t,\vec{x}] 
	&= -16 \pi \GN M 
	\int_{t'=-\infty}^{t'=t} \dd \tau \frac{(x-Y)_\alpha \dot{Y}^\mu[\tau] \dot{Y}^\nu[\tau]}{(2\pi)^{\frac{d-4}{2}}} 
	\left(\frac{1}{\dot{\sbar}} \frac{\dd}{\dd \tau}\right)^{\frac{d-2}{2}} \left( \frac{\delta[\sbar]}{4\pi} \right), \\
	\label{LinearizedGravity_Minkowski_EinsteinEquation_deDonderLW_Oddd_1grad}
	\text{Odd $d \geq 5$ : }
	\partial_\alpha \overline{\chi}^{\mu\nu}[t,\vec{x}] 
	&= -16 \pi \GN M 
	\int_{t'=-\infty}^{t'=t} \dd \tau \frac{(x-Y)_\alpha \dot{Y}^\mu[\tau] \dot{Y}^\nu[\tau]}{(2\pi)^{\frac{d-3}{2}}} 
	\left(\frac{1}{\dot{\sbar}} \frac{\dd}{\dd \tau}\right)^{\frac{d-1}{2}} \left( \frac{\Theta[\sbar]}{2\pi \sqrt{2\sbar}} \right) .
\end{align}
Note that there is no need to differentiate the upper limit $t'=t$ because the integrand evaluated at the upper limit lies outside the light cone, i.e., $\sbar \to -|\vec{x}-\vec{Y}|^2/2 < 0$, and is hence set to zero by the $\Theta[\sbar]$, $\delta[\sbar]$ and the latter's derivatives.

Beginning with the even dimensional case in equations \eqref{LinearizedGravity_Minkowski_EinsteinEquation_deDonderLW_Evend_1} and \eqref{LinearizedGravity_Minkowski_EinsteinEquation_deDonderLW_Evend_1grad}, we shall integrate by parts (IBPs) the $(\dd/\dd \tau)$'s within the integrands, to act them on the trajectories $\dot{Y}$ -- without incurring any boundary terms. For, suppose this can be done up to the $(n-1)$th IBP, then the $n$th IBP hands us, for even $d > 4$ and $1 \leq n \leq (d-4)/2$,
\begin{align}
\label{LinearizedGravity_Minkowski_EinsteinEquation_deDonderLW_Evend_2}
\overline{\chi}^{\mu\nu}[t,\vec{x}] 
&= \left[16 \pi \GN M (-)^n
\left( \left(\frac{1}{\dot{\sbar}} \frac{\dd}{\dd \tau}\right)^{ n-1 } \frac{\dot{Y}^\mu[\tau] \dot{Y}^\nu[\tau]}{(2\pi)^{\frac{d-4}{2}} \dot{\sbar}}\right) 
\left(\frac{1}{\dot{\sbar}} \frac{\dd}{\dd \tau}\right)^{ \frac{d-4}{2} - n } \left( \frac{\delta[\sbar]}{4\pi} \right) \right]_{t'=\tau=-\infty}^{t'=t} \\
& - 16 \pi \GN M (-)^n
\int_{t'=-\infty}^{t'=t} \dd \tau \frac{\dd}{\dd \tau} \left\{ \left(\frac{1}{\dot{\sbar}} \frac{\dd}{\dd \tau}\right)^{ n-1 } \left(\frac{\dot{Y}^\mu[\tau] \dot{Y}^\nu[\tau]}{(2\pi)^{\frac{d-4}{2}} \dot{\sbar}} \right) \right\}
\left(\frac{1}{\dot{\sbar}} \frac{\dd}{\dd \tau}\right)^{ \frac{d-4}{2} - n } \left( \frac{\delta[\tau - \tau_{\text{ret}}]}{4\pi |\dot{\sbar}[\tau_{\text{ret}}]|} \right) ; \nonumber
\end{align}
and for $1 \leq n \leq (d-2)/2$
\begin{align}
\label{LinearizedGravity_Minkowski_EinsteinEquation_deDonderLW_Evend_2grad}
&\partial_\alpha \overline{\chi}^{\mu\nu}[t,\vec{x}] 
= \left[ 16 \pi \GN M (-)^n
\left( \left(\frac{1}{\dot{\sbar}} \frac{\dd}{\dd \tau}\right)^{n-1} \frac{(x-Y)_\alpha \dot{Y}^\mu[\tau] \dot{Y}^\nu[\tau]}{(2\pi)^{\frac{d-4}{2}} \dot{\sbar}} \right)
\left( \frac{1}{\dot{\sbar}} \frac{\dd}{\dd \tau} \right)^{\frac{d-2}{2} - n} \left( \frac{\delta[\sbar]}{4\pi} \right) \right]_{\tau=-\infty}^{t'=t} \nonumber\\
&- 16 \pi \GN M (-)^n
\int_{t'=-\infty}^{t'=t} \dd \tau 
\frac{\dd}{\dd \tau} \left\{ \left(\frac{1}{\dot{\sbar}} \frac{\dd}{\dd \tau}\right)^{n-1} \frac{(x-Y)_\alpha \dot{Y}^\mu[\tau] \dot{Y}^\nu[\tau]}{(2\pi)^{\frac{d-4}{2}} \dot{\sbar}} \right\}
\left(\frac{1}{\dot{\sbar}} \frac{\dd}{\dd \tau}\right)^{\frac{d-2}{2} - n} \left( \frac{\delta[\tau-\tau_{\text{ret}}]}{4\pi |\dot{\sbar}[\tau_\text{ret}]|} \right) . 
\end{align}
The upper limit of the boundary term in equations \eqref{LinearizedGravity_Minkowski_EinsteinEquation_deDonderLW_Evend_2} and \eqref{LinearizedGravity_Minkowski_EinsteinEquation_deDonderLW_Evend_2grad}, as already noted, lies outside the light cone and can be set to zero immediately. The lower limit $\tau \to -\infty$ amounts to taking $\sbar \to +\infty$ -- the trajectory $Y^\mu[\tau]$ is timelike -- where again $\delta[\sbar]$ and its derivatives are zero. Note, we have also converted $\delta[\sbar] = \delta[\tau - \tau_{\text{ret}}]/|\dot{\sbar}[\tau_{\text{ret}}]|$ within the remaining integral of eq. \eqref{LinearizedGravity_Minkowski_EinsteinEquation_deDonderLW_Evend_2}, where the retarded proper time $\tau_{\text{ret}}$ is the value of $\tau$ in the past that would set $\sbar=0$. This change-of-variables would allow us to integrate over the $\delta[\tau-\tau_{\text{ret}}]$ once all the $\tau$-derivatives have been IBPs onto the $\dot{Y}$'s.

Moving on to perform similar IBPs for the odd dimensional $d \geq 5$ case, if we assume all boundary terms may be discarded up to the $(n-1)$th IBP, the $n$th IBP, for $1 \leq n \leq (d-3)/2$, would yield:
\begin{align}
\label{LinearizedGravity_Minkowski_EinsteinEquation_deDonderLW_Oddd_2}
\overline{\chi}^{\mu\nu}[t,\vec{x}] 
&= \left[ 16 \pi \GN M (-)^n
\left( \left(\frac{1}{\dot{\sbar}} \frac{\dd}{\dd \tau}\right)^{ n-1 } \frac{\dot{Y}^\mu[\tau] \dot{Y}^\nu[\tau]}{(2\pi)^{\frac{d-3}{2}} \dot{\sbar}}\right) 
\left(\frac{1}{\dot{\sbar}} \frac{\dd}{\dd \tau}\right)^{ \frac{d-3}{2} - n } \left( \frac{\Theta[\sbar]}{2\pi \sqrt{2\sbar}} \right) \right]_{t'=\tau=-\infty}^{t'=t} \\
&- 16 \pi \GN M (-)^n
\int_{t'=-\infty}^{t'=t} \dd \tau \frac{\dd}{\dd \tau} \left\{ \left(\frac{1}{\dot{\sbar}} \frac{\dd}{\dd \tau}\right)^{ n-1 } \frac{\dot{Y}^\mu[\tau] \dot{Y}^\nu[\tau]}{(2\pi)^{\frac{d-3}{2}} \dot{\sbar}} \right\}
\left(\frac{1}{\dot{\sbar}} \frac{\dd}{\dd \tau}\right)^{ \frac{d-3}{2} - n } \left( \frac{\Theta[\sbar]}{2\pi \sqrt{2\sbar}} \right) \nonumber ;
\end{align}
and for $1 \leq n \leq (d-1)/2$,
\begin{align}
\label{LinearizedGravity_Minkowski_EinsteinEquation_deDonderLW_Oddd_2grad}
&\partial_\alpha \overline{\chi}^{\mu\nu}[t,\vec{x}] 
= \left[ 16 \pi \GN M (-)^n
\left(\left(\frac{1}{\dot{\sbar}} \frac{\dd}{\dd \tau}\right)^{n-1} \frac{(x-Y)_\alpha \dot{Y}^\mu[\tau] \dot{Y}^\nu[\tau]}{(2\pi)^{\frac{d-3}{2}} \dot{\sbar}}\right) 
\left(\frac{1}{\dot{\sbar}}\frac{\dd}{\dd \tau}\right)^{\frac{d-1}{2}-n} \left( \frac{\Theta[\sbar]}{2\pi \sqrt{2\sbar}} \right) \right]_{t'=\tau=-\infty}^{t'=t} \nonumber\\
&- 16 \pi \GN M (-)^n
\int_{t'=-\infty}^{t'=t} \dd \tau \frac{\dd}{\dd \tau} \left\{ \left(\frac{1}{\dot{\sbar}} \frac{\dd}{\dd \tau}\right)^{n-1} \frac{(x-Y)_\alpha \dot{Y}^\mu[\tau] \dot{Y}^\nu[\tau]}{(2\pi)^{\frac{d-3}{2}} \dot{\sbar}} \right\}
\left(\frac{1}{\dot{\sbar}} \frac{\dd}{\dd \tau}\right)^{\frac{d-1}{2}-n} \left( \frac{\Theta[\sbar]}{2\pi \sqrt{2\sbar}} \right) . 
\end{align}
Since the upper limit gives us a strictly negative $\sbar$, it is again set to zero by the presence of $\Theta[\sbar]$, $\delta[\sbar]$ and the latter's derivatives. And since the lower limit corresponds to $\sbar \to +\infty$, the only boundary term that requires analysis is the one that does not contain $\delta$-functions, namely
\begin{align}
\label{LinearizedGravity_Minkowski_EinsteinEquation_deDonderLW_Oddd_3}
\left[ -16 \pi \GN M (-)^n \Theta[\sbar]
\left(\left(\frac{1}{\dot{\sbar}} \frac{\dd}{\dd \tau}\right)^{ n-1 } \frac{\dot{Y}^\mu[\tau] \dot{Y}^\nu[\tau]}{(2\pi)^{\frac{d-3}{2}} \dot{\sbar}}\right) 
\left(\frac{\dd}{\dd \sbar}\right)^{ \frac{d-3}{2} - n } \left( \frac{1}{2\pi \sqrt{2\sbar}} \right) \right]_{t'=\tau=-\infty} 
\end{align}
and
\begin{align}
\label{LinearizedGravity_Minkowski_EinsteinEquation_deDonderLW_Oddd_3grad}
\left[ -16 \pi \GN M (-)^n \Theta[\sbar]
\left(\left(\frac{1}{\dot{\sbar}} \frac{\dd}{\dd \tau}\right)^{ n-1 } \frac{(x-Y)_\alpha \dot{Y}^\mu[\tau] \dot{Y}^\nu[\tau]}{(2\pi)^{\frac{d-3}{2}} \dot{\sbar}}\right) 
\left(\frac{\dd}{\dd \sbar}\right)^{ \frac{d-1}{2}-n} \left( \frac{1}{2\pi \sqrt{2\sbar}} \right) \right]_{t'=\tau=-\infty} .
\end{align}
Since the $\partial_{\sbar}^{((d-3)/2) - n} (2\sbar)^{-1/2}$ and $\partial_{\sbar}^{((d-1)/2) - n} (2\sbar)^{-1/2}$ terms do tend to zero in this asymptotic past limit -- if we assume all components of the time derivatives $\{\dd^m Y^\nu/\dd \tau^m \vert m = 1,2,3,\dots \}$ remain bounded for the entire trajectory -- what needs to be examined are the remaining proper time derivatives, which take the generic forms
\begin{align}
\label{LinearizedGravity_Minkowski_EinsteinEquation_deDonderLW_Oddd_4}
\mathcal{D}_s[\{a_1,\dots,a_m\},b] \equiv \frac{1}{\dot{\sbar}^{a_m}} \frac{\dd}{\dd \tau} \left( \dots \left( \frac{1}{\dot{\sbar}^{a_2}} \frac{\dd}{\dd \tau} \left( \frac{1}{\dot{\sbar}^{a_1}} \frac{\dd}{\dd \tau} \left(\frac{1}{\dot{\sbar}^{b}}\right) \right) \right) \right) 
\end{align}
and
\begin{align}
\label{LinearizedGravity_Minkowski_EinsteinEquation_deDonderLW_Oddd_4_grad}
\mathcal{D}'_s[\{a_1,\dots,a_m\},b] \equiv \frac{1}{\dot{\sbar}^{a_m}} \frac{\dd}{\dd \tau} \left( \dots \left( \frac{1}{\dot{\sbar}^{a_2}} \frac{\dd}{\dd \tau} \left( \frac{1}{\dot{\sbar}^{a_1}} \frac{\dd}{\dd \tau} \left(\frac{(x-Y[\tau])_\alpha}{\dot{\sbar}^{b}}\right) \right) \right) \right) ,
\end{align}
for positive integers $m, a_1, \dots, a_m, b \geq 1$. Now, by mathematical induction on $m$, one may show that
\begin{align}
\label{LinearizedGravity_Minkowski_EinsteinEquation_deDonderLW_Oddd_5}
\mathcal{D}_s[\{a_1,\dots,a_m\},b] = \sum_{\ell = 1+\Sigma_m}^{m + \Sigma_m} \frac{C_\ell^{(m)}}{\dot{\sbar}^{\ell+b}} ,
\qquad\qquad
\Sigma_m \equiv \sum_{i=1}^m a_i ,
\end{align}
where $C_\ell^{(m)}$ is built out of sums and/or products of the scalars $\{ (\dd^m Y/\dd \tau^m) \cdot (Y-x) \vert m = 1,2,3,\dots \}$ and $\{ (\dd^a Y/\dd \tau^a) \cdot (\dd^b Y/\dd \tau^b) \vert a,b = 1,2,3,\dots \}$. The infinite past behavior of $\mathcal{D}_s$ in eq. \eqref{LinearizedGravity_Minkowski_EinsteinEquation_deDonderLW_Oddd_4} is therefore encoded in that of the terms in eq. \eqref{LinearizedGravity_Minkowski_EinsteinEquation_deDonderLW_Oddd_5}. By dimension analysis, because $\sbar$ itself scales as (Length)$^2 \equiv L^2$, we must have
\begin{align}
\label{LinearizedGravity_Minkowski_EinsteinEquation_deDonderLW_Oddd_6}
\Bigg[ \mathcal{D}_s[\{a_1,\dots,a_m\},b] \Bigg] = \frac{\tau^{\Sigma_m-(m-b)}}{L^{2(\Sigma_m+b)}} .
\end{align}
The $\ell$th term in the sum in eq. \eqref{LinearizedGravity_Minkowski_EinsteinEquation_deDonderLW_Oddd_5}, on the other hand, goes as $[C_\ell^{(m)}] \tau^{\ell+b}/L^{2(\ell+b)}$, which when set equal to the right-hand-side of eq. \eqref{LinearizedGravity_Minkowski_EinsteinEquation_deDonderLW_Oddd_6}, implies the numerator itself scales as
\begin{align}
[C_\ell^{(m)}] \sim \left(\frac{\dd}{\dd \tau}\right)^{\ell - \Sigma_m + m} [(\text{Length})^{2(\ell - \Sigma_m)}] .
\end{align}
Let us then exploit the fact that $(Y-x)$ is a timelike vector to deduce that $\dot{\sbar} = \dot{Y} \cdot (Y-x)$ corresponds to, in the instantaneous rest frame of the source, the (negative) time elapsed between emission and receipt, namely $Y^0 - x^0$; in particular, $\dot{\sbar} \to -\infty$ as $\tau \to -\infty$. Likewise, $Z \cdot (Y-x)$, for any $d$-vector $Z$, must be equal to the rest frame elapsed time multiplied by $-Z^0$. Because we wish to study the asymptotic behavior of $\lim_{\tau \to -\infty} C_\ell^{(m)}/\dot{\sbar}^{\ell+b}$ in eq. \eqref{LinearizedGravity_Minkowski_EinsteinEquation_deDonderLW_Oddd_4}, we therefore require the most dominant terms in $C_\ell^{(m)}$, since we already know that the denominator goes as $1/(\text{instantaneous-rest-frame elapsed time})^{\text{positive power}}$. Based on the dimensional analysis above, the largest term must come from the highest power of $(Y-x)$.
\begin{align}
\lim_{\tau \to -\infty} \frac{C_\ell^{(m)}}{\dot{\sbar}^{\ell+b}} 
&\sim \frac{\prod_{i=1}^{\ell - \Sigma_m} \left( \frac{\dd^{m_i} Y}{\dd \tau} \cdot (Y-x) \right)}{\left( \dot{Y} \cdot (Y-x) \right)^{\ell+b}}, \qquad\qquad 
\sum_{i=1}^{\ell - \Sigma_m} m_i = \ell - \Sigma_m + m . 
\end{align}
We have deduced that the dominant behavior of $\mathcal{D}_s$ in eq. \eqref{LinearizedGravity_Minkowski_EinsteinEquation_deDonderLW_Oddd_4} is given by any one of the terms in eq. \eqref{LinearizedGravity_Minkowski_EinsteinEquation_deDonderLW_Oddd_5}, because each and every term there scales the same way; it is:
\begin{align}
\lim_{\tau \to -\infty}\frac{1}{\dot{\sbar}^{a_m}} \frac{\dd}{\dd \tau} \left( \dots \left( \frac{1}{\dot{\sbar}^{a_2}} \frac{\dd}{\dd \tau} \left( \frac{1}{\dot{\sbar}^{a_1}} \frac{\dd}{\dd \tau} \left(\frac{1}{\dot{\sbar}^{b}}\right) \right) \right) \right) \sim \frac{1}{(\text{instantaneous-rest-frame elapsed time})^{\Sigma_m + b}} .
\end{align}
By assumption, $\Sigma_m + b \geq m + b$ is strictly positive, so the $\tau \to -\infty$ limit of these nested derivatives of $1/\dot{\sbar}^b$ goes to zero. All boundary terms in eq. \eqref{LinearizedGravity_Minkowski_EinsteinEquation_deDonderLW_Oddd_2} are therefore zero. 

A similar line of reasoning should lead one to deduce (recall eq. \eqref{LinearizedGravity_Minkowski_EinsteinEquation_deDonderLW_Oddd_4_grad})
\begin{align}
\lim_{\tau \to -\infty} \mathcal{D}'_s[\{a_1,\dots,a_m\},b]
= \lim_{\tau \to -\infty} &\frac{1}{\dot{\sbar}^{a_m}} \frac{\dd}{\dd \tau} \left( \dots \left( \frac{1}{\dot{\sbar}^{a_2}} \frac{\dd}{\dd \tau} \left( \frac{1}{\dot{\sbar}^{a_1}} \frac{\dd}{\dd \tau} \left(\frac{(x-Y[\tau])_\alpha}{\dot{\sbar}^{b}}\right) \right) \right) \right) \nonumber\\
& \sim \frac{1}{(\text{instantaneous-rest-frame elapsed time})^{\Sigma_m + b - 1}} .
\end{align}
By assumption, $\Sigma_m + b - 1 \geq m + b - 1$ is strictly non-negative, so the $\tau \to -\infty$ limit of these nested derivatives of $1/\dot{\sbar}^b$ does not blow up. All boundary terms in eq. \eqref{LinearizedGravity_Minkowski_EinsteinEquation_deDonderLW_Oddd_2grad} are therefore zero.

{\it Results} \qquad We have arrived at the linear Li\'{e}nard--Wiechert graviton field of a point mass $M$ moving about in a background Minkowski spacetime (i.e., eq. \eqref{StressTensor_PointMass} with $g_{\mu\nu}=\eta_{\mu\nu}$). For even spacetime dimensions $d \geq 4$,
\begin{align}
\label{LinearizedGravity_Minkowski_EinsteinEquation_deDonderLW_Evend}
\overline{\chi}^{\mu\nu}[t,\vec{x}] 
&= \left. 4 \GN M (-)^{ \frac{d-4}{2} }
\left( \left( \frac{1}{\dot{\sbar}} \frac{\dd}{\dd \tau} \right)^{ \frac{d-4}{2} } \frac{\dot{Y}^\mu[\tau] \dot{Y}^\nu[\tau]}{(2\pi)^{\frac{d-4}{2}} \dot{\sbar}} \right) \right\vert_{\tau=\tau_{\text{ret}}} , \\
\label{LinearizedGravity_Minkowski_EinsteinEquation_deDonderLW_Evend_grad}
\partial_\alpha \overline{\chi}^{\mu\nu}[t,\vec{x}] 
&= \left. 4 \GN M (-)^{ \frac{d-2}{2} }
\left( \left( \frac{1}{\dot{\sbar}} \frac{\dd}{\dd \tau} \right)^{ \frac{d-2}{2} } 
\frac{(x-Y[\tau])_\alpha \dot{Y}^\mu[\tau] \dot{Y}^\nu[\tau]}{(2\pi)^{\frac{d-4}{2}} \dot{\sbar}} \right) \right\vert_{\tau=\tau_{\text{ret}}} ;
\end{align}
and for odd spacetime dimensions $d \geq 5$,
\begin{align}
\label{LinearizedGravity_Minkowski_EinsteinEquation_deDonderLW_Oddd}
\overline{\chi}^{\mu\nu}[t,\vec{x}] 
&= 8 \GN M (-)^{ \frac{d-5}{2} }
\int_{-\infty}^{\tau_{\text{ret}}-0^+} \frac{\dd \tau}{\sqrt{2 \sbar}} 
\frac{\dd}{\dd \tau} \left( \left( \frac{1}{\dot{\sbar}} \frac{\dd}{\dd \tau} \right)^{ \frac{d-5}{2} } \frac{\dot{Y}^\mu[\tau] \dot{Y}^\nu[\tau]}{(2\pi)^{\frac{d-3}{2}} \dot{\sbar}} \right) , \\
\label{LinearizedGravity_Minkowski_EinsteinEquation_deDonderLW_Oddd_grad}
\partial_\alpha \overline{\chi}^{\mu\nu}[t,\vec{x}] 
&= 8 \GN M (-)^{ \frac{d-3}{2} }
\int_{-\infty}^{\tau_{\text{ret}}-0^+} \frac{\dd \tau}{\sqrt{2 \sbar}} 
\frac{\dd}{\dd \tau} \left( \left( \frac{1}{\dot{\sbar}} \frac{\dd}{\dd \tau} \right)^{ \frac{d-3}{2} } \frac{(x-Y[\tau])_\alpha \dot{Y}^\mu[\tau] \dot{Y}^\nu[\tau]}{(2\pi)^{\frac{d-3}{2}} \dot{\sbar}} \right) ;
\end{align}
where we record that $\dot{Y}^\mu \equiv \dd Y^\mu[\tau]/\dd\tau$ and
\begin{align}
\label{Minkowski_sbardot_sret}
\dot{\sbar}[\tau_{\text{ret}}] 
&= \left(\frac{\dd \tau_{\text{ret}}}{\dd t}\right)^{-1} \frac{\dd Y}{\dd t} \cdot \left( Y - x \right) \\
&= -\frac{1 - \vec{V} \cdot \widehat{n}}{\sqrt{1-(\dd \vec{Y}/\dd t)^2}} |\vec{x}-\vec{Y}| , \qquad\qquad
\widehat{n} \equiv \frac{\vec{x}-\vec{Y}}{| \vec{x}-\vec{Y}|} , \qquad\qquad
\vec{V} \equiv \frac{\dd\vec{Y}}{\dd t} . \nonumber
\end{align}
We also remind the reader that the retarded proper time $\tau_{\text{ret}}$ is defined as the solution to
\begin{align}
\label{Minkowski_RetardedTime}
x^0 - Y^0[\tau_{\text{ret}}] = t - Y^0[\tau_{\text{ret}}] = \left\vert \vec{x} - \vec{Y}[\tau_{\text{ret}}] \right\vert .
\end{align}
As already advertised, in even dimensions $d \geq 4$, the graviton field of eq. \eqref{LinearizedGravity_Minkowski_EinsteinEquation_deDonderLW_Evend} and its gradient in eq. \eqref{LinearizedGravity_Minkowski_EinsteinEquation_deDonderLW_Evend_grad} are the gravitational signals coming from the intersection between the past light cone of the observer at $x^\mu = (t,\vec{x})$ and the worldline of the source, i.e., from the point at $Y^\mu[\tau_{\text{ret}}]$. While in odd dimensions $d \geq 5$, the graviton field of eq. \eqref{LinearizedGravity_Minkowski_EinsteinEquation_deDonderLW_Oddd} and its gradient in eq. \eqref{LinearizedGravity_Minkowski_EinsteinEquation_deDonderLW_Oddd_grad} are the cumulative gravitational signals arising from the entire past trajectory of the source $Y^\mu[\tau < \tau_{\text{ret}}]$ lying strictly within the past light cone of the observer at $x$. That this odd $d \geq 5$ case yields pure tail signals is worth reiterating because the corresponding symmetric Green's function of eq. \eqref{MasslessScalarG_Oddd} does contain $\delta[\sbar]$ and its derivatives, and thus one might otherwise had expected some portion of the signal to travel on the null cone.

{\bf Newtonian potential of point mass} \qquad As a check of our results here, let us recover the expected $1/(\text{spatial distance})^{d-3}$ Newtonian potential of a point mass at rest. In this static case, proper time is coordinate time and from eq. \eqref{Minkowski_sbardot}:
\begin{align}
\tau=t', \qquad\qquad \dot{\sbar}=Y^0-x^0=t'-t, \qquad \text{ and } \qquad  \dot{Y}^\mu = \delta^\mu_0 .
\end{align}
Moreover, from eq. \eqref{Minkowski_RetardedTime}, the retarded time is $\tau_{\text{ret}} = t' = t-|\vec{x}-\vec{Y}|$, which tells us
\begin{align}
\dot{\sbar}[\tau_{\text{ret}}]=-|\vec{x}-\vec{Y}| .
\end{align}
With these results, we may now perform a change-of-variables $\lambda \equiv (t'-t)/|\vec{x}-\vec{Y}|$, to convert equations \eqref{LinearizedGravity_Minkowski_EinsteinEquation_deDonderLW_Evend} and \eqref{LinearizedGravity_Minkowski_EinsteinEquation_deDonderLW_Oddd} to
\begin{align}
\label{LinearizedGravity_Minkowski_EinsteinEquation_deDonderStaticLW_Evend_1}
\text{Even $d \geq 4$ : }
\overline{\chi}^{\mu\nu}[t,\vec{x}] 
&= - \frac{4 \GN M}{|\vec{x}-\vec{Y}|^{d-3}} \delta^\mu_0 \delta^\nu_0 
\frac{(d-5)!!}{(2\pi)^{\frac{d-4}{2}}} , \\
\label{LinearizedGravity_Minkowski_EinsteinEquation_deDonderStaticLW_Oddd_2}
\text{Odd $d \geq 5$ : }
\overline{\chi}^{\mu\nu}[t,\vec{x}] 
&= - \frac{8 \GN M}{|\vec{x}-\vec{Y}|^{d-3}} \delta^\mu_0 \delta^\nu_0 \frac{(d-4)!!}{(2\pi)^{\frac{d-3}{2}}}
\int_{-\infty}^{-1} \frac{\dd \lambda}{\sqrt{\lambda^2-1} \lambda^{d-3}} ;
\end{align}
For odd integer $n$, $n!! \equiv 1 \cdot 3 \cdot \dots \cdot (n-2) \cdot n$; and $(-1)!! \equiv 1$. To arrive at equations \eqref{LinearizedGravity_Minkowski_EinsteinEquation_deDonderStaticLW_Evend_1} and \eqref{LinearizedGravity_Minkowski_EinsteinEquation_deDonderStaticLW_Oddd_2}, we have made use of the following identity, valid for arbitrary positive integer $n$:
\begin{align}
\label{DifferentiationIdentity}
\left( \frac{1}{\lambda} \frac{\dd}{\dd \lambda}\right)^n \frac{1}{\lambda} = \frac{(-)^n (2n-1)!!}{\lambda^{2n+1}} .
\end{align}
The integral in eq. \eqref{LinearizedGravity_Minkowski_EinsteinEquation_deDonderStaticLW_Oddd_2} can be tackled by changing variables $\lambda \equiv -1/\lambda'^{1/2}$ and using the integral
\begin{align}
\int_{0}^{1} \dd\lambda' \lambda'^{\alpha-1} (1-\lambda')^{\beta-1} = \frac{\Gamma[\alpha] \Gamma[\beta]}{\Gamma[\alpha+\beta]}, \qquad\qquad
\text{Re}[\alpha], \ \text{Re}[\beta]>0.
\end{align}
Here, the $\Gamma[z]$'s are Gamma functions. After employing relevant $\Gamma$-function identities, the final result takes the same expression for both even and odd $d \geq 4$:
\begin{align}
\overline{\chi}^{\mu\nu}[t,\vec{x}] = 16\pi \GN M \delta^\mu_0 \delta^\nu_0 \mathcal{G}^{(\text{E})}_{d-1}[\vec{x}-\vec{Y}],
\end{align}
where $\mathcal{G}^{(\text{E})}_{d-1}$ is the Green's function of the Euclidean Laplacian in $(d-1)$-spatial dimensions,
\begin{align}
\mathcal{G}^{(\text{E})}_{d-1}[\vec{x}-\vec{Y}] 
		= \left(\frac{1}{\vec{\nabla}^2} \right)[\vec{x}-\vec{Y}]
		= -\frac{\Gamma\left[ \frac{d-3}{2} \right]}{4 \pi^{\frac{d-1}{2}} \vert \vec{x}-\vec{Y} \vert^{d-3} } .
\end{align}
A note about causal structure. Even though the static nature of the Newtonian potential guarantees it takes the same form for all relevant dimensions -- i.e., we could have obtained the same answer through Gauss' law applied to the linearized Einstein's equation -- the integral in eq. \eqref{LinearizedGravity_Minkowski_EinsteinEquation_deDonderStaticLW_Oddd_2} is a reminder that, because of the pure tail nature of the gravitational signal in odd dimensions $(d \geq 5)$, strictly speaking, Newtonian's law of gravitation requires the point mass to be still for its entire past history. In even dimensions, on the other hand, the static potential of eq. \eqref{LinearizedGravity_Minkowski_EinsteinEquation_deDonderStaticLW_Evend_1} only requires the point mass to be still in the neighborhood of the retarded time $\tau = \tau_\text{ret}$.

{\bf Graviton fields of continuous sources} \qquad Before moving on to the subject of GW memory, let us record here that, the graviton field generated by an isolated but continuous (as opposed to the discrete point mass) source in eq. \eqref{LinearizedGravity_Minkowski_EinsteinEquation_deDonderSolution} can be derived in a manner similar to that of the Li\'{e}nard--Wiechert fields above. 
\begin{align}
\label{LinearizedGravity_Minkowski_EinsteinEquation_deDonderContinuousSource_Evend}
\text{Even $d \geq 4$ : } \overline{\chi}_{\mu\nu} 
&= -4\GN \int_{\mathbb{R}^{d-1}} \dd^{d-1}\vec{x}' \left.\left( \frac{1}{2\pi(t-t')} \frac{\partial}{\partial t'} \right)^{\frac{d-4}{2}} \left( \frac{\overline{T}_{\mu\nu}[t',\vec{x}']}{t-t'} \right)\right\vert_{t-t'=|\vec{x}-\vec{x}'|} \\
\label{LinearizedGravity_Minkowski_EinsteinEquation_deDonderContinuousSource_Oddd}
\text{Odd $d \geq 5$ : } \overline{\chi}_{\mu\nu} 
&= -8\GN \int_{\mathbb{R}^{d-1}} \dd^{d-1}\vec{x}' \int_{-\infty}^{t-|\vec{x}-\vec{x}'|-0^+} \frac{\dd t'}{\sqrt{2\sbar}}
\frac{\partial}{\partial t'} \left\{ \left( \frac{1}{2\pi(t-t')} \frac{\partial}{\partial t'} \right)^{\frac{d-5}{2}} \left( \frac{\overline{T}_{\mu\nu}[t',\vec{x}']}{2\pi(t-t')} \right) \right\} 
\end{align}
The gradients are, for even $d \geq 4$: 
\begin{align}
\label{LinearizedGravity_Minkowski_EinsteinEquation_deDonderContinuousSource_Evend_grad}
\partial_\alpha \overline{\chi}_{\mu\nu} 
&= -8\pi\GN \int_{\mathbb{R}^{d-1}} \dd^{d-1}\vec{x}' \left.
\left\{ \left( \frac{1}{2\pi(t-t')} \frac{\partial}{\partial t'} \right)^{\frac{d-2}{2}} \left( \frac{(x-x')_\alpha \overline{T}_{\mu\nu}[t',\vec{x}']}{t-t'} \right)\right\} \right\vert_{t-t'=|\vec{x}-\vec{x}'|} ;
\end{align}
while for odd $d \geq 5$,
\begin{align}
\label{LinearizedGravity_Minkowski_EinsteinEquation_deDonderContinuousSource_Oddd_grad}
\partial_\alpha \overline{\chi}_{\mu\nu} 
&= -16\pi\GN \int_{\mathbb{R}^{d-1}} \dd^{d-1}\vec{x}' \int_{-\infty}^{t-|\vec{x}-\vec{x}'|-0^+} \frac{\dd t'}{\sqrt{2\sbar}}
\frac{\partial}{\partial t'} \left\{ \left( \frac{1}{2\pi(t-t')} \frac{\partial}{\partial t'} \right)^{\frac{d-3}{2}} \left( \frac{(x-x')_\alpha \overline{T}_{\mu\nu}[t',\vec{x}']}{2\pi(t-t')} \right) \right\} .
\end{align}
{\bf GW Memory from unbound orbits in even dimensions $d \geq 4$} \qquad We now consider an arbitrary number ($N \geq 2$) of compact bodies, which we will model as point masses, gravitationally scattering off each other on unbound trajectories. Specifically, in the asymptotic past, they have some (incoming) constant velocities $\{ V_{(\text{I} \vert 0)} \vert \text{I} \in 1,2,\dots,N \}$. Denoting the spacetime trajectory of the $\text{I}$-th compact object as $Y^\mu_{\text{I}}$, we have
\begin{align}
\label{PointSource_In}
\lim_{t \to -\infty} \frac{\dd Y_{\text{I}}^\mu[t]}{\dd t} 
\equiv V_{(\text{I} \vert 0)}^\mu = (1,\vec{V}_{(\text{I} \vert 0)}) \qquad\qquad \text{(constant)}.
\end{align}
These $N$ bodies interact gravitationally, and then scatter off to infinity, such that in the asymptotic future they again approach some (outgoing) constant velocities $\{ V_{(\text{I} \vert 1)} \vert \text{I} \in 1,2,\dots,N \}$, namely
\begin{align}
\label{PointSource_Out}
\lim_{t \to +\infty} \frac{\dd Y_{\text{I}}^\mu[t]}{\dd t} 
\equiv V_{(\text{I} \vert 1)}^\mu = (1,\vec{V}_{(\text{I} \vert 1)}) \qquad\qquad \text{(constant)}.
\end{align}
Note that, when $\dd Y^\mu/\dd t$ approaches a constant, so does $\dd Y^\mu/\dd \tau$; in this limit, taking a proper time derivative of eq. \eqref{Minkowski_sbardot} returns
\begin{align}
\ddot{\sbar} \equiv \frac{\dd^2 \sbar}{\dd \tau^2} = \ddot{Y} \cdot (Y-x) + \dot{Y} \cdot \dot{Y} = 1 ,
\end{align}
which in turn allows us to replace the $\dd/\dd \tau$ in eq. \eqref{LinearizedGravity_Minkowski_EinsteinEquation_deDonderLW_Evend} with $\dd/\dd \dot{\sbar}$. Employing eq. \eqref{DifferentiationIdentity}, we may invoke superposition to obtain the fields generated by the asymptotic trajectories of these $N$ bodies:
\begin{align}
\lim_{t \to \pm \infty} \overline{\chi}^{\mu\nu}[t,\vec{x}] 
&= \frac{(d-5)!!}{(2\pi)^{\frac{d-4}{2}}} \sum_{\text{I}=1}^{N} \left. 4 \GN M_{\text{I}}
\frac{\dot{Y}_{\text{I}}^\mu \dot{Y}_{\text{I}}^\nu}{\dot{\sbar}^{d-3}} \right\vert_{\tau_{\text{I}}=\tau_{(\text{I}\vert\text{ret})}} 
\end{align}
At this point, we have generalized the GW memory effect to all even dimensions $d \geq 4$. For simplicity we shall define $\vec{x}=\vec{x}'=\vec{0}$ to lie within the central region of their closest approach, and proceed to take the far zone limit $|\vec{x}-\vec{Y}_{\text{I}}| \approx |\vec{x}|$ while employing eq. \eqref{Minkowski_sbardot_sret} to convert proper-time derivatives to coordinate-time ones:
\begin{align}
\label{LinearizedGravity_Minkowski_Memory}
\Delta \overline{\chi}^{\mu\nu} \approx \,^{\text{(out)}}\overline{\chi}^{\mu\nu} - \,^{\text{(in)}}\overline{\chi}^{\mu\nu} ,
\end{align}
where the contribution to the linear graviton field from the worldlines of the point masses $\{ M_\text{I} \vert \text{I}=1,2,\dots,N \}$ in the asymptotic past is
\begin{align}
\label{LinearizedGravity_Minkowski_Memory_In}
\,^{\text{(in)}}\overline{\chi}^{\mu\nu} \approx
-\frac{(d-5)!!}{(2\pi)^{\frac{d-4}{2}}} 4 \GN \sum_{\text{I}=1}^{N} \frac{M_{\text{I}}}{|\vec{x}|^{d-3}}
\left(1-\vec{V}_{(\text{I} \vert 0)}^2\right)^{ \frac{d-5}{2} } 
\frac{ V^\mu_{(\text{I} \vert 0)} V^\nu_{(\text{I} \vert 0)} }{ \left( 1 - \vec{V}_{(\text{I} \vert 0)} \cdot \widehat{n} \right)^{d-3} } 
\end{align}
and that from the asymptotic future is
\begin{align}
\label{LinearizedGravity_Minkowski_Memory_Out}
\,^{\text{(out)}}\overline{\chi}^{\mu\nu} \approx
-\frac{(d-5)!!}{(2\pi)^{\frac{d-4}{2}}} 4 \GN \sum_{\text{I}=1}^{N} \frac{M_{\text{I}}}{|\vec{x}|^{d-3}}
\left(1-\vec{V}_{(\text{I} \vert 1)}^2\right)^{ \frac{d-5}{2} } 
\frac{ V^\mu_{(\text{I} \vert 1)} V^\nu_{(\text{I} \vert 1)} }{ \left( 1 - \vec{V}_{(\text{I} \vert 1)} \cdot \widehat{n} \right)^{d-3} } .
\end{align}
By referring to eq. \eqref{FractionalDistortion_AtConstantMinkowskiTime_FirstOrderHuman}, the contribution to the GW memory effect in even dimensions $d \geq 4$ is the TT part of eq. \eqref{LinearizedGravity_Minkowski_Memory}.

\section{Maxwell and Linearized Yang-Mills(-Wong's) Equations}
\label{Section_EMandLinearizedYM}
In section \S \eqref{Section_GaugeInvariantElectromagnetism} we solved Maxwell's equations in the spatially flat FLRW cosmology of equations \eqref{SpatiallFlatFLRW_Metric} and \eqref{SpatiallFlatFLRW_Metric_w}, using a gauge-invariant formalism. In this section we will solve Maxwell's and the linearized Yang-Mills(-Wong's) equations in the same geometry, but by fixing a gauge. We will continue to move indices with the flat metric. In particular, the zeroth component of eq. \eqref{Maxwell} can be expressed as
\begin{align}
\label{Maxwell_0}
\partial^2 A^0 - \partial^0 \partial^\sigma A_\sigma = a^4 \mathcal{J}^0 ,
\end{align}
and the spatial ones as
\begin{align}
\label{Maxwell_i}
\partial^2 A^i - \partial^i \left( \partial^\sigma A_\sigma + (d-4) \frac{\dot{a}}{a} A^0 \right) + (d-4) \frac{\dot{a}}{a} \dot{A}^i = a^4 \mathcal{J}^i .
\end{align}

\subsection{Lorenz Gauge Minkowski Spacetime: Li\'{e}nard--Wiechert Vector Fields and Memory}

In Minkowski spacetime, we set $a=1$ and $(\eta,\vec{x}) \to (t,\vec{x}) \equiv x^\mu$; choosing the Lorenz gauge
\begin{align}
\label{Minkowski_EM_LorenzGauge}
\partial^\sigma A_\sigma = 0 ,
\end{align}
Maxwell's equations \eqref{Maxwell_0} and \eqref{Maxwell_i} reduce to
\begin{align}
\partial^2 A^\mu = \mathcal{J}^\mu .
\end{align}
We may employ the Green's functions in equations \eqref{MasslessScalarG_Evend} and \eqref{MasslessScalarG_Oddd} to write down the retarded solutions:
\begin{align}
\label{EMGaugePotential_MinkowskiSolutions}
A^\mu[t,\vec{x}] = \int_{-\infty}^{t} \dd t' \int_{\mathbb{R}^{d-1}} \dd^{d-1} \vec{x}' \overline{\mathcal{G}}_d \left[ \sbar \right] \mathcal{J}^\mu[t',\vec{x}'] .
\end{align}
{\bf Li\'{e}nard--Wiechert fields} \qquad We now specialize to the current of a point electric charge $q$ in motion, sweeping out a timelike trajectory $Y^\mu$,
\begin{align}
\label{EMCurrent_PointCharge}
\mathcal{J}^\mu[x] = q \int_{-\infty}^{+\infty} \dd \tau \dot{Y}^\mu[\tau] \frac{\delta^{(d)}[x-Y[\tau]]}{\sqrt{|g[x]|}} .
\end{align}
This expression, with $\tau$ being the point charge's proper time, is valid in any curved geometry $g_{\mu\nu}$; in this section $g_{\mu\nu} = \eta_{\mu\nu}$. Inserting eq. \eqref{EMCurrent_PointCharge} into eq. \eqref{EMGaugePotential_MinkowskiSolutions}, and following a calculation very similar to the gravitational one above, one would arrive at the Lorenz gauge Li\'{e}nard--Wiechert vector potential of a point charge $q$ moving about in a background Minkowski spacetime:
\begin{align}
\label{EM_Minkowski_LorenzLW_Evend}
\text{Even $d \geq 4$ : }
A^{\mu}[t,\vec{x}] 
&= \left. \frac{q}{4\pi} (-)^{ \frac{d-2}{2} }
\left( \left( \frac{1}{\dot{\sbar}} \frac{\dd}{\dd \tau} \right)^{ \frac{d-4}{2} } \frac{\dot{Y}^\mu[\tau]}{(2\pi)^{\frac{d-4}{2}} \dot{\sbar}} \right) \right\vert_{\tau=\tau_{\text{ret}}} , \\
\label{EM_Minkowski_LorenzLW_Oddd}
\text{Odd $d \geq 5$ : }
A^{\mu}[t,\vec{x}] 
&= \frac{q}{2\pi} (-)^{ \frac{d-3}{2} }
\int_{-\infty}^{\tau_{\text{ret}}-0^+} \frac{\dd \tau}{\sqrt{2 \sbar}} 
\frac{\dd}{\dd \tau} \left( \left( \frac{1}{\dot{\sbar}} \frac{\dd}{\dd \tau} \right)^{ \frac{d-5}{2} } \frac{\dot{Y}^\mu[\tau]}{(2\pi)^{\frac{d-3}{2}} \dot{\sbar}} \right) ;
\end{align}
where $\dot{\sbar}$ evaluated at retarded proper time $\tau_{\text{ret}}$ is given in eq. \eqref{Minkowski_sbardot_sret}. The electromagnetic fields of such a point charge, encoded within the Faraday tensor $F^{\mu\nu} \equiv \partial^{[\mu} A^{\nu]}$, is given by
\begin{align}
\label{EM_Minkowski_LorenzLW_Evend_grad}
\text{Even $d \geq 4$ : }
F^{\mu\nu}[t,\vec{x}] 
&= \left. \frac{q}{4\pi} (-)^{ \frac{d}{2} }
\left( \left( \frac{1}{\dot{\sbar}} \frac{\dd}{\dd \tau} \right)^{ \frac{d-2}{2} } \frac{(x-Y[\tau])^{[\mu} \dot{Y}^{\nu]}[\tau]}{(2\pi)^{\frac{d-4}{2}} \dot{\sbar}} \right) \right\vert_{\tau=\tau_{\text{ret}}} , \\
\label{EM_Minkowski_LorenzLW_Oddd_grad}
\text{Odd $d \geq 5$ : }
F^{\mu\nu}[t,\vec{x}] 
&= \frac{q}{2\pi} (-)^{ \frac{d-1}{2} }
\int_{-\infty}^{\tau_{\text{ret}}-0^+} \frac{\dd \tau}{\sqrt{2 \sbar}} 
\frac{\dd}{\dd \tau} \left( \left( \frac{1}{\dot{\sbar}} \frac{\dd}{\dd \tau} \right)^{ \frac{d-3}{2} } \frac{(x-Y[\tau])^{[\mu} \dot{Y}^{\nu]}[\tau]}{(2\pi)^{\frac{d-3}{2}} \dot{\sbar}} \right) .
\end{align}
{\bf Electromagnetic fields of continuous sources} \qquad The electromagnetic fields generated by an isolated but continuous (as opposed to the discrete point charge) source $\mathcal{J}^\nu$ can be derived in a manner similar to that of the Li\'{e}nard--Wiechert gauge field solutions above. 
\begin{align}
\label{EM_Minkowski_LorenzLW_ContinuousSource_Evend}
\text{Even $d \geq 4$ : } A_{\mu}[t,\vec{x}] 
&= \frac{q}{4\pi} \int_{\mathbb{R}^{d-1}} \dd^{d-1}\vec{x}' \left.\left( \frac{1}{2\pi(t-t')} \frac{\partial}{\partial t'} \right)^{\frac{d-4}{2}} \left( \frac{\mathcal{J}_{\mu}[t',\vec{x}']}{t-t'} \right)\right\vert_{t-t'=|\vec{x}-\vec{x}'|} \\
\label{EM_Minkowski_LorenzLW_ContinuousSource_Oddd}
\text{Odd $d \geq 5$ : } A_{\mu}[t,\vec{x}] 
&= \frac{q}{2\pi} \int_{\mathbb{R}^{d-1}} \dd^{d-1}\vec{x}' \int_{-\infty}^{t-|\vec{x}-\vec{x}'|-0^+} \frac{\dd t'}{\sqrt{2\sbar}}
\frac{\partial}{\partial t'} \left\{ \left( \frac{1}{2\pi(t-t')} \frac{\partial}{\partial t'} \right)^{\frac{d-5}{2}} \left( \frac{\mathcal{J}_{\mu}[t',\vec{x}']}{2\pi(t-t')} \right) \right\} 
\end{align}
The Faraday tensor $F_{\alpha\beta} \equiv \partial_{[\alpha} A_{\beta]}$ is, for even $d \geq 4$,
\begin{align}
\label{EM_Minkowski_LorenzLW_ContinuousSource_Evend_grad}
F_{\alpha\beta}[t,\vec{x}] 
&= \frac{q}{2} \int_{\mathbb{R}^{d-1}} \dd^{d-1}\vec{x}' \left.
\left( \frac{1}{2\pi(t-t')} \frac{\partial}{\partial t'} \right)^{\frac{d-2}{2}} \left( \frac{(x-x')_{[\alpha} \mathcal{J}_{\beta]}[t',\vec{x}']}{t-t'} \right)  \right\vert_{t-t'=|\vec{x}-\vec{x}'|} ;
\end{align}
while for odd $d \geq 5$, it reads instead
\begin{align}
\label{EM_Minkowski_LorenzLW_ContinuousSource_Oddd_grad}
F_{\alpha\beta}[t,\vec{x}] 
&= q \int_{\mathbb{R}^{d-1}} \dd^{d-1}\vec{x}' \int_{-\infty}^{t-|\vec{x}-\vec{x}'|-0^+} \frac{\dd t'}{\sqrt{2\sbar}}
\frac{\partial}{\partial t'} \left\{ \left( \frac{1}{2\pi(t-t')} \frac{\partial}{\partial t'} \right)^{\frac{d-3}{2}} \left( \frac{(x-x')_{[\alpha} \mathcal{J}_{\beta]}[t',\vec{x}']}{2\pi(t-t')} \right) \right\} .
\end{align}
{\bf Causal structure} \qquad With the Faraday tensor, we may construct the energy-momentum-shear-stress of the electromagnetic fields $\,^{(\gamma)}T_{\mu\nu}$. It is
\begin{align}
\label{EM_StressTensor}
\,^{(\gamma)}T_{\mu\nu} = -F_{\mu\sigma} F_{\nu}^{\phantom{\nu}\sigma} + \frac{1}{4} g_{\mu\nu} F_{\sigma\rho} F^{\sigma\rho} .
\end{align}
By inspecting equations \eqref{EM_Minkowski_LorenzLW_Evend} through \eqref{EM_Minkowski_LorenzLW_ContinuousSource_Oddd_grad} we learn that, just like the linearized gravitation case, the U$_1$ gauge potential and the electromagnetic fields engendered by a spatially localized system of charges in motion transmit information solely on the light cone in even dimensions $d \geq 4$ and strictly within the null cone in odd dimensions $d \geq 5$. This statement also applies to the radiative electromagnetic Poynting vector, i.e, the $0i$ component of eq. \eqref{EM_StressTensor}, since $\,^{(\gamma)}T_{\mu\nu}$ is built solely out of the Faraday tensor $F_{\alpha\beta}$ (and the metric).

{\bf Linearized Yang-Mills-Wong's Equations} \qquad In Yang-Mills (non-Abelian) gauge theory, the vector potential $A^a_{\phantom{a}\mu}$ now acquires an additional color index $a$, and the full field strength tensor is
\begin{align}
F^a_{\phantom{a}\mu\nu} = \partial_{[\mu} A^a_{\phantom{a}\nu]} + i g f^{abc} A^b_{\phantom{b}\mu} A^c_{\phantom{c}\nu} ,
\end{align}
where $g$ measures the strength of the gauge boson self-interactions and the $f^{abc}$ are the structure constants of the gauge group of interest. In particular, if $\{ T^a \}$ are the generators of the group, normalized such that $\Tr{T^a T^b} = \delta^{ab}/2$, then the structure constants arise from the commutator $[T^a, T^b] = i f^{abc} T^c$. The analog of the electromagnetic Li\'{e}nard--Wiechert fields, i.e., the Yang-Mills gauge field produced by a point charge, is governed by Wong's equations \cite{Wong:1970fu}. Wong introduced an isospin vector $I^a$ to source the color degrees of freedom, and the resulting equations are
\begin{align}
\label{WongEquation}
D_\mu F^{a\mu\nu} = \mathcal{J}^{a\nu} , 
\end{align}
where the colored point charge current is
\begin{align}
\label{ColoredPointCharge_Current}
\mathcal{J}^{a\nu}[x] \equiv q \int_{-\infty}^{+\infty} \dd \tau I^a[\tau] \dot{Y}^\nu[\tau] \frac{\delta^{(d)}[x-Y[\tau]]}{\sqrt{|g[x]|}} . 
\end{align}
In eq. \eqref{WongEquation}, $D_\mu$ is the gauge-covariant derivative in curved spacetime. Because $D_\mu D_\nu F^{a\mu\nu} = 0$ is an identity, consistency demands that the right hand side of Wong's equations be divergence-less in the gauge-covariant sense: $D_\nu \mathcal{J}^{a\nu} = 0$. This in turn leads to the conclusion that the isospin vector must be parallel transported along the color charge's worldline, $(\dd Y^\sigma/\dd \tau) D_\sigma I^a = 0$. (This is a direct analogy to the gravitational situation, where the consistency of Einstein's equations require that the stress-energy tensor be divergence-less, $\nabla_\mu T^{\mu\nu}=0$; the geodesic equation is uncovered when this is applied to the point mass case of eq. \eqref{StressTensor_PointMass}.) Moreover, the length of $I^a$ is preserved along this same trajectory, namely $(\dd Y^\sigma/\dd \tau) D_\sigma (\delta_{ab} I^a I^b) = 0$. 

While the form of Wong's equation in eq. \eqref{WongEquation} holds in any geometry, we shall be content with $g_{\mu\nu}=\eta_{\mu\nu}$ and its linearized version. Specifically, if we impose the Lorenz gauge
\begin{align}
\partial^\sigma A^a_{\phantom{a}\sigma} = 0 ,
\end{align}
then keeping only terms linear in $A^a_{\phantom{a}\mu}$ in eq. \eqref{WongEquation} yields
\begin{align}
\label{WongEquation_Linearized}
\partial^2 A^{a\mu}[x] = q \int_{-\infty}^{+\infty} \dd \tau I^a[\tau] \dot{Y}^\mu[\tau] \delta^{(d)}[x-Y[\tau]]  .
\end{align}
This linearized Lorenz gauge Yang-Mills-Wong's equations are almost identical to its U$_1$ Maxwell counterpart (i.e., eq. \eqref{EMCurrent_PointCharge} inserted into eq. \eqref{EMGaugePotential_MinkowskiSolutions}). In fact, by comparing the right hand side of eq. \eqref{WongEquation_Linearized} to eq. \eqref{EMCurrent_PointCharge}, we see the solution to eq. \eqref{WongEquation_Linearized} can be obtained from equations \eqref{EM_Minkowski_LorenzLW_Evend} and \eqref{EM_Minkowski_LorenzLW_Oddd} by replacing $\dot{Y}^\mu \to I^a \dot{Y}^\mu$. This hands us the linearized Li\'{e}nard--Wiechert solution to eq. \eqref{WongEquation} in the Lorenz gauge:
\begin{align}
\label{YMW_Minkowski_LorenzLW_Evend}
\text{Even $d \geq 4$ : }
A^{a\mu}[t,\vec{x}] 
&= \left. \frac{q}{4\pi} (-)^{ \frac{d-2}{2} }
\left( \left( \frac{1}{\dot{\sbar}} \frac{\dd}{\dd \tau} \right)^{ \frac{d-4}{2} } \frac{I^a[\tau] \dot{Y}^\mu[\tau]}{(2\pi)^{\frac{d-4}{2}} \dot{\sbar}} \right) \right\vert_{\tau=\tau_{\text{ret}}} , \\
\label{YMW_Minkowski_LorenzLW_Oddd}
\text{Odd $d \geq 5$ : }
A^{a\mu}[t,\vec{x}] 
&= \frac{q}{2\pi} (-)^{ \frac{d-3}{2} }
\int_{-\infty}^{\tau_{\text{ret}}-0^+} \frac{\dd \tau}{\sqrt{2 \sbar}} 
\frac{\dd}{\dd \tau} \left( \left( \frac{1}{\dot{\sbar}} \frac{\dd}{\dd \tau} \right)^{ \frac{d-5}{2} } \frac{I^a[\tau] \dot{Y}^\mu[\tau]}{(2\pi)^{\frac{d-3}{2}} \dot{\sbar}} \right) ;
\end{align}
where, once again, $\dot{\sbar}$ evaluated at retarded proper time $\tau_{\text{ret}}$ is given in eq. \eqref{Minkowski_sbardot_sret}.

{\it Comparison with linearized gravity} \qquad By comparing the energy-momentum-shear-stress tensor of a point mass in eq. \eqref{StressTensor_PointMass} to that of the color charge current of Wong's equations in eq. \eqref{ColoredPointCharge_Current}, let us observe we can go from the latter to the former by replacing mass with charge and isospin vector with the $d$-proper-velocity:
\begin{align}
\label{GRYM2_I}
q \to M, \qquad\qquad I^a \to \dot{Y}^\mu . 
\end{align}
Comparing the linear graviton field generated by a point mass in equations \eqref{LinearizedGravity_Minkowski_EinsteinEquation_deDonderLW_Evend} and \eqref{LinearizedGravity_Minkowski_EinsteinEquation_deDonderLW_Oddd} to the linear non-Abelian gauge field engendered by a point charge in equations \eqref{YMW_Minkowski_LorenzLW_Evend} and \eqref{YMW_Minkowski_LorenzLW_Oddd}, we further observe that the latter can be obtained from the former by replacing
\begin{align}
\label{GRYM2_II}
-16 \pi \GN \to 1.
\end{align}
We highlight here, this sort of replacement rules -- to obtain Einstein gravity results from those of Yang-Mills (YM) theory -- finds a precise dictionary within the context of their $n$-particle scattering amplitudes. Bern, Carrasco, and Johansson showed in \cite{Bern:2008qj} how, once the scattering amplitude of $n$ on-shell non-Abelian gauge bosons are expressed in a particular form in Fourier space, the corresponding $n$ graviton amplitude can be gotten by simply replacing the YM gauge coupling with the appropriate gravitational constant and color (group theoretic) factors with their kinematic counterparts (composed of momentum and/or polarization tensor dot products).\footnote{That there is a ``squaring" relation between gravity and gauge boson scattering amplitudes was, as far as we are aware, first noticed by Kawai, Lewellen and Tye \cite{Kawai:1985xq} within the context of scattering closed versus open strings, whose massless spectrum contains respectively the spin$-2$ graviton and the spin$-1$ non-Abelian gauge boson. This is nowadays known as the ``KLT relations".} This specific set of replacement rules, nowadays known as the BCJ color-kinematics duality, is one reason for the slogan ``GR $=$ YM$^2$". An outstanding question is whether this ``double copy" relation between GR and YM extends to their classical solutions. 

To this end, recent work by Luna et al. \cite{Luna:2016due}, \cite{Luna:2015paa}, and Monteiro et al. \cite{Monteiro:2014cda} (see also Ridgway and Wise \cite{Ridgway:2015fdl}) have exploited the Kerr-Schild class of gravity solutions, which includes the physically important Schwarzschild/Kerr black hole geometries. It takes the form of ``flat metric plus perturbation," with the perturbation being proportional to two copies of the same null vector $k_\mu$:
\begin{align}
\label{KerrSchildMetric}
g_{\mu\nu} = \eta_{\mu\nu} + H k_\mu k_\nu .
\end{align}
(The $k^\mu$ is not only null with respect to both $\eta_{\mu\nu}$ and $g_{\mu\nu}$, it is also geodesic, i.e., $k^\mu \nabla_\mu k^\nu = k^\mu \partial_\mu k^\nu = 0$.) The technical significance of the Kerr-Schild class of metrics in eq. \eqref{KerrSchildMetric} is that it converts the usually nonlinear Einstein's equations to linear ones. This, in turn, is what allowed \cite{Luna:2016due}, \cite{Luna:2015paa}, and \cite{Monteiro:2014cda} to readily identify the corresponding linear YM solutions. Our replacement of the color isospin vector with the spacetime velocity eq. \eqref{GRYM2_I}, in the Yang-Mills-Wong's equation \eqref{WongEquation}, is (heuristically speaking) yet another example of this double-copy relationship between classical solutions of gravity and Yang-Mills. In fact, in the limit where our point charge is permanently at rest, it has to coincide with the Schwarzschild-Tangherlini (GR) $\leftrightarrow$ colored-Coulomb (YM) relationship identified in \cite{Monteiro:2014cda}.

Although \cite{Luna:2016due}, \cite{Luna:2015paa}, and \cite{Monteiro:2014cda} found exact relations between classical solutions of GR and YM, whereas the comparison we are making here holds only at the linearized level, we wish to suggest that -- in the spirit of ``GR $=$ YM$^2$" -- it is not necessary to only search for relations between exact classical solutions of the two theories. For one, the GR $\sim$ YM$^2$ correspondence is highly non-trivial because it involves the nonlinear self-interactions of the gravitons and gauge boson, while the classical solution relations revealed so far are strictly linear ones.\footnote{This is why, it is not clear that the relations in \cite{Luna:2016due}, \cite{Luna:2015paa}, and \cite{Monteiro:2014cda}, as well as the ones suggested in the current paper, are in fact directly connected to the BCJ color-kinematics duality \cite{Bern:2008qj} or even the KLT relations \cite{Kawai:1985xq}. We are aware that \S 4 of Luna et al. \cite{Luna:2016due} did attempt to draw such a direct link. However, even though we found the underlying physical motivation compelling, we disagree with the interpretation laid out there, that the source terms on the ``right hand side" of Einstein's and Yang-Mill's equations can be viewed, respectively, as gravitational and non-Abelian gauge boson radiation. In Einstein's gravitation, for instance, source terms must necessarily be associated with matter -- i.e., they are zero if and only if spacetime is (classically) empty.} Moreover, on the gravity side, the weak field perturbative metric generated by $n \geq 2$ point masses of the form in eq. \eqref{StressTensor_PointMass}, is in fact foundational to our understanding of astrophysical dynamics, such as that of our own Solar System or the binary neutron stars/small black holes whose GWs we hope to detect. This weak field expansion is a power series in Newton's constant $\GN$, and is known in the relativity literature as the post-Minkowskian (PN) program; while its non-relativistic version is the post-Newtonian (PN) one. We thus pose the question: \begin{quotation}
	Is there a perturbative post-colored-Coulomb program whose ``square" is that of the PN/PM one, that perhaps involves solving for the Yang-Mills gauge field engendered by $n \geq 2$ point charges governed by Wong's eq. \eqref{WongEquation} and its generalizations?\footnote{The correspondence between graviton and gauge-boson scattering amplitudes depend only on the dynamics of pure gravity and Yang-Mills theory; while, to connect their classical solutions, the relevant matter and color sources also need to be matched. For example, eq. \eqref{GRYM2_I} identifies charge with mass and isospin with proper velocity; but it is not clear how higher multipole moments, other non-trivial worldline operators, the necessary regularization and renormalization of their Wilson coefficients, etc., would be related.}
\end{quotation}To answer this question, it may be useful to first seek whether an efficient gauge for such calculations exist, such that when there is only one static point mass or charge in the system, the equations linearize automatically and the correspondence in \cite{Monteiro:2014cda} is recovered. The search for such a scheme on the gravity side, motivated primarily by the desire to improve PN/PM calculations, has been initiated recently by Harte \cite{Harte:2014ooa} and Harte and Vines \cite{Harte:2016vwo}.

{\bf Memory in even $d \geq 4$ Minkowski} \qquad We now turn to illustrate, in direct analogy to the gravitational case above, how linear Yang-Mills memory may arise from scattering of $N \geq 2$ point charges on unbound trajectories. We shall assume that some non-Abelian version of the Lorentz force law in eq. \eqref{LorentzForceLaw} holds, so that the formula for the electric memory effect in eq. \eqref{Maxwell_Memory_v1} remains applicable. Specifically, we find from the solutions in equations \eqref{YMW_Minkowski_LorenzLW_Evend} and \eqref{YMW_Minkowski_LorenzLW_Oddd},
\begin{align}
\label{LinearYMMemory}
\Delta A^{a\mu} = \,^{\text{(out)}}A^{a\mu} - \,^{\text{(in)}}A^{a\mu} ,
\end{align}
where the field generated by the color charges' configuration in the asymptotic past is
\begin{align}
\label{LinearYMMemory_In}
\,^{\text{(in)}}A^{a\mu} \approx
\frac{(d-5)!!}{(2\pi)^{\frac{d-4}{2}}} \sum_{\text{J}=1}^{N} \frac{q_{\text{J}}}{4\pi|\vec{x}|^{d-3}}
\left(1-\vec{V}_{(\text{J} \vert 0)}^2\right)^{ \frac{d-5}{2} } 
\frac{ I^a_{(\text{J} \vert 0)} V^\mu_{(\text{J} \vert 0)} }{ \left( 1 - \vec{V}_{(\text{J} \vert 0)} \cdot \widehat{n} \right)^{d-3} } ;
\end{align}
and that in the asymptotic future reads
\begin{align}
\label{LinearYMMemory_Out}
\,^{\text{(out)}}A^{a\mu} \approx
\frac{(d-5)!!}{(2\pi)^{\frac{d-4}{2}}} \sum_{\text{J}=1}^{N} \frac{q_{\text{J}}}{4\pi|\vec{x}|^{d-3}}
\left(1-\vec{V}_{(\text{J} \vert 1)}^2\right)^{ \frac{d-5}{2} } 
\frac{ I^a_{(\text{J} \vert 1)} V^\mu_{(\text{J} \vert 1)} }{ \left( 1 - \vec{V}_{(\text{J} \vert 1)} \cdot \widehat{n} \right)^{d-3} } .
\end{align}
Here, in addition to the asymptotic spacetime trajectories in equations \eqref{PointSource_In} and \eqref{PointSource_Out}, we also need to assume analogous ones for the isospin vectors
\begin{align}
\label{ColorSource_In}
\lim_{t \to -\infty} I^a_{\text{J}}[t] = I^a_{(\text{J} \vert 0)} \qquad\qquad \text{(constant)} , \\
\label{ColorSource_Out}
\lim_{t \to +\infty} I^a_{\text{J}}[t] = I^a_{(\text{J} \vert 1)} \qquad\qquad \text{(constant)} .
\end{align}
The gauge-invariant vector contribution $\Delta \alpha_i$ to eq. \eqref{Maxwell_Memory_v1} is simply the transverse portion of eq. \eqref{LinearYMMemory}. (Although we will not display them explicitly, analogous formulas for U$_1$ electromagnetic memory may be obtained from equations \eqref{LinearYMMemory}, \eqref{LinearYMMemory_In} and \eqref{LinearYMMemory_Out} by simply discarding the isospin vector $I^a$.)

{\it Double copy property} \qquad As already advertised, the gauge-invariant contributions to the memories of GR and YM, produced by $N$ point sources scattering off each other on unbound trajectories, enjoy a double copy property. Amongst the replacement rules in equations \eqref{GRYM2_I} and \eqref{GRYM2_II} that would translate between the linear GR one in equations \eqref{LinearizedGravity_Minkowski_Memory}, \eqref{LinearizedGravity_Minkowski_Memory_In}, \eqref{LinearizedGravity_Minkowski_Memory_Out} and the YM one in equations \eqref{LinearYMMemory}, \eqref{LinearYMMemory_In} and \eqref{LinearYMMemory_Out}, the replacement of the isospin vector $I^a$ with the spacetime velocity $\dot{Y}^\mu$ appears to be most closely related to the spirit of the BCJ color-kinematics duality \cite{Bern:2008qj}.

We close this subsection by wondering, does the gravitational ``spin memory" discovered in \cite{Pasterski:2015tva} have a Yang-Mills ``double copy" analog?

\subsection{Generalized Lorenz Gauge: Spatially Flat FLRW with Constant EoS $w$}

For the second part of this section -- to complement the gauge-invariant formalism of \S \eqref{Section_GaugeInvariantElectromagnetism} -- we turn to solving Maxwell's equations \eqref{Maxwell_0} and \eqref{Maxwell_i} in the generalized Lorenz gauge
\begin{align}
\label{Maxwell_FlatFLRW_Gauge}
\partial \cdot A + (d-4) \frac{\dot{a}}{a} A^0 = 0 .
\end{align}
The zeroth component is now
\begin{align}
\label{Maxwell_Lorenz_0}
\left( \partial^2 - \frac{d-4}{4} \left((d-2) \left(\frac{\dot{a}}{a}\right)^2 - 2 \frac{\ddot{a}}{a} \right) \right) \left(a^{\frac{d-4}{2}} A^0\right) 
= a^{\frac{d+4}{2}} \mathcal{J}^0 ;
\end{align}
while the spatial component becomes
\begin{align}
\label{Maxwell_Lorenz_i}
\left(\partial^2 - \frac{d-4}{4} \left(2 \frac{\ddot{a}}{a} + (d-6) \left(\frac{\dot{a}}{a}\right)^2 \right) \right) \left(a^{\frac{d-4}{2}} A^i\right) 
= a^{\frac{d+4}{4}} \mathcal{J}^i .
\end{align}
To make progress we shall now take the scale factor to describe a cosmology with a constant equation of state $w$, given by eq. \eqref{SpatiallFlatFLRW_Metric_w}. Maxwell's equations with the gauge in eq. \eqref{Maxwell_FlatFLRW_Gauge} now read
\begin{align}
\label{Maxwell_Lorenz_0_w}
\left( \partial^2 - (d-4) \frac{2d-7 + (d-1)w}{q_w^2} \frac{1}{\eta^2} \right)\left( a[\eta]^{\frac{d-4}{2}} A^0[\eta,\vec{x}]\right) 
&= a[\eta]^{\frac{d+4}{2}} \mathcal{J}^0[\eta,\vec{x}] , \\
\left( \partial^2 + (d-4) \frac{1+(d-1) w}{q_w^2} \frac{1}{\eta^2} \right) \left( a[\eta]^{\frac{d-4}{2}} A^i[\eta,\vec{x}]\right) 
\label{Maxwell_Lorenz_i_w}
&= a[\eta]^{\frac{d+4}{2}} \mathcal{J}^i[\eta,\vec{x}] .
\end{align}
{\bf Solutions} \qquad The retarded solutions to the gauge potential subject to the generalized Lorenz gauge of eq. \eqref{Maxwell_FlatFLRW_Gauge} are given by
\begin{align}
a[\eta]^{\frac{d-4}{2}} A^0[\eta,\vec{x}]
&= \int_{\eta_\circ}^{\eta} \dd \eta' \int_{\mathbb{R}^{d-1}} \dd^{d-1} \vec{x}' a[\eta']^{\frac{d+4}{2}}
\mathcal{G}^{(\gamma \vert \text{time})}[\eta,\eta';\sbar] \mathcal{J}^0[\eta',\vec{x}'] , \\
a[\eta]^{\frac{d-4}{2}} A^i[\eta,\vec{x}]
&= \int_{\eta_\circ}^{\eta} \dd \eta' \int_{\mathbb{R}^{d-1}} \dd^{d-1} \vec{x}' a[\eta']^{\frac{d+4}{2}}
\mathcal{G}^{(\gamma \vert \text{space})}[\eta,\eta';\sbar] \mathcal{J}^i[\eta',\vec{x}'] ;
\end{align}
where the symmetric Green's functions in all relevant even dimensions are
\begin{align}
\mathcal{G}^{(\gamma \vert \text{time})}_{\text{even $d \geq 4$}}[\eta,\eta';\sbar]
&= \frac{1}{(2\pi)^{\frac{d-2}{2}}} \left(\frac{\partial}{\partial \sbar}\right)^{\frac{d-2}{2}} \left( \frac{\Theta[\sbar]}{2} P_{-\frac{2d-7 + (d-1)w}{q_w}}[1+s] \right) , \\
\mathcal{G}^{(\gamma \vert \text{space})}_{\text{even $d \geq 4$}}[\eta,\eta';\sbar]
&= \frac{1}{(2\pi)^{\frac{d-2}{2}}} \left(\frac{\partial}{\partial \sbar}\right)^{\frac{d-2}{2}} \left( \frac{\Theta[\sbar]}{2} P_{-\frac{d-4}{q_w}}[1+s] \right) ;
\end{align}
and their counterparts in all relevant odd dimensions are
\begin{align}
\mathcal{G}^{(\gamma \vert \text{time})}_{\text{odd $d \geq 3$}}[\eta,\eta';\sbar]
&= \left( \frac{1}{2\pi} \frac{\partial}{\partial \sbar}\right)^{\frac{d-3}{2}} \left( 
\frac{\Theta[\sbar]}{4\pi} \frac{1 + \left(s+\sqrt{s (s+2)}+1\right)^{\frac{3 d - 11 + (d - 1) w}{q_w}}}{\sqrt{\sbar (s+2)} \left(s+\sqrt{s (s+2)}+1\right)^{\frac{3 d - 11 + (d - 1) w}{2 q_w}}} \right) , \\
\mathcal{G}^{(\gamma \vert \text{space})}_{\text{odd $d \geq 3$}}[\eta,\eta';\sbar]
&= \left( \frac{1}{2\pi} \frac{\partial}{\partial \sbar}\right)^{\frac{d-3}{2}} \left( 
\frac{\Theta[\sbar]}{4\pi} \frac{1 + \left(s+\sqrt{s (s+2)}+1\right)^{\frac{d - 5 - (d - 1) w}{q_w}}}{\sqrt{\sbar (s+2)} \left(s+\sqrt{s (s+2)}+1\right)^{\frac{d - 5 - (d - 1) w}{2 q_w}}} \right) .
\end{align}

\section{Proper Spatial Distance Between Two Geodesic Timelike Test Particles}
\label{Section_GeodesicDistance}
{\bf Setup} \qquad In this section, we re-visit the derivation of eq. (5) in \cite{Chu:2016qxp}, which describes the fractional distortion between two timelike worldlines in cosmological geometries. Specifically, we wish to derive an integral formula for the geodesic spatial distance, on a constant time $(d-1)$-surface, between two point test masses at rest (aka ``co-moving") in some cosmological spacetime. To this end, it is very useful to employ the the perturbed cosmological geometry in synchronous gauge in eq. \eqref{SpatiallFlatFLRW_PerturbedMetric_SynchronousGauge}. For, in this synchronous gauge coordinate system, co-moving observers do not change their spatial coordinates. Denoting their trajectories as $Y^\alpha[s]$ and $Z^\alpha[s]$, we shall define their constant spatial components as
\begin{align}
\label{GeodesicEquation_InitialConditions}
\vec{Y}[s] = \vec{Y}_0 \qquad \qquad \text{ and } \qquad \qquad \vec{Z}[s] = \vec{Z}_0 ;
\end{align}
whereas their time components satisfy the proper time condition $\dd s = \dd \eta \sqrt{g_{00} (\dd Y^0/\dd\eta)^2} = \dd \eta \sqrt{g_{00} (\dd Z^0/\dd\eta)^2}$, namely
\begin{align}
\frac{\dd Y^0[s]}{\dd s} = \frac{\dd Z^0[s]}{\dd s} = \frac{\dd \eta}{\dd s} = \frac{1}{a[\eta]} .
\end{align}
As already stated in \S \eqref{Section_PT_GWsAndMemory}, one may check readily that the relevant geodesic equations are satisfied to all orders in $\chi_{ij}$:
\begin{align}
\frac{\dd^2 Y^\alpha}{\dd s^2} + \Gamma^\alpha_{\phantom{\alpha}00} \left(\frac{\dd Y^0}{\dd s}\right)^2
= \frac{\dd^2 Z^\alpha}{\dd s^2} + \Gamma^\alpha_{\phantom{\alpha}00} \left(\frac{\dd Z^0}{\dd s}\right)^2
= 0 .
\end{align}
{\bf Spacelike geodesic distance} \qquad Now, the geodesic spatial distance $L[\eta]$ between the pair of test masses on a constant$-\eta$ hypersurface can be expressed through the positive square root of the integral
\begin{align}
\label{WorldFunction_AtConstantCosmicTime}
L[\eta]^2 \equiv a[\eta]^2 
\int_{0}^{1} \left( \delta_{ij} - \chi_{ij} \right) \frac{\dd W^i}{\dd \lambda} \frac{\dd W^j}{\dd \lambda} \dd \lambda .
\end{align}
The $W^i$, in turn, obeys the spatial geodesic equation
\begin{align}
\label{GeodesicEquation_SpacelikeTestMasses}
\frac{D^2 W^i}{\dd \lambda^2} &\equiv \frac{\dd^2 W^i}{\dd \lambda^2} + \gamma^i_{\phantom{i}ab} \frac{\dd W^a}{\dd \lambda} \frac{\dd W^b}{\dd \lambda} = 0 , \\
\gamma^i_{\phantom{i}ab} & \approx -\frac{1}{2} \left( \partial_{a} \chi_{b i} + \partial_{b} \chi_{a i} - \partial_{i} \chi_{ab} \right) ,
\end{align}
subject to the boundary conditions
\begin{align}
\label{GeodesicEquation_SpacelikeTestMasses_BC}
W^i[\lambda=0] = Y^i[\eta] = Y^i_0
\qquad\qquad \text{ and } \qquad\qquad
W^i[\lambda=1] = Z^i[\eta] = Z^i_0 .
\end{align}
Because eq. \eqref{WorldFunction_AtConstantCosmicTime} is the action whose variation with respect to $W^i$ yields the geodesic equation of eq. \eqref{GeodesicEquation_SpacelikeTestMasses}, to first order in $\chi_{ij}$, we may simply replace $W^i$ with its solution if the spatial metric were conformally flat (i.e., if $\chi_{ij}=0$). In other words, if in eq. \eqref{WorldFunction_AtConstantCosmicTime} we replaced $W^i$ with a straight line joining $\vec{Y}[\eta]=\vec{Y}_0$ to $\vec{Z}[\eta]=\vec{Z}_0$,
\begin{align}
\label{GeodesicEquation_SpacelikeTestMasses_ZerothOrder}
W^i \to \overline{W}^i[\lambda] \equiv Y_0^i + \lambda (Z_0-Y_0)^i ,
\end{align}
the error incurred would scale higher than $\mathcal{O}[\chi_{ij}]$. To see this more explicitly, we vary $W^i \to W^i + \delta W^i$ in eq. \eqref{WorldFunction_AtConstantCosmicTime}, taking into account $\delta W^i[\lambda = 0] = \delta W^i[\lambda = 1] = 0$ (from eq. \eqref{GeodesicEquation_SpacelikeTestMasses_BC}):
\begin{align}
\delta_{\vec{W}} L[\eta]^2 = -2 a[\eta]^2 
\int_{0}^{1} \delta W^i \left( \delta_{ij} - \chi_{ij} \right) \frac{D^2 W^j}{\dd \lambda^2} \dd \lambda 
+ \mathcal{O}[(\delta\vec{W})^2] .
\end{align}
When the spatial geodesic equation in eq. \eqref{GeodesicEquation_SpacelikeTestMasses} is satisfied by $W^i$, the integrand vanishes at first order in $\chi_{ij}$. 

At this point, making the replacement in eq. \eqref{GeodesicEquation_SpacelikeTestMasses_ZerothOrder} and noting its $\lambda$-derivative returns $\dot{\overline{W}}^i[\eta] = (Z_0-Y_0)^i$, reduces eq. \eqref{WorldFunction_AtConstantCosmicTime} to
\begin{align}
\label{WorldFunction_AtConstantCosmicTime_v2}
L[\eta]^2 \approx a[\eta]^2 \left( (\vec{Z}_0-\vec{Y}_0)^2
- (Z_0-Y_0)^i (Z_0-Y_0)^j \int_{0}^{1} \chi_{ij}\left[ \eta, \overline{W}^l[\lambda] \right] \dd \lambda 
\right) .
\end{align}
(The variational argument here has previously been used in Pfenning and Poisson \cite{Pfenning:2000zf} and Chu and Starkman \cite{Chu:2011ip} to derive an expression, similar in spirit to eq. \eqref{WorldFunction_AtConstantCosmicTime_v2}, for the geodesic distance between any two points in a weakly curved spacetime.) 

{\bf Results} \qquad The $\mathcal{O}[\chi_{ij}]$-accurate spatial geodesic distance between the pair of test masses at $\vec{Y}_0$ and $\vec{Z}_0$, at constant $\eta$, is now provided by the integral
\begin{align}
\label{WorldFunction_AtConstantCosmicTime_FirstOrderResult}
L[\eta] \approx a[\eta] \left\vert \vec{Z}_0-\vec{Y_0} \right\vert \left( 
1 - \frac{\widehat{n}^i \widehat{n}^j}{2} \int_{0}^{1} \chi_{ij}\left[ \eta, \vec{Y}_0 + \lambda \left(\vec{Z}_0-\vec{Y}_0\right) \right] \dd \lambda 
\right), \qquad\qquad
\widehat{n} \equiv \frac{\vec{Z}_0-\vec{Y}_0}{|\vec{Z}_0-\vec{Y}_0|} .
\end{align}
Suppose we began our hypothetical experiment at some time $\eta'$. At some later time $\eta > \eta'$, the fractional distortion induced on the pair of test masses is
\begin{align}
\left(\frac{\delta L}{L_0}\right)[\eta > \eta'] \equiv \frac{L[\eta]}{L[\eta']} - 1 ,
\end{align}
which according to eq. \eqref{WorldFunction_AtConstantCosmicTime_FirstOrderResult} translates to
\begin{align}
\label{FractionalDistortion_AtConstantCosmicTime_FirstOrderResult}
\left(\frac{\delta L}{L_0}\right)[\eta > \eta'] 
= \left( \frac{a[\eta]}{a[\eta']} -  1 \right)
- \frac{a[\eta]}{a[\eta']} \frac{\widehat{n}^i \widehat{n}^j}{2} \int_{0}^{1} \Delta \chi_{ij} \dd \lambda + \mathcal{O}[(\chi_{\mu\nu})^2] ,
\end{align}
where the $\Delta \chi_{ij}$ refers to the total change in the spatial metric perturbations over the duration $[\eta',\eta]$ (cf. eq. \eqref{FractionalDistortion_AtConstantCosmicTime_Deltachiij}).

To summarize: eq. \eqref{FractionalDistortion_AtConstantCosmicTime_FirstOrderResult} is valid up at first order in metric perturbations in any $d$-dimensional background spatially flat FLRW cosmology. However, over the timescales of human experiments (i.e., $\mathcal{O}[\text{decades}]$), the size of the universe will not change appreciably. We may thus Taylor expand the scale factor in $a[\eta]/a[\eta'] - 1 \approx (\dot{a}[\eta']/a[\eta']) (\eta-\eta') \ll 1$ and arrive at eq. \eqref{FractionalDistortion_AtConstantCosmicTime_FirstOrderHuman}.

{\bf Remark I} \qquad Previously, in \cite{Chu:2016qxp}, because we did not explicitly invoke the synchronous gauge form of the metric in eq. \eqref{SpatiallFlatFLRW_PerturbedMetric_SynchronousGauge}, what we derived there was really a comparison of the proper spatial distances between the pair of test masses with and without the metric perturbations at a fixed time. (We also, implicitly, set $\chi_{ij}[\eta',\vec{x}]=0$.) It is the desired fractional distortion result in eq. \eqref{FractionalDistortion_AtConstantCosmicTime_FirstOrderHuman} if the spatial coordinates of the test masses do not change with time, as is the case in the synchronous gauge employed here. 

{\bf Remark II} \qquad We also comment on the misleading ``transverse-traceless gauge" (TT-gauge) often invoked to derive GW observables such as the proper distance formula in eq. \eqref{WorldFunction_AtConstantCosmicTime_FirstOrderResult}. The TT-gauge amounts to setting to zero all components of $\chi_{\mu\nu}$ in eq. \eqref{SpatiallFlatFLRW_PerturbedMetric}, except the spatial ones $\chi_{ij}^{\text{TT}}$ obeying $\partial_i \chi_{ij}^{\text{TT}} = \delta^{ij} \chi_{ij}^{\text{TT}} = 0$. This is a legitimate choice of coordinates if spacetime were completely empty; but as we have seen in both the perturbed cosmological and Minkowski cases, the ``TT-gauge" really does not exist once there is matter -- including the very astrophysical source of GWs responsible, in the first place, for the fluctuations in proper distances measured by our laser interferometers. For, the presence of a GW source would yield non-zero scalar $(\Phi,\Psi)$ and vector $(V_i)$ metric perturbations; and since they are gauge-invariant, it is not possible to coordinate transform them to zero.\footnote{In appendix A of \cite{Flanagan:2005yc}, a supposed proof of the existence of the TT-gauge was provided for ``local vacuum regions". We believe the proof there essentially assumed its conclusions and is hence invalid. (See its equations (A.1)--(A.4); as well as eq. (A.13) and the preceding paragraph.) Specifically, equations (A.1)--(A.4) of \cite{Flanagan:2005yc} amounted to the imposition of the vanishing of the scalar ($\overline{\Phi}$, $\overline{\Psi}$) and vector ($\overline{V}_i$) gauge invariant quantities on the spatial boundary of some local vacuum region of spacetime, which in turn imply they have to vanish throughout its interior. Physically speaking, the boundary values of these scalar, vector and tensor perturbations really ought to be determined by the matter/GW sources external to the region considered, and one should not be free to choose them arbitrarily.}

Why, then, is one able to get away with adopting this inconsistent ``TT-gauge" for making GW predictions? It is likely due to the dynamics of General Relativity itself. To understand this, and as a complementary approach to the Synge's world-function based one used to derive eq. \eqref{FractionalDistortion_AtConstantCosmicTime_FirstOrderHuman}, we shall examine the geodesic deviation equation that encodes the tidal forces exerted upon a GW detector. If $U^\sigma$ is tangent to the timelike geodesic swept out by one component of LISA, for instance; and $\epsilon^\mu$ is the vector joining it to a nearby component; from eq. \eqref{GeodesicDeviation_Riemann},
\begin{align}
\label{eLISA_GeodesicDeviation}
U^\sigma U^\rho \nabla_\sigma \nabla_\rho \epsilon^\mu = -R^\mu_{\phantom{\mu}\nu \alpha\beta} U^\nu \epsilon^\alpha U^\beta .
\end{align}
If the Riemann tensor were zero, $\epsilon^\mu$ is parallel transported along $U^\alpha$ and there would be no tidal forces. 

It is convenient to work with the synchronous gauge metric in eq. \eqref{SpatiallFlatFLRW_PerturbedMetric_SynchronousGauge}, because it allows us to make the exact statement $U^\sigma = \delta^\sigma_0/a$. By a direct calculation, one may then verify that this choice of gauge eliminates the $\mu = 0$ component of eq. \eqref{eLISA_GeodesicDeviation} because both $U^\sigma U^\rho \nabla_\sigma \nabla_\rho \epsilon^0$ and $R^0_{\phantom{0}\nu \alpha\beta} U^\nu \epsilon^\alpha U^\beta$ are individually zero. 

{\it Minkowski background} \qquad Let us, for a start, assume the detector is operating nearly at rest in a background Minkowski spacetime. Since the background Riemann tensor vanishes, its first order counterpart is gauge-invariant. In fact, the spatial components of the right hand side of eq. \eqref{eLISA_GeodesicDeviation} can be re-cast in terms of the gauge-invariant variables,
\begin{align}
\label{eLISA_GeodesicDeviation_v2}
U^\sigma U^\rho \nabla_\sigma \nabla_\rho \epsilon^i = -R^i_{\phantom{i}\nu \alpha\beta} U^\nu \epsilon^\alpha U^\beta
\approx (R \vert 1)_{i0 j0} \epsilon^j ,
\end{align}
where
\begin{align}
(R \vert 1)_{i0 j0}
= \frac{1}{2} \left( - \frac{1}{2} \partial_i \partial_j \overline{\Psi} - \delta_{ij} \ddot{\overline{\Phi}}
+ \partial_{\{i} \dot{\overline{V}}_{j\}}
- \ddot{\overline{D}}_{ij} \right) .
\end{align}
Without the solutions to linearized General Relativity at hand, one may be tempted to surmise that all scalar, vector and tensor perturbations produced by a distant astrophysical system contribute to the tidal squeezing and stretching of a GW detector. But if we do in fact assume General Relativity holds, so that equations \eqref{LinearizedGravity_MatterConservation}, \eqref{LinearizedGravity_SVTDecomposition_Einstein_Scalars}, and \eqref{LinearizedGravity_Minkowski_EinsteinSolution_Scalar} through \eqref{LinearizedGravity_Minkowski_EinsteinSolution_Tensor} may be exploited,
\begin{align}
(R \vert 1)_{i0 j0}
&= 8\pi \GN \left( \left(\partial_i \partial_j - \frac{\delta_{ij}}{d-1}\Laplacian\right) \Upsilon + \partial_{\{i} \sigma_{j\}} - \delta_{ij} \frac{\sigma}{(d-1)(d-2)} \right) \nonumber\\
&\qquad\qquad- \frac{1}{2} \left( \ddot{\overline{D}}_{ij} + (d-3) \partial_i \partial_j \overline{\Phi} \right) .
\end{align}
By referring to eq. \eqref{LinearizedGravity_SVTDecomposition_Matter} we recognize the first line to contain various irreducible spatial components of the matter stress-energy tensor. If we assume the GW source is isolated and well separated from the observer, this first line vanishes identically at the GW detector's location. As for the second line, by recalling the ``{\it Coulomb vs. Waves}" discussion in \S \eqref{Section_LinearizedEinstein_GaugeInvariant}, we may estimate the scalar-to-tensor amplitude to scale as ((characteristic timescale of GW source)/(observer-source spatial distance))$^2$. Hence, in the far zone where the GW detector is located,
\begin{align}
U^\sigma U^\rho \nabla_\sigma \nabla_\rho \epsilon^i 
		\approx - \frac{1}{2} \ddot{\overline{D}}_{ij} \epsilon^j \qquad\qquad \text{(far zone)} .
\end{align}
To reiterate: it is only after the dynamics of General Relativity has been incorporated, eq. \eqref{eLISA_GeodesicDeviation} reduces to the statement that the acceleration of the deviation vector in spatial directions is driven primarily by the tensor mode filling the space around a distant GW detector.

{\it Cosmological background} \qquad We move on to discuss the case where GWs are journeying across a cosmological background. We shall invoke the form of the Riemann tensor in eq. \eqref{Riemann_WeylCCMatter}, and for technical simplicity focus only on the Weyl contribution to tidal forces. Now, the Weyl tensor $C^\alpha_{\phantom{\alpha}\beta \mu\nu}$, with one upper index, is conformally invariant. That means, when linearized about a perturbed spatially flat FLRW universe, it must be equal to its flat spacetime counterpart. Moreover, since it vanishes about the background homogeneous and isotropic universe, the linearized Weyl tensor $(C \vert 1)^\alpha_{\phantom{\alpha}\beta \mu\nu}$ must itself be gauge-invariant. A direct calculation then shows the components of Weyl relevant for the $i$th component of the geodesic deviation eq. \eqref{eLISA_GeodesicDeviation} are
\begin{align}
\label{LinearizedWeyl_GaugeInvariant}
(C \vert 1)^i_{\phantom{i}0 j0}
= \frac{1}{2} \Bigg\{ 
&-\frac{a^2}{d-2} \overline{\Box}^{\text{(S)}} D_{ij}
+ \frac{\partial_0 \left( a \cdot \dot{D}_{ij} \right)}{a} \nonumber\\
&\qquad\qquad
- \frac{d-3}{d-2} \left( \partial_{\{i} \dot{V}_{j\}} 
- \left\{ \partial_i \partial_j - \frac{\delta_{ij}}{d-1} \vec{\nabla}^2 \right\} \left( \Phi + \Psi \right)
\right)
\Bigg\} ,
\end{align}
where the first term (from the left) involves the scalar spacetime Laplacian, $\overline{\Box}^{\text{(S)}} D_{ij} \equiv \partial_\mu \left( a^{d-2} \eta^{\mu\nu} \partial_\nu D_{ij} \right)/a^d$. Since $(C \vert 1)^i_{\phantom{i}0 j0}$ actually equals its linearized counterpart about Minkowski, all scale factors really cancel out. But putting the Weyl tensor in this form allows the EoM for $D_{ij}$, $V_i$, $\Phi$ and $\Psi$ -- specifically, equations \eqref{FirstOrderPTEqns_Scalar4of4}, \eqref{FirstOrderPTEqns_Tensor}, and \eqref{FirstOrderPTEqns_Vector_SolnIIofII} -- to be employed. If we further make the approximation that the GW experiment does not run over cosmological timescales, so that all the ensuing $\dot{a}/a$ terms may be neglected, eq. \eqref{LinearizedWeyl_GaugeInvariant} could then be re-expressed as
\begin{align}
\label{LinearizedWeyl_GaugeInvariant_WithMatter}
(C \vert 1)^i_{\phantom{i}0 j0}
&\approx 8\pi\GN \left\{ 
\frac{\sigma_{ij}}{d-2} 
- \frac{d-3}{d-2} \left( \partial_{\{i} \sigma_{j\}} 
+ \left\{ \partial_i \partial_j - \frac{\delta_{ij}}{d-1} \vec{\nabla}^2 \right\} \Upsilon
\right)
\right\} \nonumber\\
&\qquad\qquad
+ \frac{1}{2} \left( \ddot{D}_{ij} + (d-3) \left\{ \partial_i \partial_j - \frac{\delta_{ij}}{d-1} \vec{\nabla}^2 \right\} \Phi \right)
\end{align}
Much like the Minkowski case above, the first line vanishes identically at the GW-detector's location because it contains various $\,^{(\text{a})}T_{\mu\nu}$ components of the isolated astrophysical system. The subtlety in the cosmological case at hand arises on the second line: even though $D_{ij}$ admits wave solutions and therefore always plays a key role in exerting tidal forces on the GW detector; $\Phi$ does so only when $0 < w \leq 1$. For, it is within this range of EoS that $\Phi$ obeys a wave equation (namely, eq. \eqref{FirstOrderPTEqns_Scalar1of4_v2}) and the $\partial_i$s acting on it in eq. \eqref{LinearizedWeyl_GaugeInvariant_WithMatter} can be treated on roughly the same footing as the time derivative $\partial_0 \sim 1/$(characteristic timescale of GW source).

We close this section by suggesting that the ``Conformal Fermi Normal Coordinates" recently developed in Dai, Pajer and Schmidt \cite{Dai:2015rda} (which built upon earlier work by Pajer, Schmidt and Zaldarriaga \cite{Pajer:2013ana}) for cosmological Large Scale Structure applications, could perhaps also be used to integrate the geodesic deviation equation to study GW observables and potential memory effects in our universe.

\section{Solution To A Space-Translation-Invariant PDE}
\label{Section_PDESolution}
In this section, for the reader's reference, we will provide a summary of appendix (B) of \cite{Chu:2016qxp}, which delineates how to solve the primary partial differential equation (PDE) encountered there and in this paper:
\begin{align}
\label{MasterPDE}
\left( \partial^2 - \frac{\kappa(\kappa+1)}{\eta^2} \right) \psi[\eta,\vec{y}] = \mathcal{S}[\eta,\vec{y}] , 
\qquad\qquad
\partial^2 \equiv \eta^{\mu\nu} \partial_\mu \partial_\nu ,
\end{align}
where the source $\mathcal{S}$ is assumed to be some externally prescribed function of spacetime. In the main text, and depending on the context, the coordinates are either $y^\alpha \equiv (\eta,\vec{x})$ or $y^\alpha \equiv (\eta,\vec{x}/\sqrt{w})$ (for EoS $0 < w \leq 1$); here, we will just use $y^\alpha = (\eta,\vec{y})$. The $\kappa(\kappa+1)$ in the $1/\eta^2$ term is a convenient choice; for our purposes it is to be treated as an arbitrary real constant. The key object we wish to compute is the symmetric Green's function $\mathcal{G}$ obeying
\begin{align}
\label{MasterPDE_G}
\left( \partial_{\eta,\vec{y}}^2 - \frac{\kappa(\kappa+1)}{\eta^2} \right) \mathcal{G}
= \left( \partial_{\eta',\vec{y}'}^2 - \frac{\kappa(\kappa+1)}{\eta'^2} \right) \mathcal{G} 
= 2 \delta[\eta-\eta'] \delta^{(d-1)}[\vec{y}-\vec{y}'] .
\end{align}
Once $\mathcal{G}$ is known, the solution to eq. \eqref{MasterPDE} is
\begin{align}
\label{MasterPDE_Soln}
\psi[\eta,\vec{y}]
= \int_{\mathbb{R}^{d-1}} \dd^{d-1}\vec{y}' \int_{\eta_\circ}^{\eta} \dd\eta' 
		\mathcal{G}\left[ \eta,\vec{y}; \eta',\vec{y}' \right] \mathcal{S}[\eta',\vec{y}'] .
\end{align}
{\bf Dimension reduction \& Nariai's ansatz} \qquad The Green's function equation in eq. \eqref{MasterPDE_G} may be tackled via dimension reduction because of the space-translation symmetry and parity invariance of the wave operator in eq. \eqref{MasterPDE}.\footnote{This dimension reduction reasoning can be found in the discussion enveloping equations (B5)--(B12) of \cite{Chu:2016qxp}; for an operator based approach, but specialized to flat spacetime, see section \eqref{Section_DimRaisingOperator} below.} Once the 2D $\mathcal{G}_2$ and 3D $\mathcal{G}_3$ are known, the higher dimensional Green's functions can be obtained by repeated differentiation with respect to Synge's world function in Minkowski spacetime $\sbar \equiv (1/2)((\eta-\eta')^2 - (\vec{y}-\vec{y}')^2)$:
\begin{align}
\label{G_EvenD}
\mathcal{G}_{\text{even $d \geq 2$}}[\eta,\eta';\sbar] 
&= \left(\frac{1}{2\pi} \frac{\partial}{\partial\sbar}\right)^{ \frac{d-2}{2} } \mathcal{G}_2[\eta,\eta';\sbar] , \\
\label{G_OddD}
\mathcal{G}_{\text{odd $d \geq 3$}}[\eta,\eta';\sbar] 
&= \left(\frac{1}{2\pi} \frac{\partial}{\partial\sbar}\right)^{ \frac{d-3}{2} } \mathcal{G}_3[\eta,\eta';\sbar] .
\end{align}
Both the 2D and 3D cases can be gotten by exploiting Nariai's ansatz \cite{Nariai} -- in 2D, we postulate that the tail function depends on spacetime solely through a single object $s$:
\begin{align}
\label{G2}
\mathcal{G}_2[\eta,\eta';\sbar] = \frac{\Theta[\sbar]}{2} g_2[s], \qquad\qquad s \equiv \frac{\sbar}{\eta\eta'} .
\end{align}
This $g_2[s]$ obeys the homogeneous version of eq. \eqref{MasterPDE}, and Nariai's ansatz turns the PDE into the ordinary differential equation
\begin{align}
\left( \partial_{\eta,\vec{y}}^2 - \frac{\kappa(\kappa+1)}{\eta^2} \right) g_2[s]
= \frac{s (2 + s) g_2''[s] + 2 (1 + s) g_2'[s] - \kappa (1 + \kappa) g_2[s]}{\eta^2} = 0 .
\end{align}
Following that, one needs to impose the null cone boundary condition $g_2[\sbar=0]=1$. This identifies the unique solution to be the Legendre function
\begin{align}
\label{G2_Tail}
g_2[s] = P_\kappa[1+s] .
\end{align}
In 3D, we postulate that
\begin{align}
\label{G3}
\mathcal{G}_3[\eta,\eta';\sbar] = \frac{\Theta[\sbar]}{2\pi} \frac{\widehat{g}_3[s]}{\sqrt{\eta\eta'}} .
\end{align}
This $\widehat{g}_3[s]/\sqrt{\eta\eta'}$ obeys the homogeneous version of eq. \eqref{MasterPDE}, and Nariai's ansatz turns the PDE into the ordinary differential equation
\begin{align}
\left( \partial_{\eta,\vec{y}}^2 - \frac{\kappa(\kappa+1)}{\eta^2} \right) \frac{g_3[s]}{\sqrt{\eta\eta'}}
= \frac{4 s (2 + s) \widehat{g}_3''[s] + 12 (1 + s) \widehat{g}_3'[s] +  (3 - 4 \kappa (1 + \kappa)) \widehat{g}_3[s]}{4 \eta^2 \sqrt{\eta\eta'}}
= 0 .
\end{align}
Following that, one needs to impose the null cone boundary condition $\widehat{g}_3[\sbar=0]/\sqrt{\eta\eta'}=1/\sqrt{2\sbar}$. The unique solution is
\begin{align}
\label{G3_Tail}
\frac{\widehat{g}_3[s]}{\sqrt{\eta\eta'}}
= \frac{\left( s + \sqrt{s(s+2)} + 1 \right)^{1+2\kappa} + 1}{2 \sqrt{\sbar(s+2)} \left( s + \sqrt{s(s+2)} + 1 \right)^{\frac{1}{2}(1+2\kappa)}} .
\end{align}
{\bf Results} \qquad Inserting the results in equations \eqref{G2}, \eqref{G2_Tail}, \eqref{G3}, \eqref{G3_Tail} into equations \eqref{G_EvenD} and \eqref{G_OddD}, the symmetric Green's function solutions to equation \eqref{MasterPDE_G} are thus
\begin{align}
\label{GResults_EvenD}
\mathcal{G}_{\text{even $d \geq 2$}}[\eta,\eta';\sbar]
&= \left( \frac{1}{2\pi} \frac{\partial}{\partial\sbar}\right)^{ \frac{d-2}{2} } \left(\frac{\Theta[\sbar]}{2} P_\kappa[1+s]\right) ,  \\
\label{GResults_OddD}
\mathcal{G}_{\text{odd $d \geq 3$}}[\eta,\eta';\sbar]
&= \left( \frac{1}{2\pi} \frac{\partial}{\partial\sbar}\right)^{ \frac{d-3}{2} } 
\left( \frac{\Theta[\sbar]}{4\pi} \frac{\left( s + \sqrt{s(s+2)} + 1 \right)^{1+2\kappa} + 1}{\sqrt{\sbar(s+2)} \left( s + \sqrt{s(s+2)} + 1 \right)^{\frac{1}{2}(1+2\kappa)}} \right) , \\
\sbar &\equiv \frac{(\eta-\eta')^2-(\vec{y}-\vec{y}')^2}{2}, \qquad\qquad s \equiv \frac{\sbar}{\eta\eta'} .
\end{align}
Note that the retarded Green's function $\mathcal{G}^+$ propagates signals strictly in the future of the source, so to obtain it from its symmetric cousin, we simply multiply it by a $\Theta$-function of elapsed time:
\begin{align}
\mathcal{G}^+\left[ \eta,\eta';|\vec{y}-\vec{y}'| \right]
= \Theta[\eta-\eta'] \mathcal{G}\left[ \eta,\eta'; \sbar \right] .
\end{align}
{\bf Caution} \qquad In this paper, whenever $2/q_w > 0$ in eq. \eqref{SpatiallFlatFLRW_Metric_w}, the wave solutions in eq. \eqref{MasterPDE_Soln} are globally incomplete because of the presence of a particle horizon -- they vanish outside of the light (or, acoustic) cone of the GW source at the Big Bang $\eta = \eta_\circ = 0$. The symbol $\Theta_p$ appearing throughout the main text reminds us of this fact. Specifically, whenever $q_w < 0$, $\Theta_p \equiv 1$. Whereas, when $q_w > 0$, we shall define $\Theta_p = 1$ whenever the past light (acoustic) cone of the observer at $(\eta,\vec{x})$ has a non-trivial intersection with the spacetime world tube of the GW source; and $\Theta_p = 0$ otherwise.

This global incompleteness of the wave solutions may lead to the violation of Gauss' law in both linearized Einstein's equations and Maxwell's electromagnetism, since the relevant flux integrals would vanish outside the particle horizon at a given time $\eta$. We hope to return to this issue in future work; for now, see Higuchi and Lee \cite{Higuchi:2008fu} for a discussion. 

{\bf Remark} \qquad It has not escaped our attention that the $s \equiv \sbar/(\eta\eta')$ occurring throughout this paper can be viewed as a proper Lorentz SO$_{d,1}$ invariant object in a $(d+1)$-dimensional Minkowski spacetime. (SO$_{d,1}$ is also isomorphic to the conformal group in $d-1$ spatial dimensions; see, for instance, Di Francesco et al. \cite{DiFrancesco:1997nk} for a discussion.) Even though we are not working exclusively in de Sitter spacetime ($w=-1$), we may pretend for this purpose that our observer at $y$ and source at $y'$ lie on a fictitious hyperboloid of size $1/H$, for some $H>0$, embedded in a $(d+1)$-dimensional Minkowski spacetime. In this higher dimensional flat spacetime, let us define the $d$ Cartesian coordinates $\{X^{\mathfrak{A}}[\eta,\vec{y}]\}$ to be
\begin{align*}
X^0 = \frac{1}{2\eta} \left( \eta^2 - \vec{y}^2 - \frac{1}{H^2} \right), \qquad
X^d = \frac{1}{2\eta} \left( -\eta^2 + \vec{y}^2 - \frac{1}{H^2} \right), \\
X^i = \frac{y^i}{H \eta}, \qquad i = 1,2,\dots,d-1, 
\end{align*}
and make a similar definition for $\{X'^{\mathfrak{A}}[\eta',\vec{y}']\}$. (Unlike the de Sitter case, where $\eta$ is non-positive, the sign of $\eta$ here depends on the sign of $q_w$.) If $\eta_{\mathfrak{A}\mathfrak{B}} = \text{diag}[1,-1,\dots,-1]$ is the metric in the ambient Minkowski spacetime, we not only obtain the equation of the hyperboloid
\begin{align}
\eta_{\mathfrak{A}\mathfrak{B}} X^\mathfrak{A} X^{\mathfrak{B}} 
= \eta_{\mathfrak{A}\mathfrak{B}} X'^{\mathfrak{A}} X'^{\mathfrak{B}}
= -\frac{1}{H^2} ;
\end{align}
we also have the proper Lorentz invariant object
\begin{align}
-1-H^2 \eta_{\mathfrak{A}\mathfrak{B}} X^\mathfrak{A} X'^{\mathfrak{B}} = \frac{\sbar}{\eta\eta'} \equiv s .
\end{align}

\section{$(-2\pi r)^{-1} \partial_r$ As ``Dimension-Raising Operator"}
\label{Section_DimRaisingOperator}
In a $d$-dimensional geometry that enjoys spatial translation symmetry and parity invariance, the Green's function $G_d$ of its massive wave operator $\partial^2 + m^2$ may be obtained from its counterpart $2$ dimensions lower, i.e., from $G_{d-2}$ through the formula
\begin{align}
\label{2DPlaneSource}
G_d\left[ \eta,\eta';R \equiv |\vec{x}-\vec{x}'|_{d-1} \right]
= -\frac{1}{2\pi R} \frac{\partial}{\partial R} G_{d-2} \left[ \eta,\eta';R \equiv |\vec{x}-\vec{x}'|_{d-3} \right] ,
\end{align}
where on the left hand side $|\vec{x}-\vec{x}'|_{d-1}$ is the Euclidean distance in $(d-1)$-space and on the right hand side $|\vec{x}-\vec{x}'|_{d-3}$ is that in $(d-3)$-space. This can be understood -- see, for instance, \cite{SoodakTiersten} and \cite{Chu:2015yua} -- by viewing the $(d-2)$-dimensional Green's function $G_{d-2}$ as being sourced by an appropriately defined ``plane source" extending into two extra spatial dimensions. In this section, we will specialize to Minkowski spacetime 
\begin{align}
\label{Minkowski_SphericalCoordinates}
\dd s^2 = \dd t^2 - \dd r^2 - r^2 \dd \Omega_{d-2}^2 
\end{align}
-- where $\dd \Omega_{d-2}^2$ is the metric on the $(d-2)$-sphere -- and show how the recursion relation eq. \eqref{2DPlaneSource} can also be understood from an algebraic perspective. 

To begin, let us note that, by the Poinc\'{a}re invariance of Minkowski spacetime, we expect the Green's function to depend on spacetime through the time elapsed $t-t'$ and the Euclidean distance $|\vec{x}-\vec{x}'|$, where in Cartesian coordinates $x \equiv (t,\vec{x})$ is the location of the observer and $x' \equiv (t',\vec{x}')$ that of the spacetime point source. Again by Poinc\'{a}re invariance, we may place the source at $x' = (0,\vec{0})$ without loss of generality. This means the Green's functions now depend on spacetime coordinates as $G[t,r \equiv |\vec{x}|]$. Let us define, in the coordinate system of eq. \eqref{Minkowski_SphericalCoordinates},
\begin{align}
\label{RaisingOperator}
\mathcal{D}_r \equiv -\frac{1}{2\pi r} \frac{\partial}{\partial r} .
\end{align}
If we apply the $d$-dimensional wave operator $\partial^2_d$ of the Minkowski metric in eq. \eqref{Minkowski_SphericalCoordinates} to any function $F[t,r]$ that depends only on $t$ and $r$, we would find
\begin{align}
\partial^2_d F[t,r] = \partial_t^2 F - \frac{1}{r^{d-2}} \partial_r \left( r^{d-2} \partial_r F \right) .
\end{align}
The key algebraic observation here is that, applying the $d$-dimensional wave operator followed by the $\mathcal{D}_r$ defined in eq. \eqref{RaisingOperator} is the same as applying $\mathcal{D}_r$ first followed by the $(d+2)$-dimensional wave operator:
\begin{align}
\label{RaisingOperator_IofII}
\mathcal{D}_r \left(\partial^2_d + m^2\right) F[t,r] = \left(\partial^2_{d+2} + m^2\right) \mathcal{D}_r F[t,r] .
\end{align}
Furthermore, if we use the identity $(r \delta[r])' = 0$ to replace $\delta'[r] \to -\delta[r]/r$, we find that $\mathcal{D}_r$ converts the $d$-dimensional $\delta$-function measure associated with a spacetime point source at the origin into one in $(d+2)$ dimensions, namely
\begin{align}
\label{RaisingOperator_IIofII}
\mathcal{D}_r \left(\frac{\delta[t] \delta[r]}{\Omega_{d-2} r^{d-2}}\right) = \frac{\delta[t] \delta[r]}{\Omega_{d} r^{d}} ,
\end{align}
where $\Omega_{d-2} = 2 \pi^{(d-1)/2}/\Gamma[(d-1)/2]$ is the solid angle subtended by a unit radius sphere embedded in a $(d-1)$-dimensional Euclidean space. At this point, suppose we had solved the $(d-2)$-dimensional Green's function, which obeys
\begin{align}
\label{RaisingOperator_dminus2}
\left(\partial_{d-2}^2 + m^2\right)G_{d-2}[t,r] = \frac{\delta[t]\delta[r]}{\Omega_{d-4} r^{d-4}} ,
\end{align}
we may proceed to apply $\mathcal{D}_r$ on both sides of eq. \eqref{RaisingOperator_dminus2}. Utilizing eq. \eqref{RaisingOperator_IofII} on the left hand side and eq. \eqref{RaisingOperator_IIofII} on the right then leads us to
\begin{align}
\left(\partial_{d}^2 + m^2\right)\mathcal{D}_r G_{d-2}[t,r] = \frac{\delta[t]\delta[r]}{\Omega_{d-2} r^{d-2}} .
\end{align}
Once $G_{d-2}[t,r]$ is known, $\mathcal{D}_r G_{d-2}[t,r]$ would solve the $d$-dimensional massive Green's function equation -- this is equivalent to eq. \eqref{2DPlaneSource} by the translation invariance of space.\footnote{There is an analogous algebraic perspective that relates the massive scalar Green's function in $(d-2)$-de Sitter spacetime to that in $d$-de Sitter; see footnote 6 of \cite{Chu:2013xca}.}

{\bf Questions} \qquad That the minimally coupled massless scalar Green's function has non-zero tails in all odd dimensional Minkowski spacetimes, suggests spherical symmetry is not compatible with strictly-null propagation in the latter, since the Green's function is itself an outgoing spherical wave. Could operator methods, like the one deployed here, be used to sharpen/quantify this incompatibility? And could that lead, in turn, to insight into why is it that, while tails in 4D curved spacetimes are often associated with low frequency waves interacting with the curvature of the background geometry, no such interpretation appears to be available in the zero curvature of odd dimensional flat spacetimes?


\begin{thebibliography}{99}

\bibitem{Abbott:2016blz} 
B.~P.~Abbott {\it et al.} [LIGO Scientific and Virgo Collaborations],
``Observation of Gravitational Waves from a Binary Black Hole Merger,''
Phys.\ Rev.\ Lett.\  {\bf 116}, no. 6, 061102 (2016)
doi:10.1103/PhysRevLett.116.061102
[arXiv:1602.03837 [gr-qc]].

\bibitem{Abbott:2016nmj} 
B.~P.~Abbott {\it et al.} [LIGO Scientific and Virgo Collaborations],
``GW151226: Observation of Gravitational Waves from a 22-Solar-Mass Binary Black Hole Coalescence,''
Phys.\ Rev.\ Lett.\  {\bf 116}, no. 24, 241103 (2016)
doi:10.1103/PhysRevLett.116.241103
[arXiv:1606.04855 [gr-qc]].

\bibitem{Poisson:2011nh} 
E.~Poisson, A.~Pound and I.~Vega,
``The Motion of point particles in curved spacetime,''
Living Rev.\ Rel.\  {\bf 14}, 7 (2011)
doi:10.12942/lrr-2011-7
[arXiv:1102.0529 [gr-qc]].

\bibitem{Hadamard}
J. Hadamard, 
``Lectures on Cauchy's problem: In linear partial differential equations," 
Dover Publications (2003)

\bibitem{Casals:2016qyj} 
M.~Casals and B.~Nolan,
``Global Hadamard form for the Green Function in Schwarzschild space-time,''
arXiv:1606.03075 [gr-qc].

\bibitem{Zel’Dovich1974}
Zel'Dovich Y B and Polnarev A G 1974 Astron. Zh. 51 30 [Sov. Astron. 18, 17 (1974)]

\bibitem{Braginsky1985}
Braginsky V B and Grishchuk L P 1985 Zh. Eksp. Teor. Fiz. 89 744–750 [Sov. Phys. JETP 62, 427 (1985)]

\bibitem{BraginskyThorne}
V.B.~Braginsky and K.S.~Thorne, Nature (London) {\bf 327}, 123 (1987)

\bibitem{Christodoulou:1991cr} 
D.~Christodoulou,
``Nonlinear nature of gravitation and gravitational wave experiments,''
Phys.\ Rev.\ Lett.\  {\bf 67}, 1486 (1991).
doi:10.1103/PhysRevLett.67.1486

\bibitem{Thorne1992}
K.S.~Thorne,
``Gravitational-wave bursts with memory: The Christodoulou effect"
Phys.\ Rev.\ D {\bf 45}, 520 (1992)

\bibitem{Chu:2015yua} 
Y.~Z.~Chu,
``Transverse traceless gravitational waves in a spatially flat FLRW universe: Causal structure from dimensional reduction,''
Phys.\ Rev.\ D {\bf 92}, no. 12, 124038 (2015)
doi:10.1103/PhysRevD.92.124038
[arXiv:1504.06337 [gr-qc]].

\bibitem{Lifshitz1946}
E.M.~Lifshitz E M (1946) J. Phys. (USSR), 10, 116

\bibitem{Mukhanov:2005sc} 
V.~Mukhanov,
``Physical Foundations of Cosmology,''
Cambridge University Press (2005)

\bibitem{Weinberg:2008zzc} 
S.~Weinberg,
``Cosmology,''
Oxford, UK: Oxford Univ. Pr. (2008) 593 p

\bibitem{Pasterski:2015tva} 
S.~Pasterski, A.~Strominger and A.~Zhiboedov,
``New Gravitational Memories,''
JHEP {\bf 1612}, 053 (2016)
doi:10.1007/JHEP12(2016)053
[arXiv:1502.06120 [hep-th]].

\bibitem{Chu:2016qxp} 
Y.~Z.~Chu,
``Gravitational Wave Memory In dS$_{4+2n}$ and 4D Cosmology,''
Class.\ Quant.\ Grav.\  {\bf 34}, no. 3, 035009 (2017)
doi:10.1088/1361-6382/34/3/035009
[arXiv:1603.00151 [gr-qc]].

\bibitem{Bieri:2015jwa} 
L.~Bieri, D.~Garfinkle and S.~T.~Yau,
``Gravitational wave memory in de Sitter spacetime,''
Phys.\ Rev.\ D {\bf 94}, no. 6, 064040 (2016)
doi:10.1103/PhysRevD.94.064040
[arXiv:1509.01296 [gr-qc]].

\bibitem{Bieri:2013ada} 
L.~Bieri and D.~Garfinkle,
``Perturbative and gauge invariant treatment of gravitational wave memory,''
Phys.\ Rev.\ D {\bf 89}, no. 8, 084039 (2014)
doi:10.1103/PhysRevD.89.084039
[arXiv:1312.6871 [gr-qc]].

\bibitem{Ashtekar:2015lxa} 
A.~Ashtekar, B.~Bonga and A.~Kesavan,
``Asymptotics with a positive cosmological constant: III. The quadrupole formula,''
Phys.\ Rev.\ D {\bf 92}, no. 10, 104032 (2015)
doi:10.1103/PhysRevD.92.104032
[arXiv:1510.05593 [gr-qc]].

\bibitem{Date:2015kma} 
G.~Date and S.~J.~Hoque,
``Gravitational waves from compact sources in a de Sitter background,''
Phys.\ Rev.\ D {\bf 94}, no. 6, 064039 (2016)
doi:10.1103/PhysRevD.94.064039
[arXiv:1510.07856 [gr-qc]].

\bibitem{Kehagias:2016zry} 
A.~Kehagias and A.~Riotto,
``BMS in Cosmology,''
JCAP {\bf 1605}, no. 05, 059 (2016)
doi:10.1088/1475-7516/2016/05/059
[arXiv:1602.02653 [hep-th]].

\bibitem{Strominger:2014pwa} 
A.~Strominger and A.~Zhiboedov,
``Gravitational Memory, BMS Supertranslations and Soft Theorems,''
JHEP {\bf 1601}, 086 (2016)
doi:10.1007/JHEP01(2016)086
[arXiv:1411.5745 [hep-th]].

\bibitem{Iliopoulos:1998wq} 
J.~Iliopoulos, T.~N.~Tomaras, N.~C.~Tsamis and R.~P.~Woodard,
``Perturbative quantum gravity and Newton's law on a flat Robertson-Walker background,''
Nucl.\ Phys.\ B {\bf 534}, 419 (1998)
doi:10.1016/S0550-3213(98)00528-8
[gr-qc/9801028].

\bibitem{Tolish:2014oda} 
A.~Tolish, L.~Bieri, D.~Garfinkle and R.~M.~Wald,
``Examination of a simple example of gravitational wave memory,''
Phys.\ Rev.\ D {\bf 90}, no. 4, 044060 (2014)
doi:10.1103/PhysRevD.90.044060
[arXiv:1405.6396 [gr-qc]].

\bibitem{Wong:1970fu} 
S.~K.~Wong,
``Field and particle equations for the classical Yang-Mills field and particles with isotopic spin,''
Nuovo Cim.\ A {\bf 65}, 689 (1970).
doi:10.1007/BF02892134

\bibitem{Nariai} 
H.~Nariai,
``On the Green's Function in an Expanding Universe and Its Role in the Problem of Mach's Principle,''
Prog.\ Theor.\ Phys.\  {\bf 40}, 49 (1968).

\bibitem{Chu:2013xca} 
Y.~Z.~Chu,
``A line source in Minkowski for the de Sitter spacetime scalar Green’s function: massive case,''
Class.\ Quant.\ Grav.\  {\bf 32}, no. 13, 135008 (2015)
doi:10.1088/0264-9381/32/13/135008
[arXiv:1310.2939 [gr-qc]].

\bibitem{Dubovsky:2011sj} 
S.~Dubovsky, L.~Hui, A.~Nicolis and D.~T.~Son,
``Effective field theory for hydrodynamics: thermodynamics, and the derivative expansion,''
Phys.\ Rev.\ D {\bf 85}, 085029 (2012)
doi:10.1103/PhysRevD.85.085029
[arXiv:1107.0731 [hep-th]].

\bibitem{Ballesteros:2012kv} 
G.~Ballesteros and B.~Bellazzini,
``Effective perfect fluids in cosmology,''
JCAP {\bf 1304}, 001 (2013)
doi:10.1088/1475-7516/2013/04/001
[arXiv:1210.1561 [hep-th]].

\bibitem{Andersson:2006nr} 
N.~Andersson and G.~L.~Comer,
``Relativistic fluid dynamics: Physics for many different scales,''
Living Rev.\ Rel.\  {\bf 10}, 1 (2007)
doi:10.12942/lrr-2007-1
[gr-qc/0605010].

\bibitem{Bardeen:1980kt} 
J.~M.~Bardeen,
``Gauge Invariant Cosmological Perturbations,''
Phys.\ Rev.\ D {\bf 22}, 1882 (1980).
doi:10.1103/PhysRevD.22.1882

\bibitem{PrivateCommunicationTW}
Private communication with Alexander Tolish and Robert M. Wald, February 2017.

\bibitem{Jackson:2002rj} 
J.~D.~Jackson,
``From Lorenz to Coulomb and other explicit gauge transformations,''
Am.\ J.\ Phys.\  {\bf 70}, 917 (2002)
doi:10.1119/1.1491265
[physics/0204034].

\bibitem{G&S}
I.~S.~Gradshteyn and I.~M.~Ryzhik, 
``Table of Integrals, Series, and Products" 
Edited by A. Jeffrey and D. Zwillinger, Academic Press, New York, 7th edition, 2007

\bibitem{Bieri:2013hqa} 
L.~Bieri and D.~Garfinkle,
``An electromagnetic analogue of gravitational wave memory,''
Class.\ Quant.\ Grav.\  {\bf 30}, 195009 (2013)
doi:10.1088/0264-9381/30/19/195009
[arXiv:1307.5098 [gr-qc]].

\bibitem{Susskind:2015hpa} 
L.~Susskind,
``Electromagnetic Memory,''
arXiv:1507.02584 [hep-th].

\bibitem{DeWitt:1960fc} 
B.~S.~DeWitt and R.~W.~Brehme,
``Radiation damping in a gravitational field,''
Annals Phys.\  {\bf 9}, 220 (1960).
doi:10.1016/0003-4916(60)90030-0

\bibitem{Burko:2002ge} 
L.~M.~Burko, A.~I.~Harte and E.~Poisson,
``Mass loss by a scalar charge in an expanding universe,''
Phys.\ Rev.\ D {\bf 65}, 124006 (2002)
doi:10.1103/PhysRevD.65.124006
[gr-qc/0201020].

\bibitem{Mathematica}
Wolfram Research, Inc., Mathematica, Version 10.3.1.0, Champaign, IL (2015).

\bibitem{xAct}
http://www.xact.es/

\bibitem{Flanagan:2005yc} 
E.~E.~Flanagan and S.~A.~Hughes,
``The Basics of gravitational wave theory,''
New J.\ Phys.\  {\bf 7}, 204 (2005)
doi:10.1088/1367-2630/7/1/204
[gr-qc/0501041].

\bibitem{Bern:2008qj} 
Z.~Bern, J.~J.~M.~Carrasco and H.~Johansson,
``New Relations for Gauge-Theory Amplitudes,''
Phys.\ Rev.\ D {\bf 78}, 085011 (2008)
doi:10.1103/PhysRevD.78.085011
[arXiv:0805.3993 [hep-ph]].

\bibitem{Kawai:1985xq} 
H.~Kawai, D.~C.~Lewellen and S.~H.~H.~Tye,
``A Relation Between Tree Amplitudes of Closed and Open Strings,''
Nucl.\ Phys.\ B {\bf 269}, 1 (1986).
doi:10.1016/0550-3213(86)90362-7

\bibitem{Luna:2016due} 
A.~Luna, R.~Monteiro, I.~Nicholson, D.~O'Connell and C.~D.~White,
``The double copy: Bremsstrahlung and accelerating black holes,''
JHEP {\bf 1606}, 023 (2016)
doi:10.1007/JHEP06(2016)023
[arXiv:1603.05737 [hep-th]].

\bibitem{Luna:2015paa} 
A.~Luna, R.~Monteiro, D.~O'Connell and C.~D.~White,
``The classical double copy for Taub–NUT spacetime,''
Phys.\ Lett.\ B {\bf 750}, 272 (2015)
doi:10.1016/j.physletb.2015.09.021
[arXiv:1507.01869 [hep-th]].

\bibitem{Monteiro:2014cda} 
R.~Monteiro, D.~O'Connell and C.~D.~White,
``Black holes and the double copy,''
JHEP {\bf 1412}, 056 (2014)
doi:10.1007/JHEP12(2014)056
[arXiv:1410.0239 [hep-th]].

\bibitem{Ridgway:2015fdl} 
A.~K.~Ridgway and M.~B.~Wise,
``Static Spherically Symmetric Kerr-Schild Metrics and Implications for the Classical Double Copy,''
Phys.\ Rev.\ D {\bf 94}, no. 4, 044023 (2016)
doi:10.1103/PhysRevD.94.044023
[arXiv:1512.02243 [hep-th]].

\bibitem{Harte:2016vwo} 
A.~I.~Harte and J.~Vines,
``Generating exact solutions to Einstein’s equation using linearized approximations,''
Phys.\ Rev.\ D {\bf 94}, no. 8, 084009 (2016)
doi:10.1103/PhysRevD.94.084009
[arXiv:1608.04359 [gr-qc]].

\bibitem{Harte:2014ooa} 
A.~I.~Harte,
``Taming the Nonlinearity of the Einstein Equation,''
Phys.\ Rev.\ Lett.\  {\bf 113}, no. 26, 261103 (2014)
doi:10.1103/PhysRevLett.113.261103
[arXiv:1409.4674 [gr-qc]].

\bibitem{Dai:2015rda} 
L.~Dai, E.~Pajer and F.~Schmidt,
``Conformal Fermi Coordinates,''
JCAP {\bf 1511}, no. 11, 043 (2015)
doi:10.1088/1475-7516/2015/11/043
[arXiv:1502.02011 [gr-qc]].

\bibitem{Pajer:2013ana} 
E.~Pajer, F.~Schmidt and M.~Zaldarriaga,
``The Observed Squeezed Limit of Cosmological Three-Point Functions,''
Phys.\ Rev.\ D {\bf 88}, no. 8, 083502 (2013)
doi:10.1103/PhysRevD.88.083502
[arXiv:1305.0824 [astro-ph.CO]].

\bibitem{Pfenning:2000zf} 
M.~J.~Pfenning and E.~Poisson,
``Scalar, electromagnetic, and gravitational selfforces in weakly curved space-times,''
Phys.\ Rev.\ D {\bf 65}, 084001 (2002)
doi:10.1103/PhysRevD.65.084001
[gr-qc/0012057].

\bibitem{Chu:2011ip} 
Y.~Z.~Chu and G.~D.~Starkman,
``Retarded Green's Functions In Perturbed Spacetimes For Cosmology and Gravitational Physics,''
Phys.\ Rev.\ D {\bf 84}, 124020 (2011)
doi:10.1103/PhysRevD.84.124020
[arXiv:1108.1825 [astro-ph.CO]].

\bibitem{SoodakTiersten}
H.~Soodak and M.~S.~Tiersten,
``Wakes and waves in N dimensions," 
Am. J. Phys. {\bf 61}, 395 (1993)

\bibitem{Higuchi:2008fu} 
A.~Higuchi and Y.~C.~Lee,
``How to use retarded Green's functions in de Sitter spacetime,''
Phys.\ Rev.\ D {\bf 78}, 084031 (2008)
doi:10.1103/PhysRevD.78.084031
[arXiv:0808.0642 [gr-qc]].

\bibitem{DiFrancesco:1997nk} 
P.~Di Francesco, P.~Mathieu and D.~Senechal,
``Conformal Field Theory,''
doi:10.1007/978-1-4612-2256-9

\end{thebibliography}
\end{document}